\documentclass[twocolumn, twocolappendix]{aastex631}

\usepackage{tabularx}
\usepackage{amsmath}
\usepackage{tablefootnote}
\usepackage{placeins}

\newcommand{\lya}{Ly$\alpha$}

\newcommand{\HI}{\rm H\,{\textsc {i}}}
\newcommand{\HII}{\rm H\,{\textsc {ii}}}

\newcommand{\MgII}{\rm Mg\,{\textsc {ii}}}

\begin{document}

\title{Synergistic Radiative Transfer Modeling of $\rm Mg\,{\textsc {II}}$ and \lya\ Emission in Multiphase, Clumpy Galactic Environments: Application to Low-Redshift Lyman Continuum Leakers}

\author[0000-0001-5113-7558]{Zhihui Li}
\affiliation{Center for Astrophysical Sciences, Department of Physics \& Astronomy, Johns Hopkins University, Baltimore, MD 21218, USA}
\affiliation{Cahill Center for Astronomy and Astrophysics, California Institute of Technology, 1200 E California Blvd, MC 249-17, Pasadena, CA 91125, USA}
\email{zli367@jh.edu}

\author[0000-0003-2491-060X]{Max Gronke}
\affiliation{Max-Planck Institute for Astrophysics, Karl-Schwarzschild-Str. 1, D-85741 Garching, Germany}

\author[0000-0001-6670-6370]{Timothy Heckman}
\affiliation{Center for Astrophysical Sciences, Department of Physics \& Astronomy, Johns Hopkins University, Baltimore, MD 21218, USA}

\author[0000-0002-9217-7051]{Xinfeng Xu}
\affiliation{Department of Physics and Astronomy, Northwestern University, 2145 Sheridan Road, Evanston, IL, 60208, USA}
\affiliation{Center for Interdisciplinary Exploration and Research in Astrophysics (CIERA), Northwestern University, 1800 Sherman Avenue, Evanston, IL, 60201, USA}

\author[0000-0002-6586-4446]{Alaina Henry}
\affiliation{Center for Astrophysical Sciences, Department of Physics \& Astronomy, Johns Hopkins University, Baltimore, MD 21218, USA}
\affiliation{Space Telescope Science Institute, 3700 San Martin Drive, Baltimore, MD 21218, USA}

\author[0000-0003-4166-2855]{Cody Carr}
\affiliation{Center for Cosmology and Computational Astrophysics, Institute for Advanced Study in Physics, Zhejiang University, Hangzhou 310058,
China}
\affiliation{Institute of Astronomy, School of Physics, Zhejiang University, Hangzhou 310058, China}

\author[0000-0002-0302-2577]{John Chisholm}
\affiliation{Department of Astronomy, The University of Texas at Austin, 2515 Speedway, Stop C1400, Austin, TX 78712, USA}

\author[0000-0002-2724-8298]{Sanchayeeta Borthakur}
\affiliation{School of Earth \& Space Exploration, Arizona State University, Tempe, AZ 85287, USA}

\author[0000-0001-8442-1846]{Rui Marques-Chaves}
\affiliation{Department of Astronomy, University of Geneva, 51 Chemin Pegasi, 1290 Versoix, Switzerland}

\author[0000-0001-7144-7182]{Daniel Schaerer}
\affiliation{Department of Astronomy, University of Geneva, 51 Chemin Pegasi, 1290 Versoix, Switzerland}
\affiliation{CNRS, IRAP, 14 Avenue E. Belin, 31400 Toulouse, France}

\author[0000-0002-6085-5073]{Floriane Leclercq}
\affiliation{Université Lyon, Université Lyon 1, ENS de Lyon, CNRS, Centre de Recherche Astrophysique de Lyon UMR 5574, 69230 Saint-Genis-Laval,
France}

\author[0000-0002-4153-053X]{Danielle A. Berg}
\affiliation{Department of Astronomy, The University of Texas at Austin, 2515 Speedway, Stop C1400, Austin, TX 78712, USA}



\begin{abstract}

We conducted systematic radiative transfer (RT) modeling of the \MgII\ doublet line profiles for 33 low-redshift Lyman continuum (LyC) leakers, and \lya\ modeling for a subset of six objects, using a multiphase, clumpy circumgalactic medium (CGM) model. Our RT models successfully reproduced the \MgII\ line profiles for all 33 galaxies, revealing a necessary condition for strong LyC leakage: high maximum clump outflow velocity ($v_{\rm MgII,\,max} \gtrsim 390\,\rm km\,s^{-1}$) and low total \MgII\ column density ($N_{\rm MgII,\,tot} \lesssim 10^{14.3}\,\rm cm^{-2}$). We found that the clump outflow velocity and total \MgII\ column density have the most significant impact on \MgII\ spectra and emphasized the need for full RT modeling to accurately extract the CGM gas properties. In addition, using archival HST COS/G160M data, we modeled \lya\ profiles for six objects and found that their spectral properties do not fully align with the conventional LyC leakage criteria, yet no clear correlation was identified between the modeled parameters and observed LyC escape fractions. We inferred LyC escape fractions based on \HI\ properties from \lya\ RT modeling and found that LyC leakage is primarily governed by the number of optically thick \HI\ clumps per sightline ($f_{\rm cl}$). Intriguingly, two galaxies with relatively low observed LyC leakage exhibited the highest RT-inferred LyC escape fractions due to their lowest $f_{\rm cl}$ values, driven by the strong blue peaks of their \lya\ emission. Future high-resolution, spatially resolved observations are crucial for resolving this puzzle. Overall, our results support a ``picket fence'' geometry over a ``density-bounded'' scenario for the CGM, where a combination of high \MgII\ outflow velocities and low \MgII\ column densities may be correlated with the presence of more low-density \HI\ channels that facilitate LyC escape. 

\end{abstract}

\keywords{Circumgalactic medium (1879) --- Interstellar medium (847) --- Galactic winds (572) --- Reionization (1383) -- Ultraviolet spectroscopy (2284)}


\section{Introduction} \label{sec:intro}

In recent years, the study of the \MgII\ $\rm \lambda\lambda$2796, 2803 doublet has emerged as a prominent research topic. Similar to \lya, \MgII\ is a resonant line with an ionization potential comparable to \HI\ ($\sim$15.0 eV and 13.6 eV, respectively). Both \MgII\ and \lya\ are valuable for probing the properties of the ``cool'' ($T \sim 10^4$\,K), neutral gas phase in the circumgalactic medium (CGM), yet the optical depths experienced by \MgII\ photons are typically much lower than those of \lya, due to the generally low Mg abundance ([Mg/H] $\sim 10^{-5}$ in the Milky Way, \citealt{Jenkins09}). Since neutral gas regulates the escape of ionizing photons from galaxies, \MgII\ serves as an excellent complementary tracer to \lya\ in understanding this process, offering valuable insights into the epoch of reionization.

To date, the \MgII\ doublet has been observed in a wide range of galaxies \citep[e.g.,][]{Weiner2009, Rubin2010, Rubin2011, Giavalisco2011, Martin2012, Erb2012, Kornei2013, Rigby2014, Schroetter2015, Finley2017, Feltre2018, Huang2021, Xu2022, Xu2023}. In addition, recent advancements in integral field unit (IFU) spectrograph technology have made it possible to measure spatially resolved \MgII\ emission within the CGM of galaxies \citep[e.g.,][]{Rupke2019, Chisholm2020, Burchett2021, Zabl2021, Shaban2022, Seive2022, Dutta2023, Pessa2024}. A number of studies have explored the use of the \MgII\ doublet as a probe for LyC escape, particularly through empirical indicators such as equivalent width ratios or line ratios of the \MgII\ doublet (e.g., \citealt{Henry18, Chisholm2020, Izotov2022, Seive2022, Xu2023}). Other theoretical works have utilized idealized models or cosmological simulations to examine the radiative transfer (RT) process of \MgII\ emission \citep[e.g.,][]{Prochaska2011, Burchett2021, Chang24, Seon24, Carr24}. However, up to now, there has been little systematic effort to reproduce observed \MgII\ line profiles for a large galaxy sample using RT models that realistically represent the physical conditions of galactic environments.

In this work, we perform systematic RT modeling for the \MgII\ doublet line profiles of 33 low-$z$ LyC leakers, drawn from the Low-redshift Lyman Continuum Survey (LzLCS; \citealt{Flury22}). We employ a multiphase, clumpy CGM model to reproduce the \MgII\ line profiles presented in \citet{Xu2023}, aiming to accurately determine the underlying CGM gas properties. Moreover, we apply the same CGM model to the \lya\ emission line profiles for a subset of the sample, leveraging archival HST COS/G160M data. By comparing the results from both \MgII\ and \lya\ modeling, we seek to gain new insights into the relationship between \MgII\ and \lya\ emission and their connection to the LyC leakage in low-$z$ LyC leakers.

The structure of this paper is as follows. In Section \ref{sec:mgII_RT}, we introduce the RT model used to analyze the \MgII\ doublet emission line profiles. Section \ref{sec:classification} provides a brief summary of the observational data for \MgII\ emission and LyC leakage. In Section \ref{sec:mgII_modeling}, we present our \MgII\ RT modeling results and demonstrate the effect of each individual parameter on the \MgII\ spectra. Section \ref{sec:lya_modeling} models the \lya\ profiles for a subset of our sample using the same CGM model and infers their LyC leakage. In Section \ref{sec:comparison}, we discuss the implications of our modeling results and compare with previous work. Finally, we summarize and conclude in Section \ref{sec:conclusion}.

\section{Radiative Transfer Modeling of Mg\,{\textsc {ii}} Emission}\label{sec:mgII_RT}

We model the \MgII\ doublet emission by adapting the 3D \lya\ Monte Carlo RT code, \texttt{tlac} \citep{Gronke14}. The atomic fine structures of the \MgII\ and \lya\ doublet emission are intrinsically very similar, except that \MgII\ has a much larger, non-negligible level splitting. Following convention, we will refer to the 2796 \AA\ transition as K and the 2803 \AA\ transition as H for the \MgII\ doublet in the rest of this work. To model the \MgII\ doublet emission, two major modifications to the code have been implemented: the gas scattering cross section (see Eq. 1 in \citealt{Chang24}) and the expression for the parallel velocity component of the scattering atoms (see Eqs. 5–7 in \citealt{Seon24}). For further technical details, we refer interested readers to recent theoretical studies (e.g., \citealt{Chang24, Seon24}). Additionally, the physical constants for \lya\ should be replaced with those for \MgII\ (such as atomic mass and Einstein coefficients).

\subsection{Summary of Major $\rm Mg\,{\textsc {ii}}$ Emission Mechanisms}

In general, there are four primary mechanisms responsible for \MgII\ doublet emission in the ISM / CGM \citep[e.g.,][]{Burchett2021, Chang24, Seon24}:

(1) Recombination in central \HII\ regions: High-energy UV photons ($\gtrsim$ 15 eV) from massive stars (e.g., O and B types) can doubly ionize magnesium atoms, creating Mg$^{2+}$ ions. These ions can then recombine to produce \MgII\ doublet emission, exhibiting a doublet ratio ($F_{\rm 2796} / F_{\rm 2803}$) of approximately 2:1.

(2) Collisional excitation: Mg$^{+}$ ions can be collisionally excited by electrons near the \HII\ regions, followed by de-excitation, which results in \MgII\ doublet emission. This process also yields a doublet ratio close to 2:1.

(3) Continuum pumping: UV continuum photons near 2800 \AA\ can excite Mg$^{+}$ ions, leading to de-excitation and subsequent \MgII\ emission. In this scenario, the spectrum often displays a P-Cygni profile, characterized by an absorption trough and an emission peak atop the continuum.

(4) Ionization by diffuse FUV radiation: FUV continuum radiation in the diffuse warm neutral medium (WNM) at around 1620 \AA\ can ionize magnesium atoms to form Mg$^{+}$ ions. These ions, when collisionally excited and then de-excited by electrons, produce \MgII\ doublet emission, again with a doublet ratio close to 2:1.

In this work, we focus primarily on the first three mechanisms that produce \MgII\ doublet emission from the central galaxy within the ISM, which either generate line emission with a 2:1 ratio or continuum emission. Our subsequent modeling suggests that such a choice is sufficient to reproduce nearly all of the observed \MgII\ emission spectra. Therefore, in line with Occam’s Razor, we believe that while additional mechanisms may contribute, their impact is likely minor.

\subsection{Configuration of the RT Model}

We model a multiphase, clumpy gaseous medium that resembles the physical structure of the CGM of low-$z$ LyC leakers, as illustrated in Figure 1.  We assume that \MgII\ and \lya\ emission are produced by the central galaxy and propagate outward through such a two-phase CGM, which consists of cool ($\sim 10^4 \,\rm K$), outflowing clumps and a hot ($\sim 10^6 \,\rm K$), diffuse, outflowing inter-clump medium (ICM).  The Mg$^+$ ions are assumed to reside solely in the cool clumps, as they are likely to be fully ionized in the hot medium. In contrast, \HI\ atoms are assumed to exist both in the clumps, with high column densities ($\gtrsim 10^{17}\,\rm cm^{-2}$), and in the ICM, with much lower column densities ($\sim 10^{15}\,\rm cm^{-2}$). We note that although in reality, Mg$^+$ and \HI\ may be relatively well-mixed in the cool clumps, their properties (such as outflow velocities and column densities) will be modeled independently -- we will focus on \MgII\ RT in Section \ref{sec:mgII_modeling} and \lya\ RT in Section \ref{sec:lya_modeling}, respectively.

In each model, we further assume that both the clumps and the CGM halo are spherical, and set the clump radius to $R_{\rm cl} = 100$ pc and halo radius to $R_{\rm halo} = 10$ kpc\footnote{Note that these physical sizes are generally rescalable in idealized RT simulations.}, based on the typical physical extent of the observed \lya\ and \MgII\ emission for this sample \citep{Henry15, Henry18}. The clumps are distributed following a power-law $n_{\rm cl} \propto r^{-2}$, with the total number determined by the clump volume filling factor, $F_{\rm V}$. This volume filling factor can be converted into the clump covering factor (i.e., the average number of clumps per sightline), $f_{\rm cl}$, using the relation $f_{\rm cl} = \frac{3}{4} F_{\rm V} R_{\rm halo}/{R_{\rm cl}}$ \citep{Li24}. In our standard model, the clumps are assumed to have a constant \MgII\ column density, $N_{\rm MgII,\,cl}$, and the average total \MgII\ column density along a given sightline is $N_{\rm MgII,\,tot} = \frac{4}{3} f_{\rm cl} N_{\rm MgII,\,cl}$ \citep{Dijkstra2012, Gronke16_model}.

\begin{figure}
\centering
\includegraphics[width=0.50\textwidth]{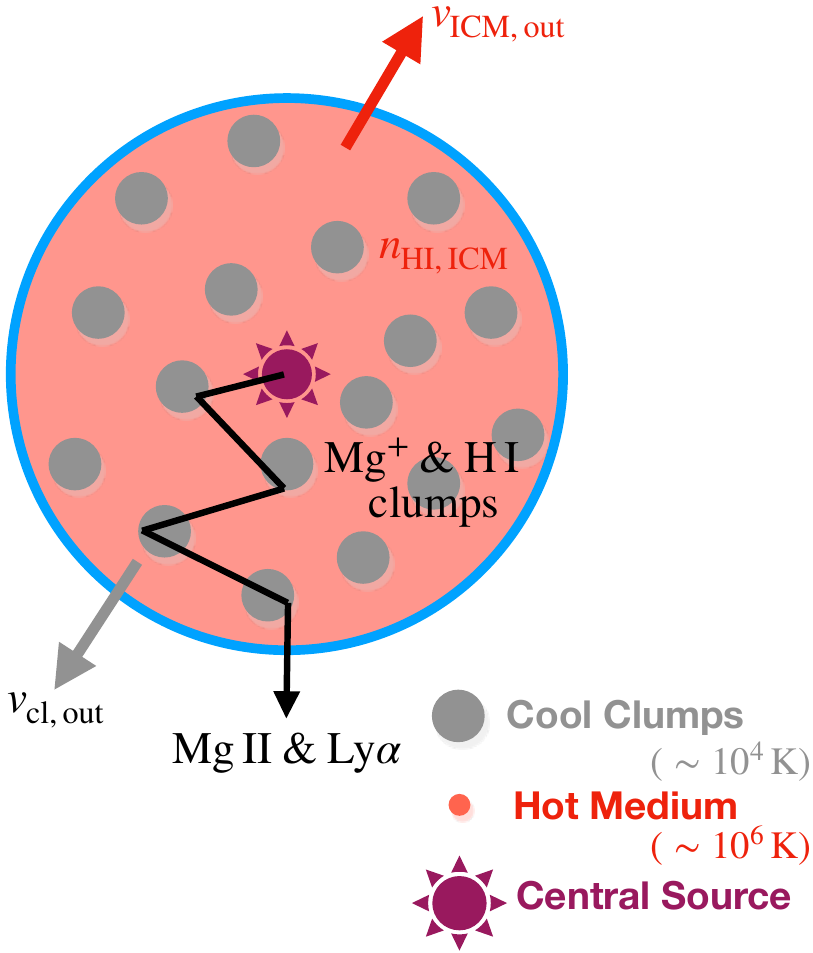}
    \caption{\textbf{Schematic of the multiphase, clumpy RT model.} \MgII\ and \lya\ emission are assumed to originate from the central galaxy and propagate outward through a two-phase CGM, consisting of cool ($\sim 10^4\,\rm K$) outflowing clumps and a hot ($\sim 10^6\,\rm K$), diffuse, outflowing inter-clump medium (ICM). The Mg$^{+}$ ions are confined to the cool clumps, whereas \HI\ atoms are present in both the clumps and the ICM. The properties of Mg$^+$ and \HI\ gas are constrained separately via \MgII\ and \lya\ RT modeling (see Section \ref{sec:mgII_modeling} and \ref{sec:lya_modeling}, respectively). }
    \label{fig:schematic}
\end{figure}

The motion of the clumps is characterized by two modes: a microscopic turbulent motion\footnote{Such internal turbulence can result from momentum transfer from the hot, ambient medium (see e.g., \citealt{Nikolis2024}).}, represented by the clump Doppler parameter $b_{\rm D,\,cl}$, and a macroscopic radial outflow. For the latter mode, instead of using simplistic velocity functions (e.g., linear or power-law), we adopt a realistic clump velocity profile where the clumps are accelerated by an $r^{-\alpha}$ force and decelerated by the gravitational pull from an NFW-type dark matter halo. Specifically, the kinematic equation governing an outflowing clump is given by:

\begin{equation}
\frac{\mathrm{d}v_{\rm cl,\,out}(r)}{\mathrm{d}t}=-\frac{GM(r)}{r^2} + Ar^{-\alpha}
\label{eq:clump_momentum}
\end{equation}

The solution to this equation is (see \citealt{Li24} for details):

\begin{align}
v_{\rm cl,\,out}(r) = & \Bigg\{\frac{2GM_{\rm vir}}{{\rm ln}(1 + c) - \frac{c}{1 + c}} 
\left[\frac{{\rm ln}(1+r/r_{\rm s})}{r} - \frac{{\rm ln}(1+r_{\rm min}/r_{\rm s})}{r_{\rm min}}\right] \nonumber \\
& + \mathcal{V}^2_{\infty} \left[1 - \left(\frac{r}{r_{\rm min}}\right)^{1-\alpha}\right] \Bigg\}^{1/2}
\label{eq:v}
\end{align}
where $M_{\rm vir}$ is the virial mass of the dark matter halo, $r_{\rm vir}$ is the virial radius, $c$ is the halo concentration parameter, $r_{\rm s} = r_{\rm vir} / c$ is the halo scale radius, $r_{\rm min}$ is the clump launch radius (we have fixed it at 1 kpc), and $\mathcal{V_{\infty}}$ is the asymptotic outflow velocity in the absence of gravitational deceleration. For simplicity, in our modeling, we adopt a typical dark matter halo mass of $M_{\rm vir} = 10^{11} M_{\odot}$ and an average redshift of $z = 0.2$ for our sample. Consequently, the outflow velocity profile depends on only two free parameters: $\mathcal{V_{\infty}}$ and $\alpha$. Varying combinations of $\mathcal{V_{\infty}}$ and $\alpha$ yield different velocity profiles and clump maximum outflow velocities ($v_{\rm cl,\,max}$). We present several example $v_{\rm cl,\,out}(r)$ profiles in Appendix \ref{sec:appendix}.

In addition to the resonant scattering of \MgII, we account for the effect of dust absorption and scattering within the clumps. The dust scattering albedo near $\sim 2800$\,\AA\ is approximately $\alpha_{\rm d}$ = 0.57, with an asymmetry factor of $g = 0.55$ in the Henyey-Greenstein scattering phase function \citep{Draine03}. We use the dust absorption optical depth in each clump, $\tau_{\rm d,\,cl}$, to characterize the amount of dust absorption\footnote{Note that the difference in the dust absorption optical depth of the K and H transitions are negligible. We have also accounted for the fact that the ratio of $\tau_{\rm d,\,cl}$ for \lya\ and \MgII\ is 9.26 in the SMC dust model (this ratio varies across different dust models; see \citealt{Chang24}).}. We also employ a fiducial Small Magellanic Cloud (SMC) dust model, as described by \citet{Draine03, Draine03b}.

\begin{table*}
\centering
    \scriptsize \caption{Parameter values of the fiducial model grid used for \MgII\ RT modeling.}
    \label{tab:params}
    \setlength{\tabcolsep}{1pt}
    \begin{tabular}{ccc}
    \hline\hline
    Parameter & Definition & Values\\ 
     (1)  & (2) & (3)\\
    \hline
    $F_{\rm V}$ & Clump volume filling factor & (0.005, 0.01, 0.02, ..., 0.06)\\
     ${\rm log}\,N_{\rm MgII,\,cl}$ & Clump {\rm Mg\,{\textsc {ii}}} column density & (12.0, 12.5, ..., 14.5) log cm$^{-2}$\\
     $b_{\rm D,\,cl}$ & Clump Doppler parameter & (8, 15, 26, 47, 83)\tablenotemark{a} km\,s$^{-1}$ \\
     $\mathcal{V}_{\infty}$ & Clump asymptotic outflow velocity & (200, 400, 600, 800) km\,s$^{-1}$ \\
     $\alpha$ & Clump acceleration power-law index & (1.1, 1.5, 1.9) \\
     $R_{\rm line}$ & Line-to-continuum photon ratio & (0.1, 0.16, 0.25, 0.40, 0.63, 1.0)\\
     $\tau_{\rm d,\,cl}$ & Clump dust absorption optical depth & (0, 0.03, 0.05, 0.1)\\
     $b_{\rm max}$ & Maximum photon impact parameter& (0.5, 1, 1.5, 2, 4, ..., 10) kpc\\
     $\Delta v$ & Velocity shift relative to systemic $z$ & [-100,\,100] km\,s$^{-1}$  (continuous)\\
     $f_{\rm scale}$ & Continuum scaling factor & [0.8,\,1.2]  (continuous)\\
    \hline\hline
    \end{tabular}
    \tablenotetext{}{\textbf{Notes.} The parameter values of the fiducial model grid used for fitting the \MgII\ profiles. The columns are: (1) parameter name; (2) parameter definition; (3) parameter values on the grid.}
    \tablenotetext{a}{This parameter is varied in increments of 10$^{0.25}$ on the fiducial model grid.}
\end{table*}

\subsection{Additional Parameters \& Fitting Pipeline}

Before fitting the observed \MgII\ spectra with the RT models, we need to introduce several additional parameters. First of all, since most of the observed \MgII\ spectra exhibit positive equivalent widths (EW) that suggest net emission, we introduce the ratio of line photons to continuum photons, $R_{\rm line} = N_{\rm line}/N_{\rm continuum}$\footnote{$R_{\rm line}$ can be converted to an intrinsic emission EW using the following relation: ${\rm EW}_{\rm int,\,K+H} = R_{\rm line} 2\Delta v {\lambda_0} / c \simeq 28 R_{\rm line} \rm (\AA)$.}, to characterize the source function of the \MgII\ emission. For each model configuration, two separate RT simulations are performed -- one with 10,000 line photons (in a 2:1 ratio for the K and H transitions) and another with 10,000 continuum photons near 2800 \AA. In the first case, the intrinsic line emission is modeled as two Gaussian functions centered at the K and H line centers with $\sigma = 100\,\rm km\,s^{-1}$. In the second case, the intrinsic emission is assumed to be a flat continuum within $\Delta v = \pm 1500\,\rm km\,s^{-1}$ from the K line center\footnote{Here ``intrinsic'' refers to the line profiles emerging from the ISM, meaning that we essentially assume that all RT effects come from the outflowing gas in the CGM, while the ISM only contributes to the initial line broadening.}. Since the RT of each photon is independent, we can construct a composite model spectrum by combining the continuum photons with a fraction of the line photons, characterized by $R_{\rm line}$\footnote{Our subsequent modeling shows that all galaxies, except J1648+4957, require $R_{\rm line} < 1$. }. We also introduce an aperture correction factor to account for aperture losses, which may arise due to the limited slit size of spectrographs used in \MgII\ observations (see e.g., \citealt{Scarlata2015}). To incorporate this effect, we define a parameter, $b_{\rm max}$, representing the maximum impact parameter within which photons are included in constructing the model spectra\footnote{Note that our treatment of aperture loss is somewhat idealized, as the apertures of slit spectrographs are typically non-circular in reality.}.

In addition, we incorporate a velocity shift parameter, $\Delta v$, to account for any potential offset between the systemic redshift of \MgII\ emission in the model and the observed systemic redshift derived from nebular emission lines. We also include a scaling factor, $f_{\rm scale}$, to provide the model flexibility in matching the continuum level of the data. Our results show that $\Delta v$ is typically small and falls within the observational uncertainties, while $f_{\rm scale}$ is always close to 1, as both the models and the data are normalized prior to comparison. As a result, our modeling has 10 free parameters in total: the clump volume filling factor $F_{\rm V}$, the clump \MgII\ column density $N_{\rm MgII,\,cl}$, the clump Doppler parameter $b_{\rm D,\,cl}$, the clump asymptotic outflow velocity $\mathcal{V_{\infty}}$, the clump acceleration power-law index $\alpha$, the ratio of line photons to continuum photons $R_{\rm line}$, the photon maximum impact parameter $b_{\rm max}$, the clump dust absorption optical depth $\tau_{\rm d,\,cl}$, the velocity shift $\Delta v$, and the continuum scaling factor $f_{\rm scale}$. 

Our fitting pipeline utilizes the \texttt{python} nested sampling package \texttt{dynesty} \citep{Skilling04, Skilling06, Speagle20}. To avoid the computational expense of real-time RT calculations, the fitting process relies on a pre-computed grid of \MgII\ RT models. For each parameter space point visited, the model spectrum is computed through a parameter-weighted multi-dimensional linear interpolation the grid of \MgII\ RT model spectra. The resulting spectrum is then convolved with a Gaussian function (FWHM = $50\,\rm km\,s^{-1}$) to simulate the finite instrumental resolution before being compared to the observed \MgII\ spectrum\footnote{For the four objects observed with HET, the spectral resolution is relatively low \citep{Xu2023}. We therefore apply a convolution with FWHM = $125\,\rm km\,s^{-1}$ in these cases.}.

For each object, we start by performing a test fit using a fiducial model grid. After examining the posterior distribution, we expand the grid along different dimensions as needed for individual objects. This approach allows us to account for the wide range of best-fit parameters across the parameter space. By customizing the model grid for each object, we ensure that the best-fit parameter values remain within the initial prior bounds. We present the parameter values (i.e., prior ranges) for the fiducial model grid in Table \ref{tab:params}.

\section{Observational Data for Mg\,{\textsc {ii}} Emission and LyC Leakage}\label{sec:classification}

In this work, we use the \MgII\ spectra presented in \citet{Xu2023}, where high- to medium-resolution \MgII\ line profiles were obtained for 34 low-$z$ LyC leakers using MMT, VLT, and HET. After inspecting the \MgII\ profiles individually, we determined that 33 are reproducible by our current model, with one exception -- J0957+2357, which shows an unusual pure absorption line profile with the absorption trough located between the K and H transitions. For now, we have decided to exclude this object and focus on modeling the remaining 33 galaxies.

We further utilize the LyC flux measurements presented in \citet{Flury22}. While the LyC fluxes can be converted into LyC escape fractions, this process is inevitably model-dependent and subject to uncertainties. Therefore, for our purposes here, we use the original LyC flux measurements near $\lambda_{\rm rest} = 912\rm\, \AA$ and categorize the 33 galaxies into the following four groups:

(1) Strong leakers: LyC flux is detected with a significance greater than 4$\sigma$, and the flux ratio $F_{\rm \lambda LyC} / F_{\rm \lambda 1100} \geq 0.05$.

(2) Moderate leakers: LyC flux is detected with a significance greater than 4$\sigma$, but the flux ratio $F_{\rm \lambda LyC} / F_{\rm \lambda 1100} < 0.05$.

(3) Potential leakers: LyC flux is detected, but the detection significance is less than 4$\sigma$.

(4) Non-leakers: LyC flux is undetected, and only upper limits are determined.

\begin{figure*}
\centering
\includegraphics[width=0.80\textwidth]{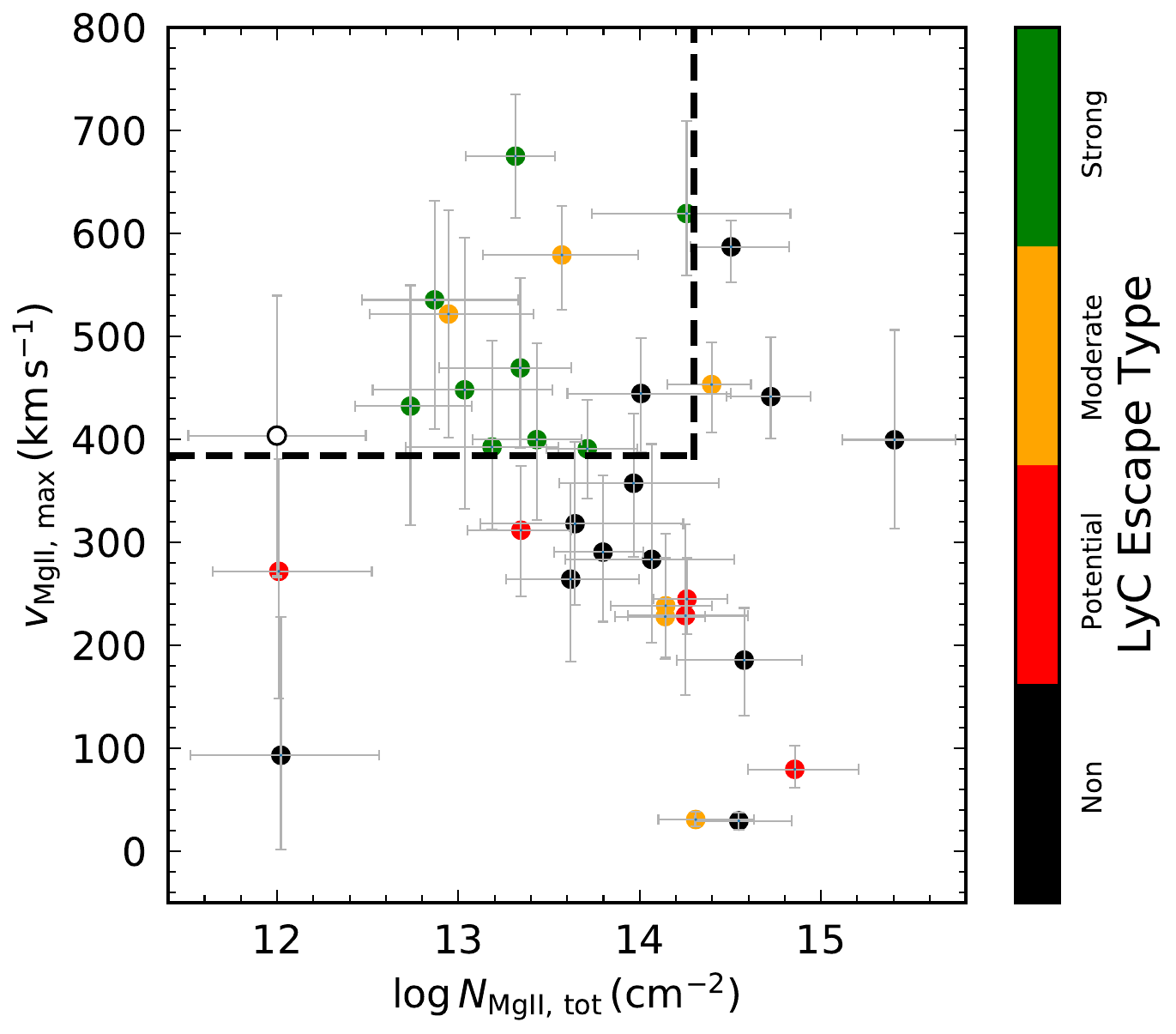}
    \caption{\textbf{Relation between the best-fit maximum clump radial outflow velocity and {\MgII} total column density inferred from {\MgII} emission RT modeling.} The different types of LyC leakers are color-coded as follows: black for non-leakers, red for potential leakers, orange for moderate leakers, and green for strong leakers (see our definition in Section \ref{sec:classification}). The strong LyC leakers all occupy the upper left corner, suggesting a necessary (though not sufficient) condition for a LyC leaker to be a strong leaker: a high maximum clump radial outflow velocity ($v_{\rm MgII,\,max} \gtrsim 390\,\rm km\,s^{-1}$) {\emph{and}} a low total $\rm Mg\,{\textsc {ii}}$ column density ($N_{\rm MgII,\,tot} \lesssim 10^{14.3}\,\rm cm^{-2}$).}
    \label{fig:vmax_mgII}
\end{figure*}

We believe the classification scheme used here is reliable, particularly for distinguishing between strong leakers and non-leakers. We note that \citet{Flury22} also derived two additional $f_{\rm LyC,\,esc}$ estimates based on either H$\beta$ or the UV SED of the galaxies. By our definition, a strong leaker has at least either $f_{\rm LyC,\,esc}(\rm H\beta)$ or $f_{\rm LyC,\,esc}(\rm UV) > 5\%$ (or both), whereas a non-leaker has upper limits for both $f_{\rm LyC,\,esc}(\rm H\beta)$ and $f_{\rm LyC,\,esc}(\rm UV)$. For moderate and potential leakers, the classification may vary depending on the $f_{\rm LyC,\,esc}$ metric used, but this does not affect the main result presented in the next section -- a necessary condition for a LyC leaker to be classified as a strong leaker.

Based on the criteria above, 9, 6, 5, and 13 objects fall into categories (1) through (4), respectively. In our subsequent analysis, we will model the observed \MgII\ spectra and attempt to establish connections between the inferred underlying gas properties and the LyC leakage characteristics for our sample.

\section{Results of $\rm Mg\,{\textsc {ii}}$ RT Modeling}\label{sec:mgII_modeling}

Our \MgII\ RT modeling has successfully reproduced the \MgII\ emission line profiles of nearly all 33 galaxies\footnote{One notable exception is J0826+1820, where the \MgII\ spectrum is too noisy to yield any reasonable constraints on the model parameters.} when the data quality is sufficient. The posterior distributions from the fitting indicate that most parameters are well-constrained within the priors, with two notable degeneracies: 
the anti-correlation between $\{F_{\rm V}, N_{\rm MgII,\,cl}\}$ and between $\{\mathcal{V}_{\infty}, \alpha\}$. The first degeneracy arises because primarily the total \MgII\ column density along a sightline is proportional to $F_{\rm V}N_{\rm MgII,\,cl}$, while the second is due to the competition between the acceleration and deceleration forces in the clumps' radial velocity profile. We therefore focus on two physical parameters that are not susceptible to these degeneracies from now on: the total \MgII\ column density $N_{\rm MgII,\,tot} (= F_{\rm V}N_{\rm MgII,\,cl}R_{\rm halo}/R_{\rm cl}$), and the maximum clump outflow velocity $v_{\rm MgII,\,max}$ (see Appendix \ref{sec:appendix}). 

We hereby present our \MgII\ RT modeling results and examine the differences in \MgII\ gas properties among various types of LyC leakers.

\subsection{The $v_{\rm MgII,\,max}$ -- $N_{\rm MgII,\,tot}$ Plane}\label{sec:plane}

\begin{figure*}
\centering
\includegraphics[width=0.328\textwidth]{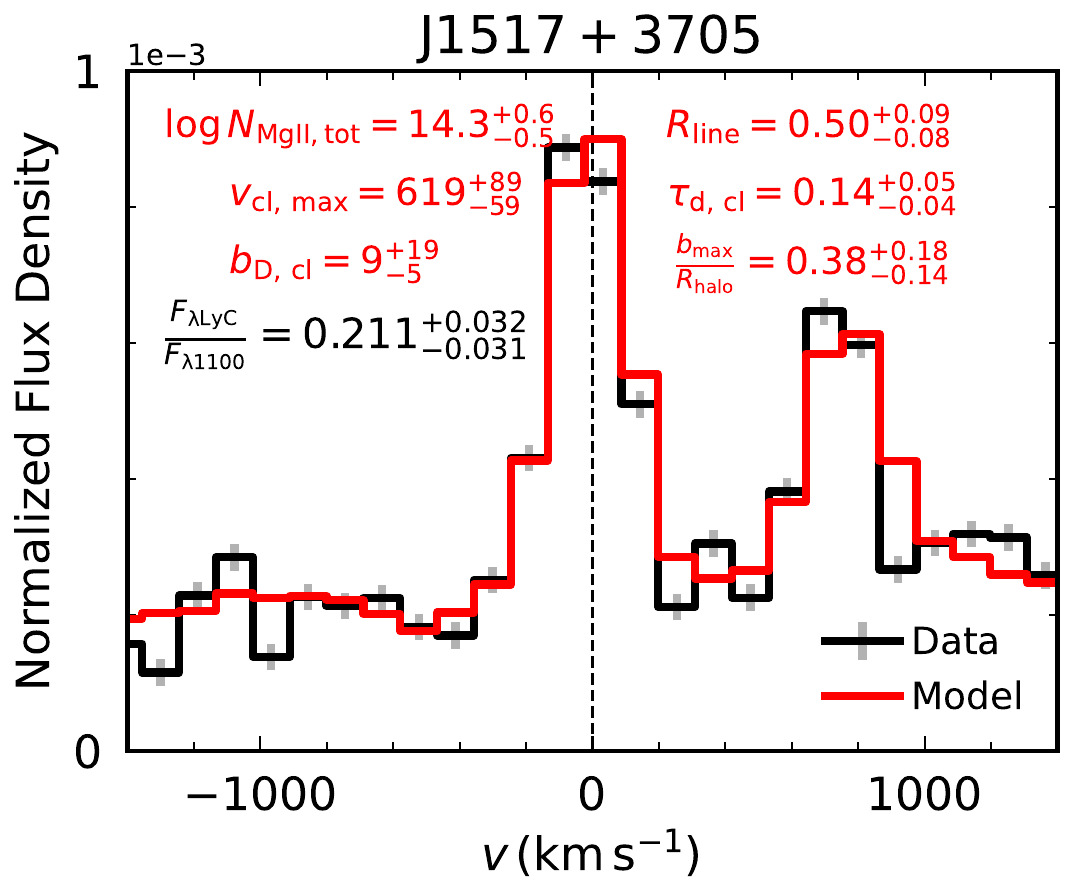}
\includegraphics[width=0.328\textwidth]{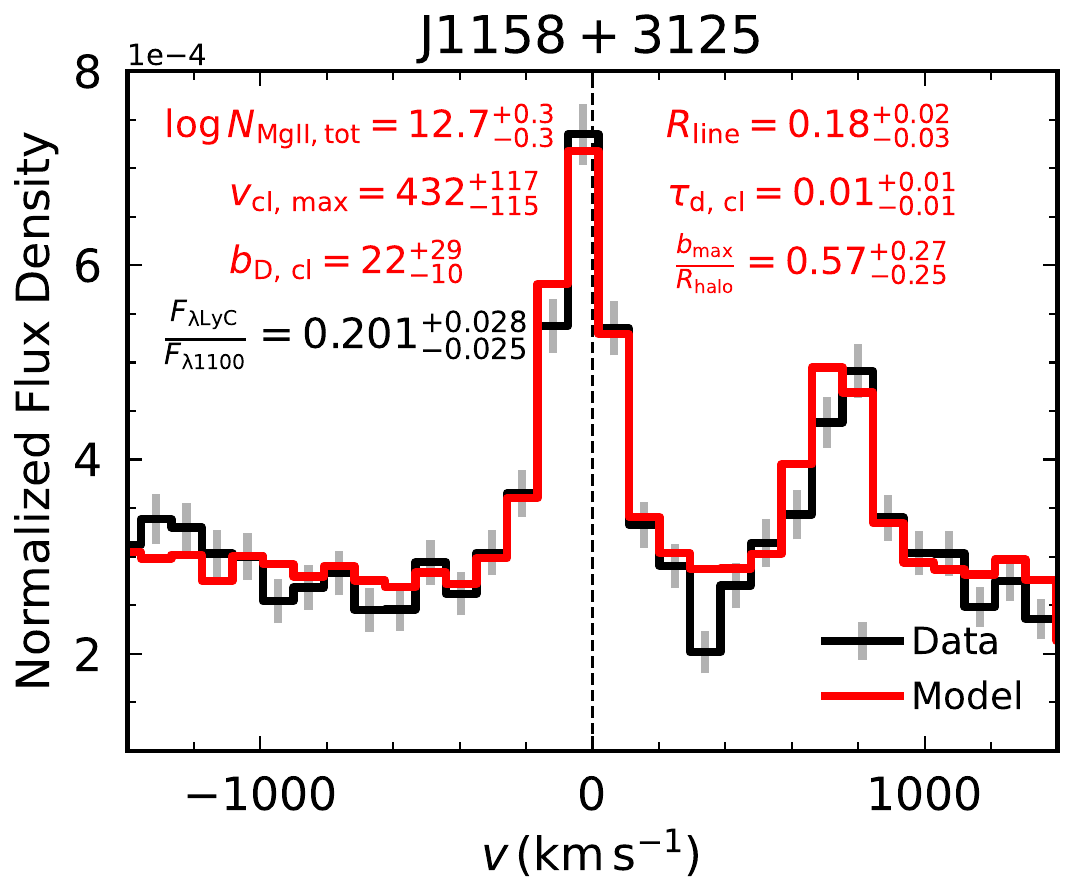}
\includegraphics[width=0.328\textwidth]{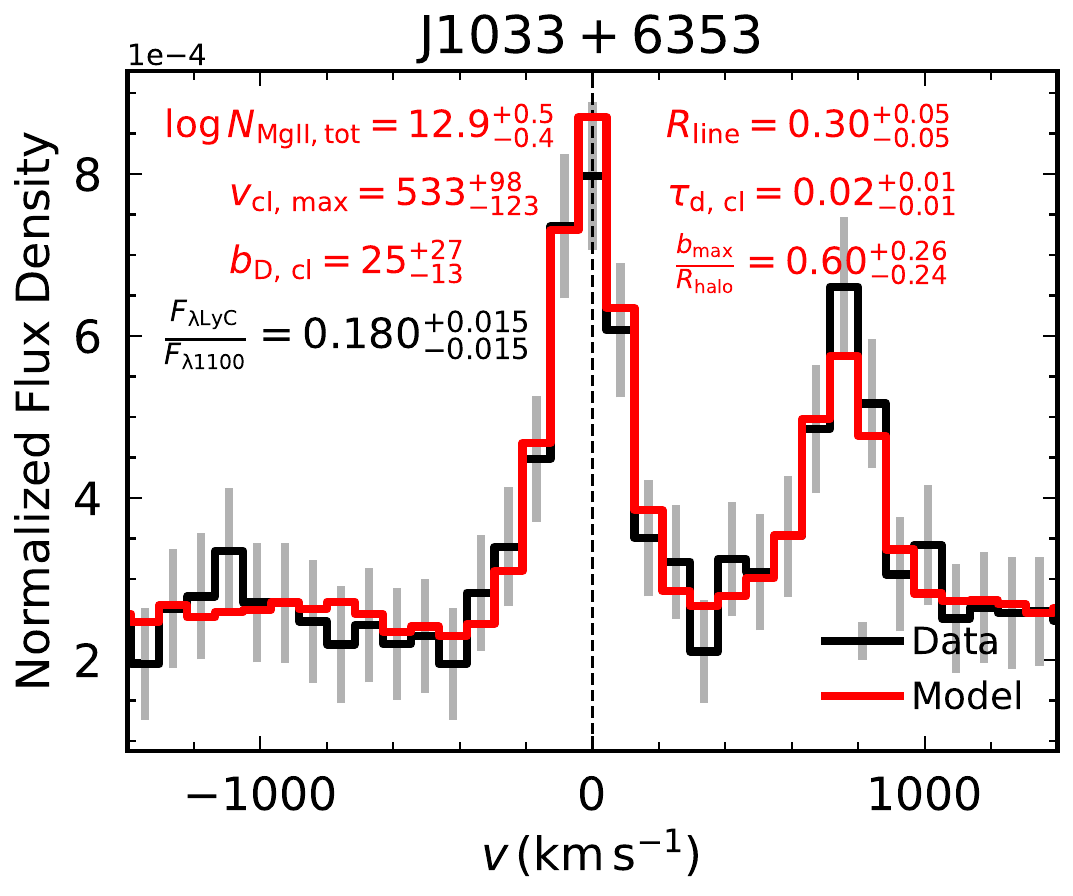}
\includegraphics[width=0.328\textwidth]{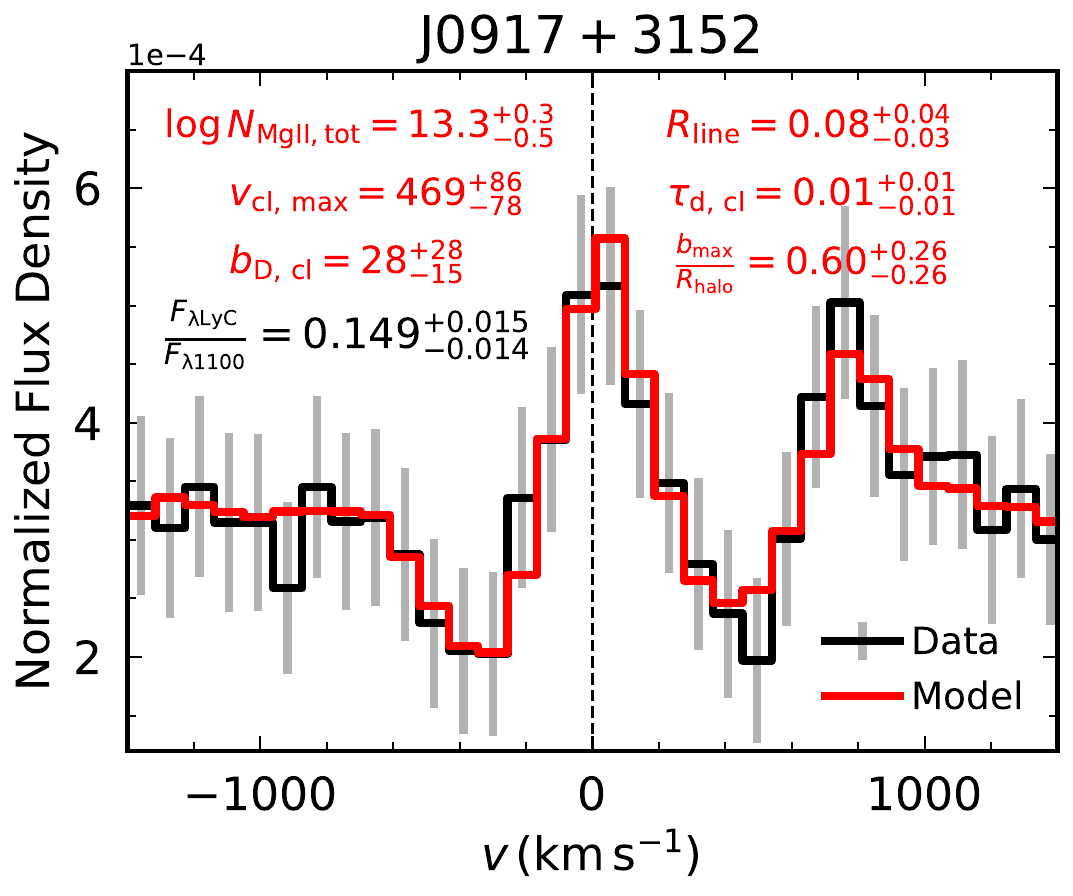}
\includegraphics[width=0.328\textwidth]{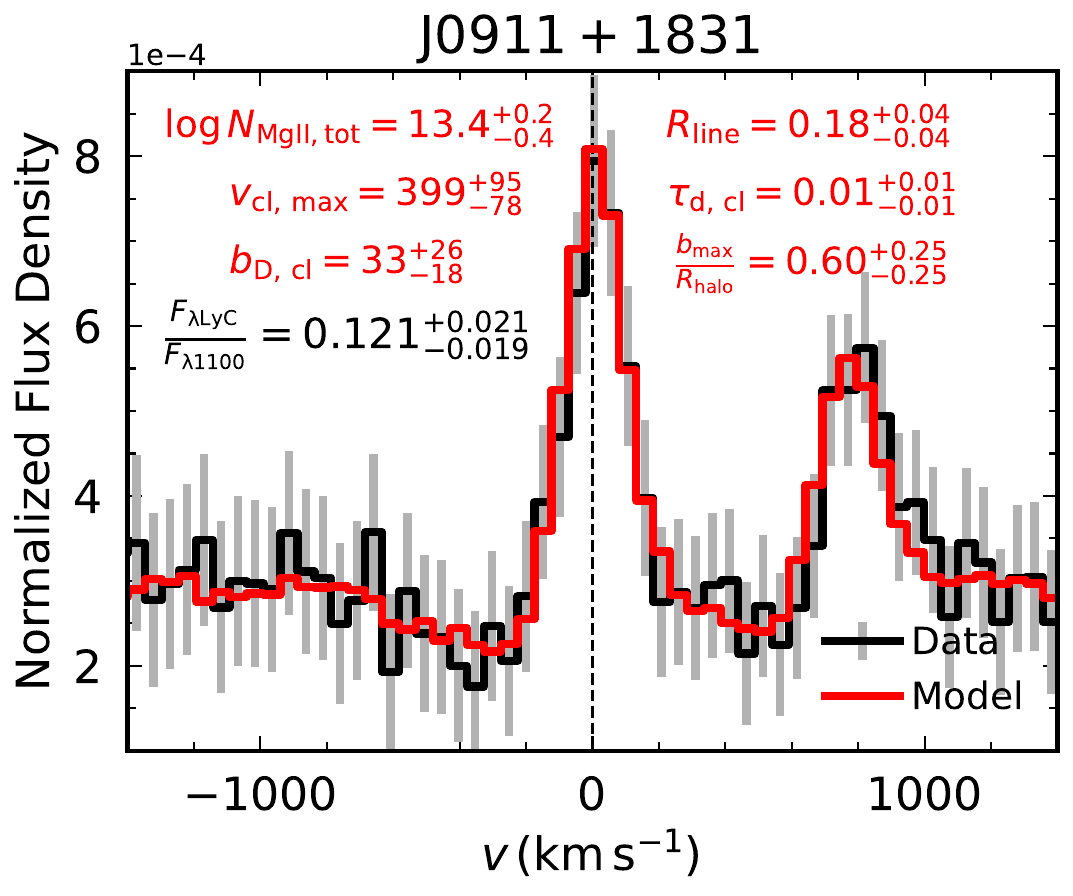}
\includegraphics[width=0.328\textwidth]{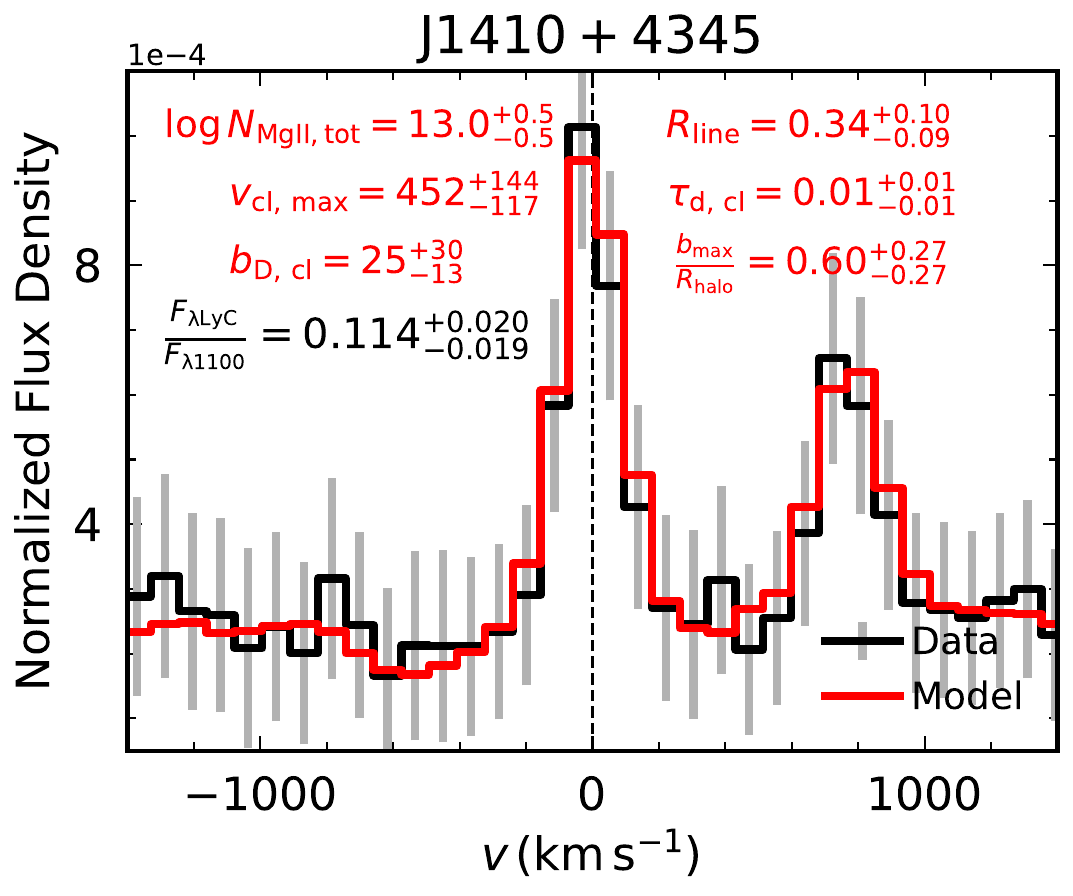}
\includegraphics[width=0.328\textwidth]{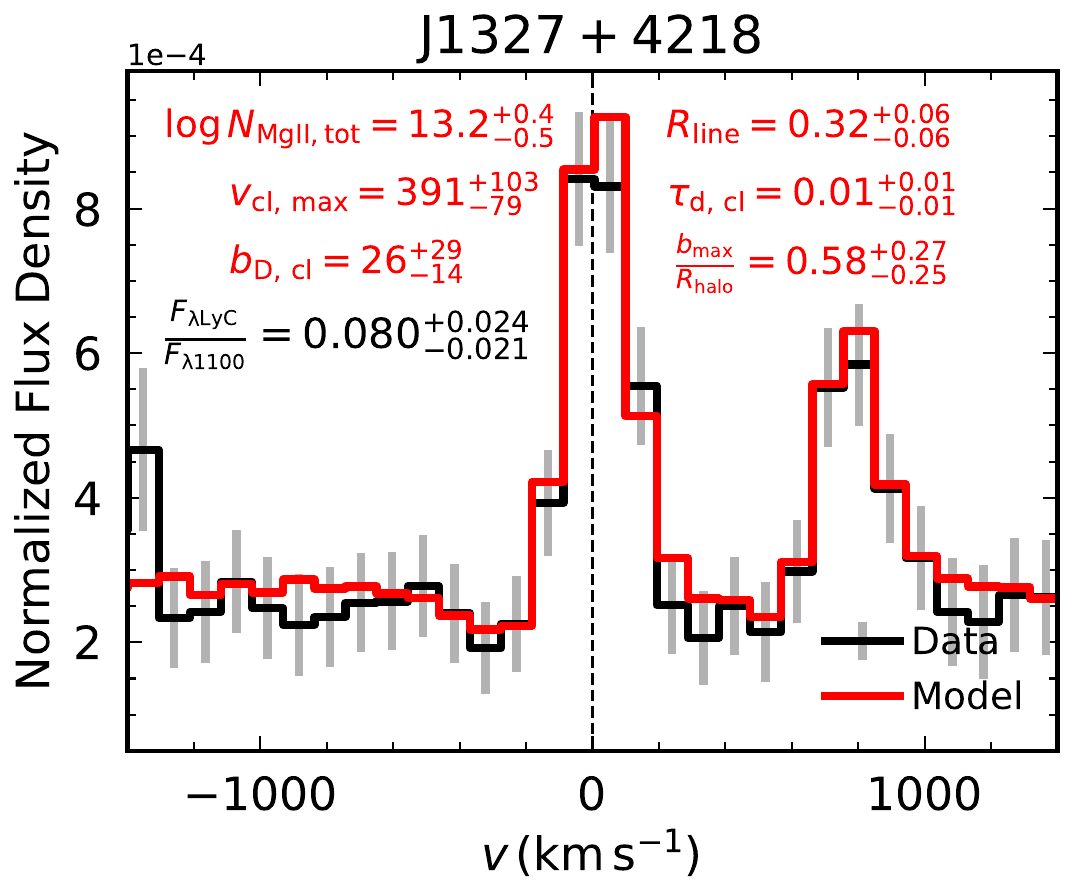}
\includegraphics[width=0.328\textwidth]{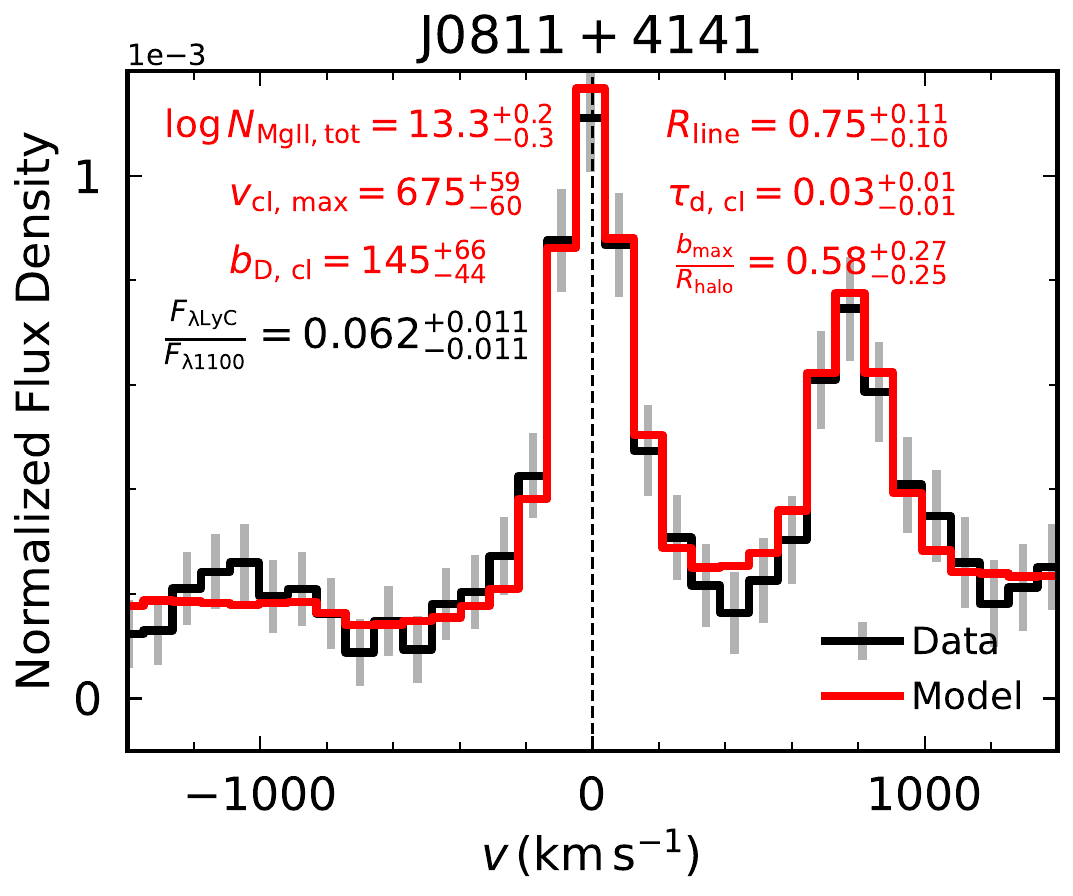}
\includegraphics[width=0.328\textwidth]{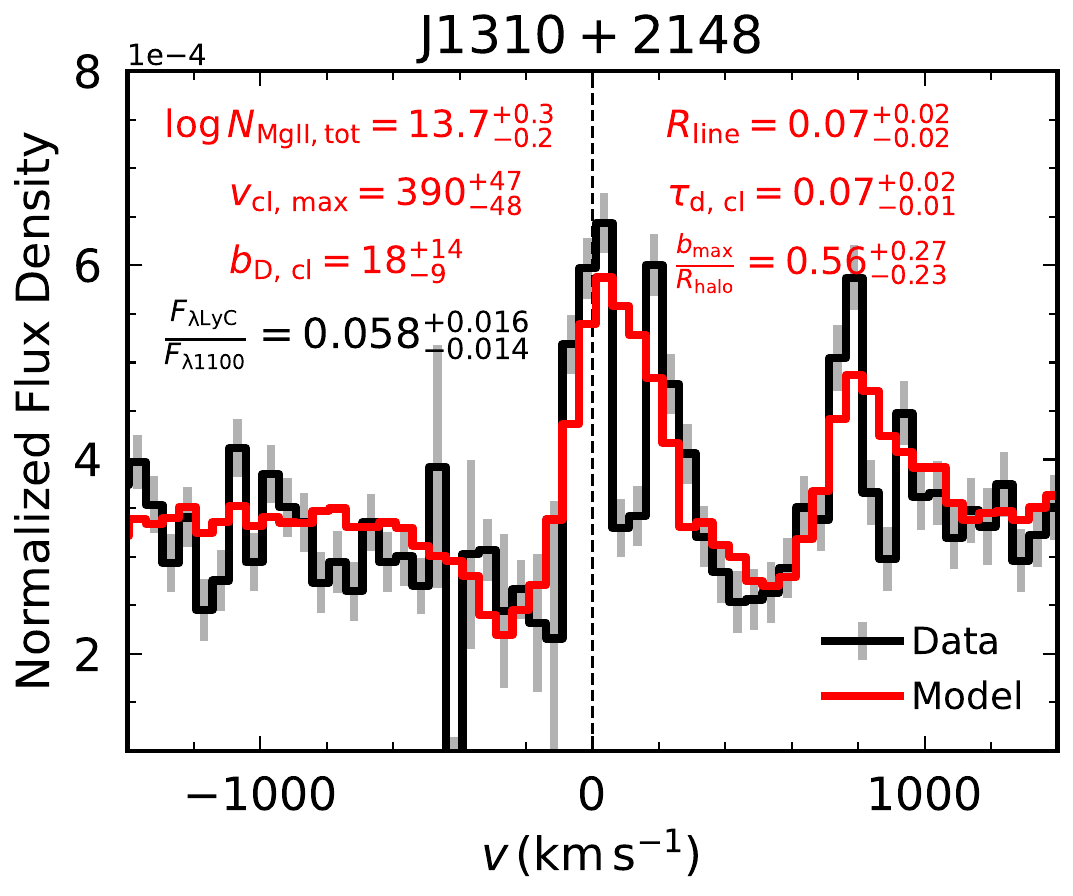}
    \caption{\textbf{{\MgII} best-fits for nine strong LyC leakers obtained by RT modeling.} The galaxies are presented in decreasing order of their $F_{\rm \lambda LyC} / F_{\rm \lambda 1100}$ ratios, with the best-fit parameters shown in red in each panel. The observed \MgII\ line profiles are displayed as black curves (with 1-$\sigma$ error bars in grey), while the best-fit RT model is shown in red. Note that the parameter $v_{\rm cl,\,max}$ refers to the same physical quantity as $v_{\rm MgII,\,max}$ in Figure \ref{fig:vmax_mgII}. 
    \label{fig:strong_leakers}}
\end{figure*}

\begin{figure*}
\centering
\includegraphics[width=0.328\textwidth]{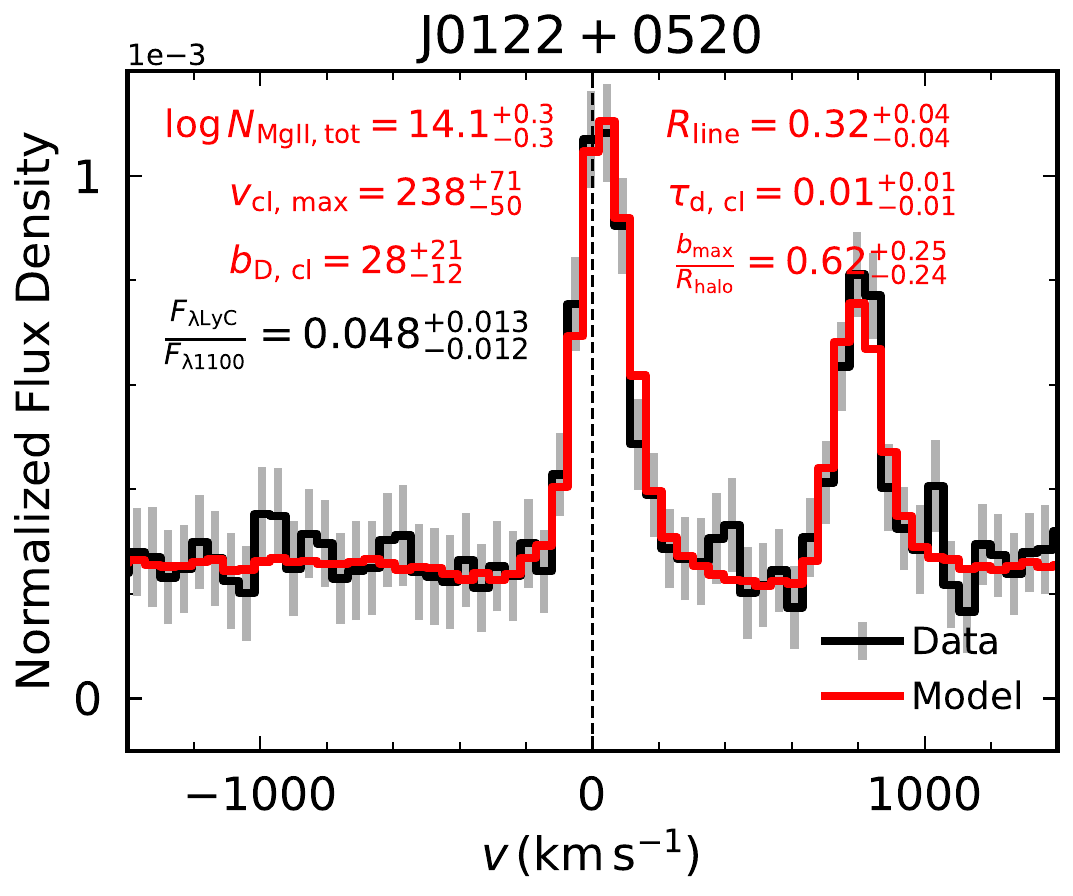}
\includegraphics[width=0.328\textwidth]{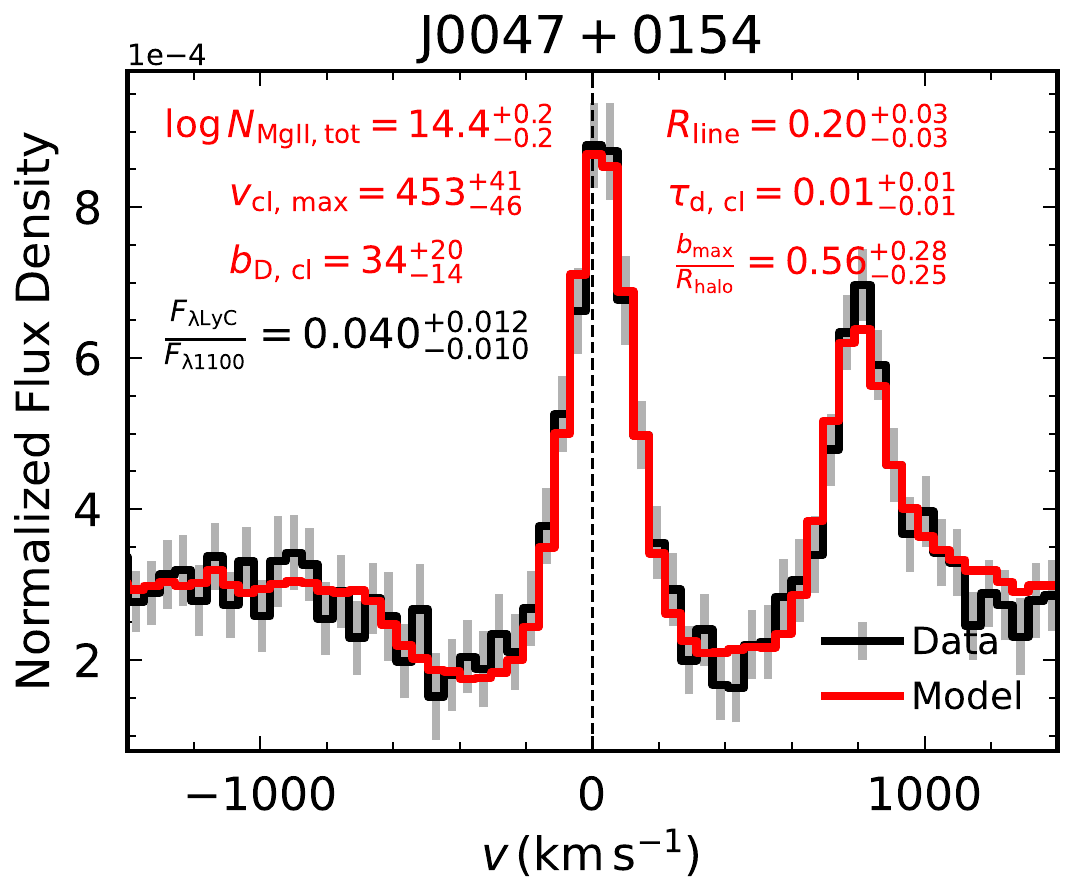}
\includegraphics[width=0.328\textwidth]{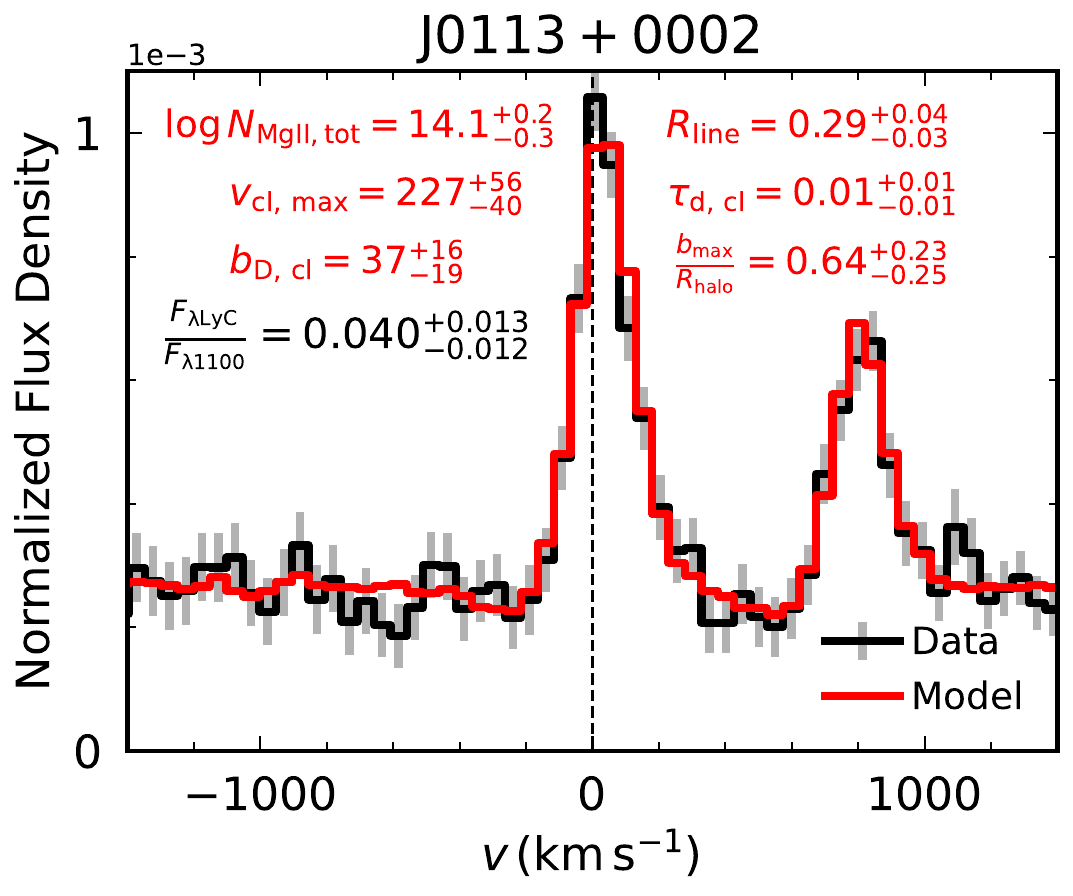}
\includegraphics[width=0.328\textwidth]{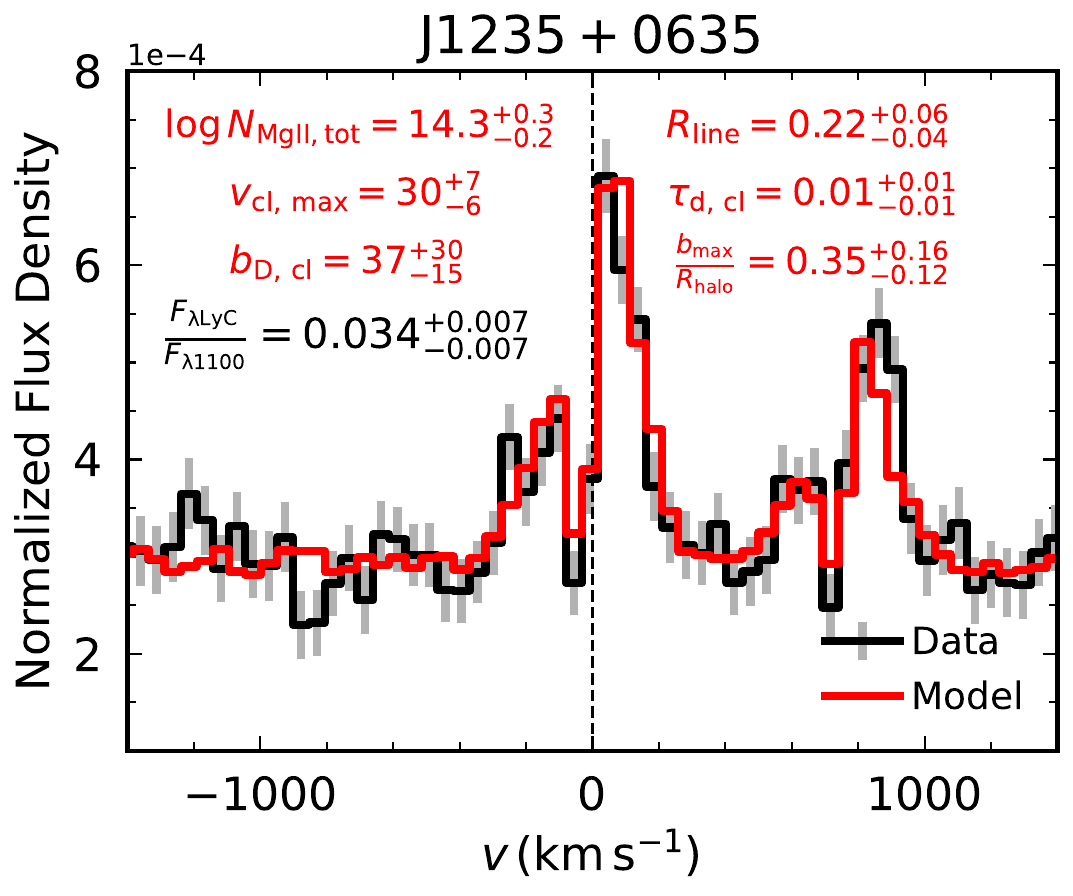}
\includegraphics[width=0.328\textwidth]{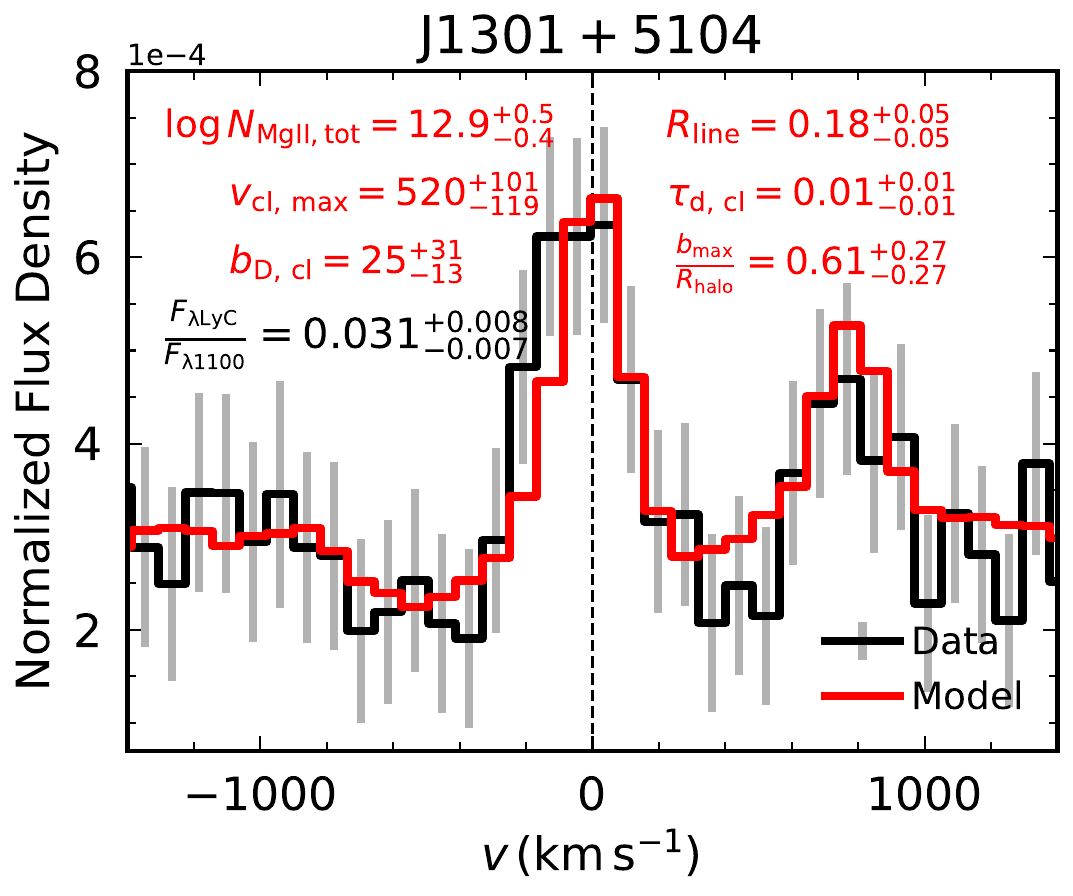}
\includegraphics[width=0.328\textwidth]{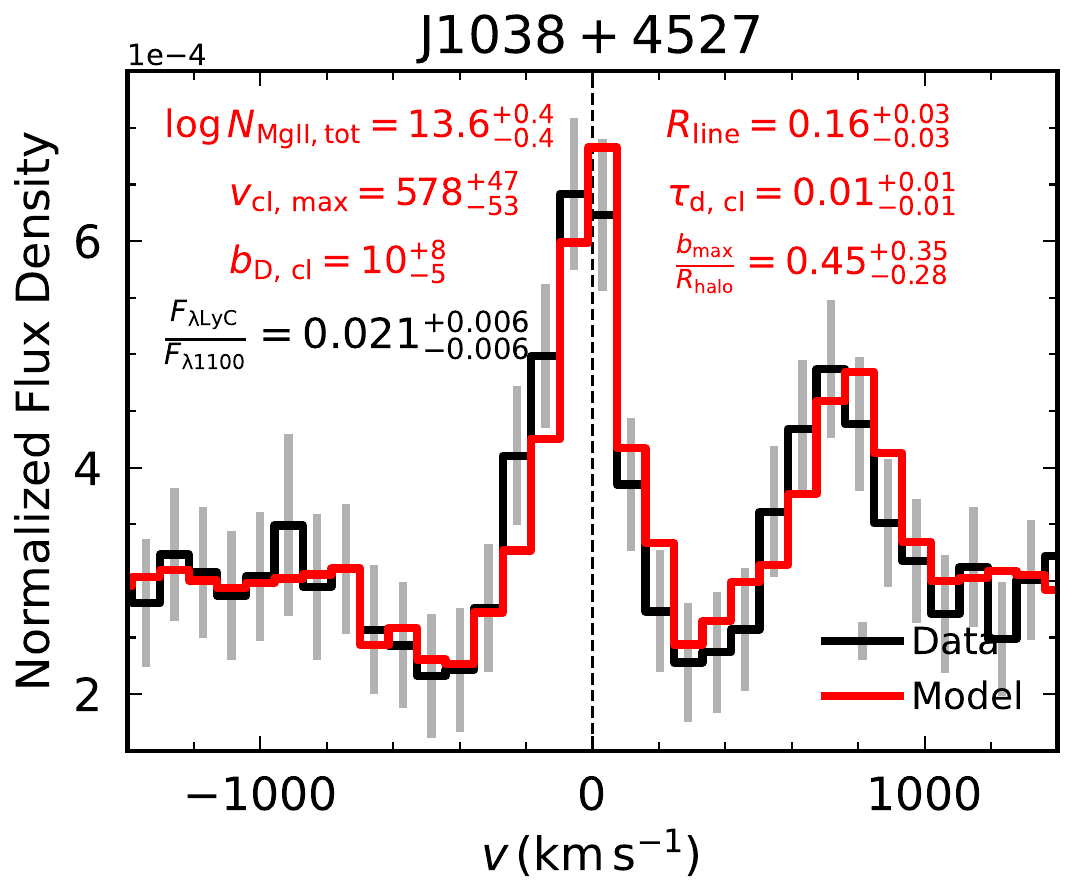}
    \caption{\textbf{Same as Figure \ref{fig:strong_leakers}, but for six moderate LyC leakers.} 
    \label{fig:moderate_leakers}}
\end{figure*}

\begin{figure*}
\centering
\includegraphics[width=0.328\textwidth]{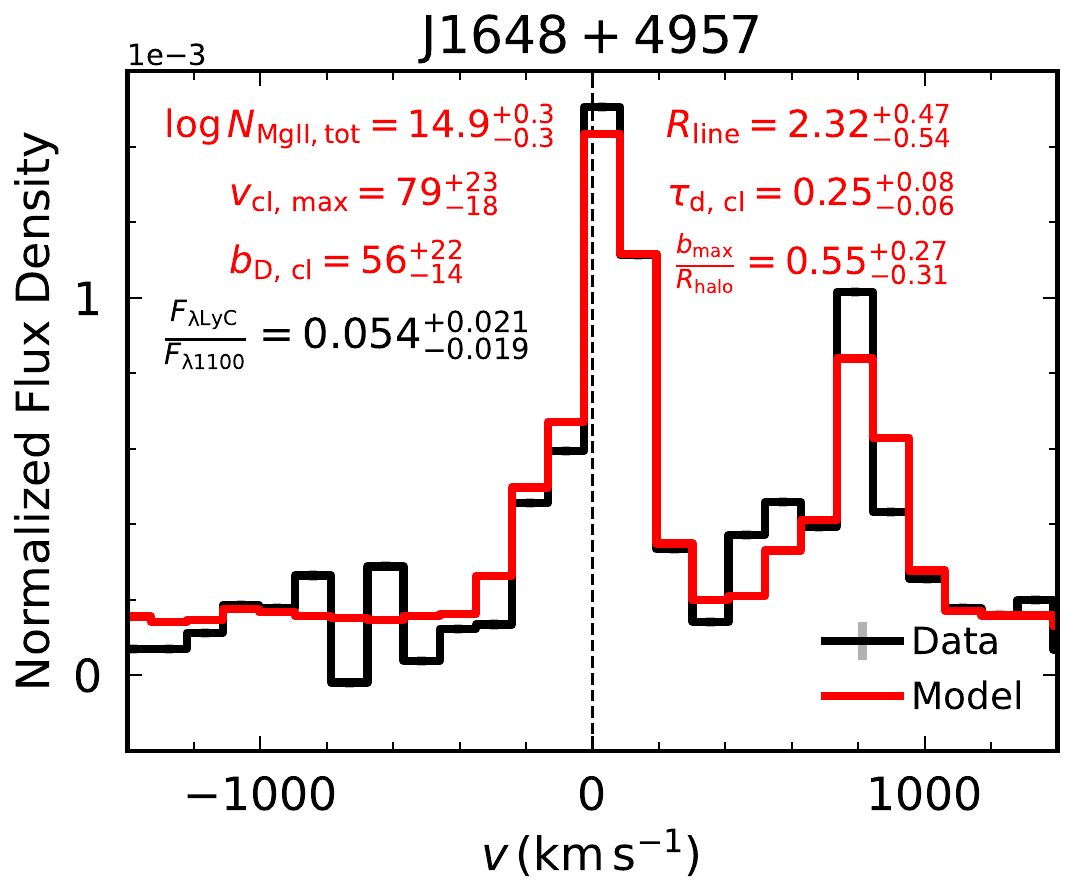}
\includegraphics[width=0.328\textwidth]{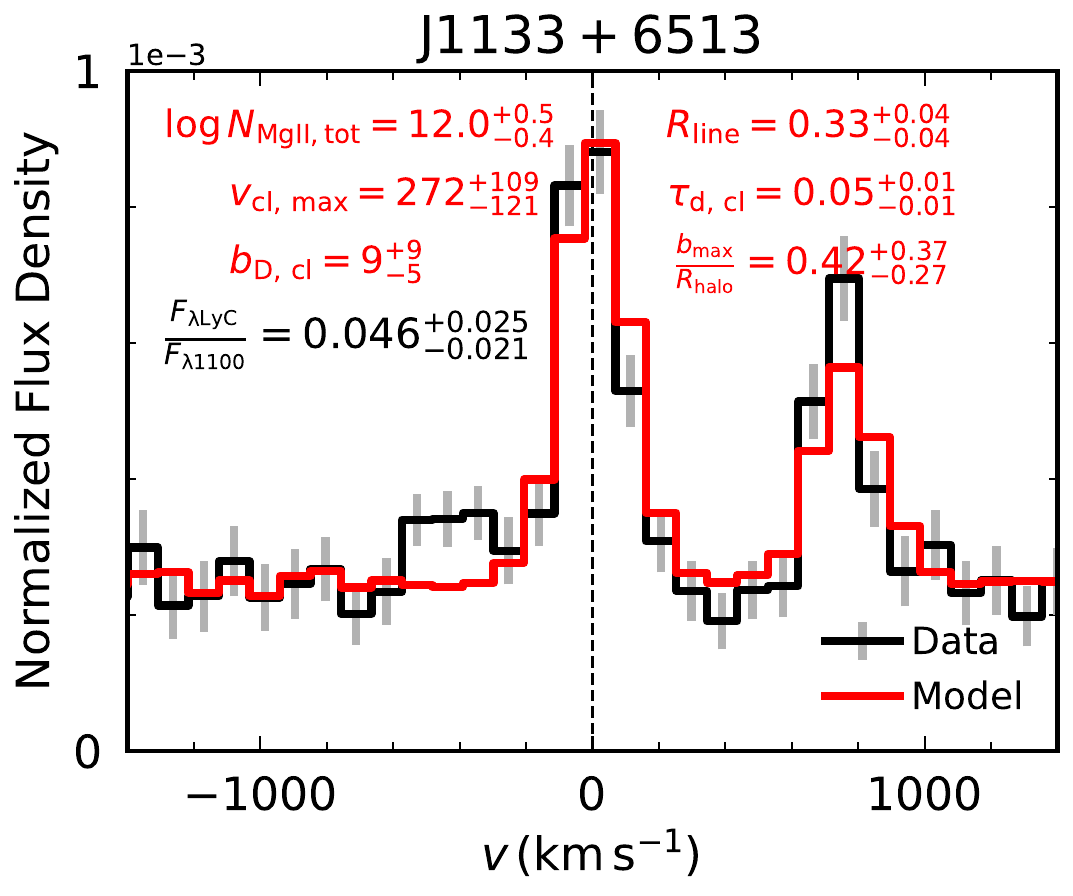}
\includegraphics[width=0.328\textwidth]{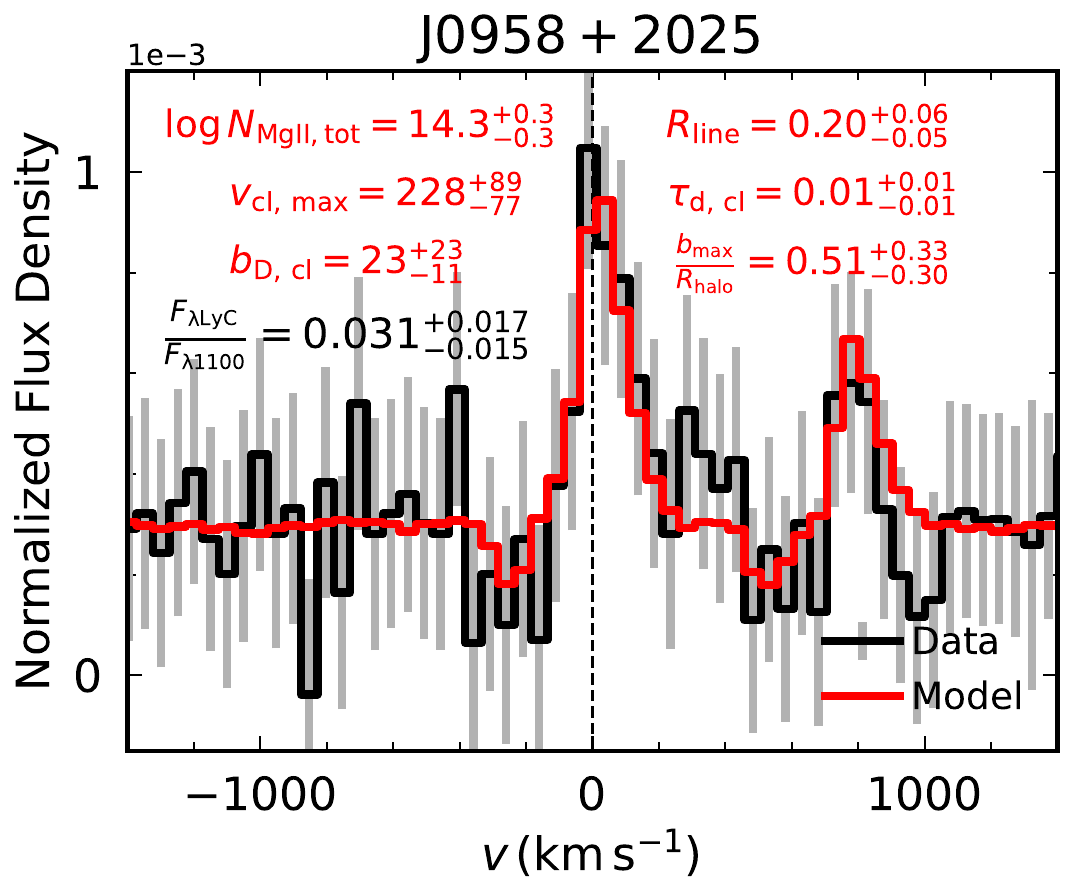}
\includegraphics[width=0.328\textwidth]{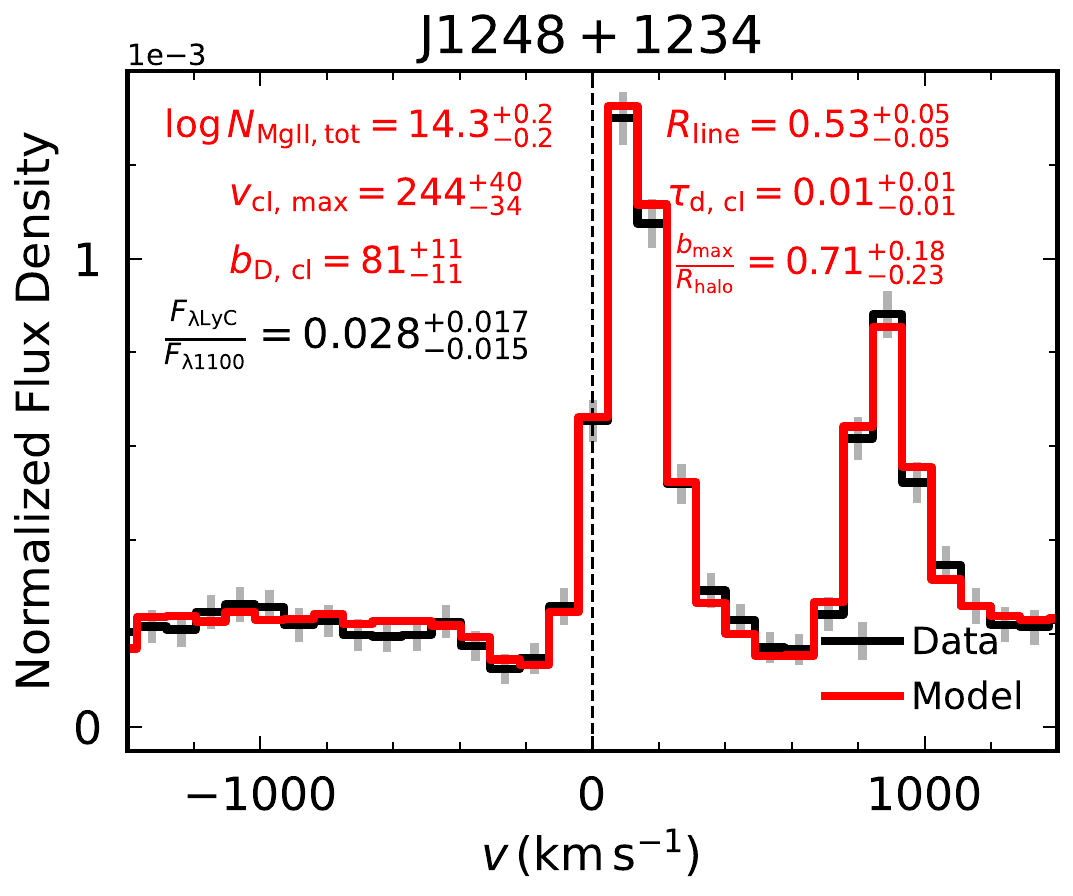}
\includegraphics[width=0.328\textwidth]{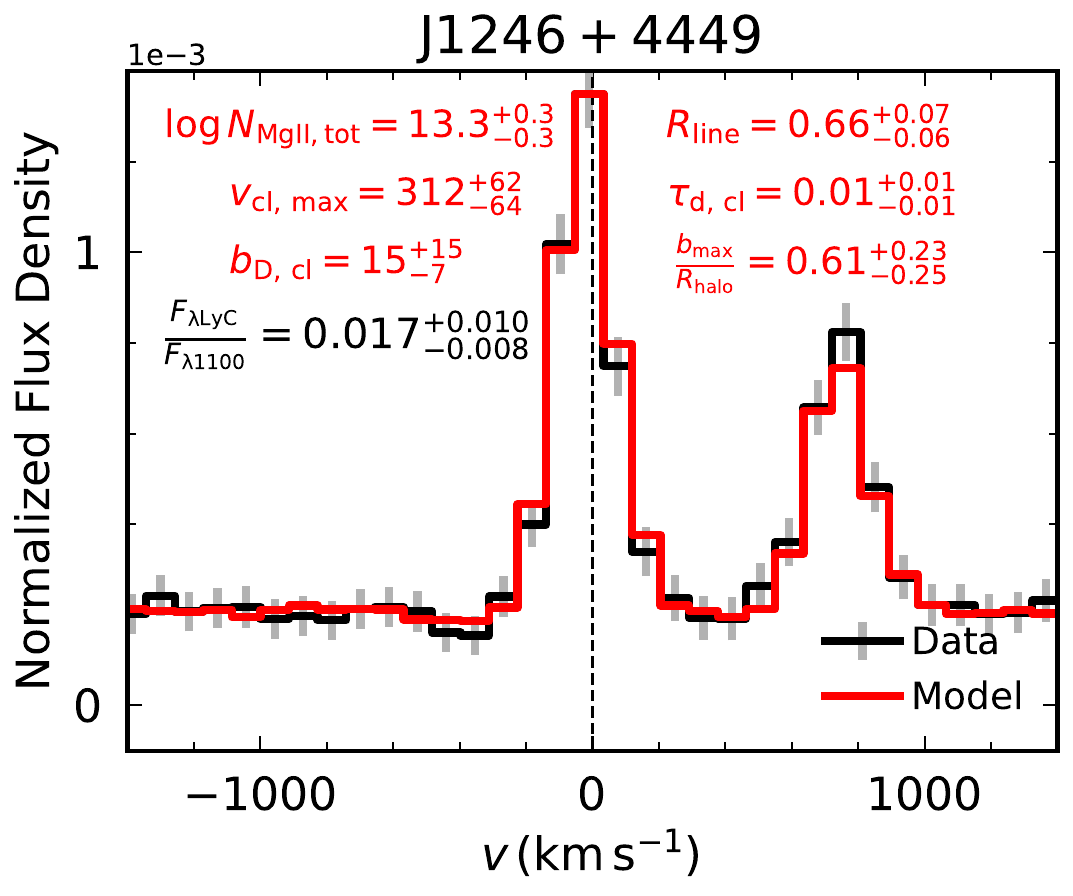}
    \caption{\textbf{Same as Figure \ref{fig:strong_leakers}, but for five potential LyC leakers.} 
    \label{fig:potential_leakers}}
\end{figure*}

\begin{figure*}
\centering
\includegraphics[width=0.328\textwidth]{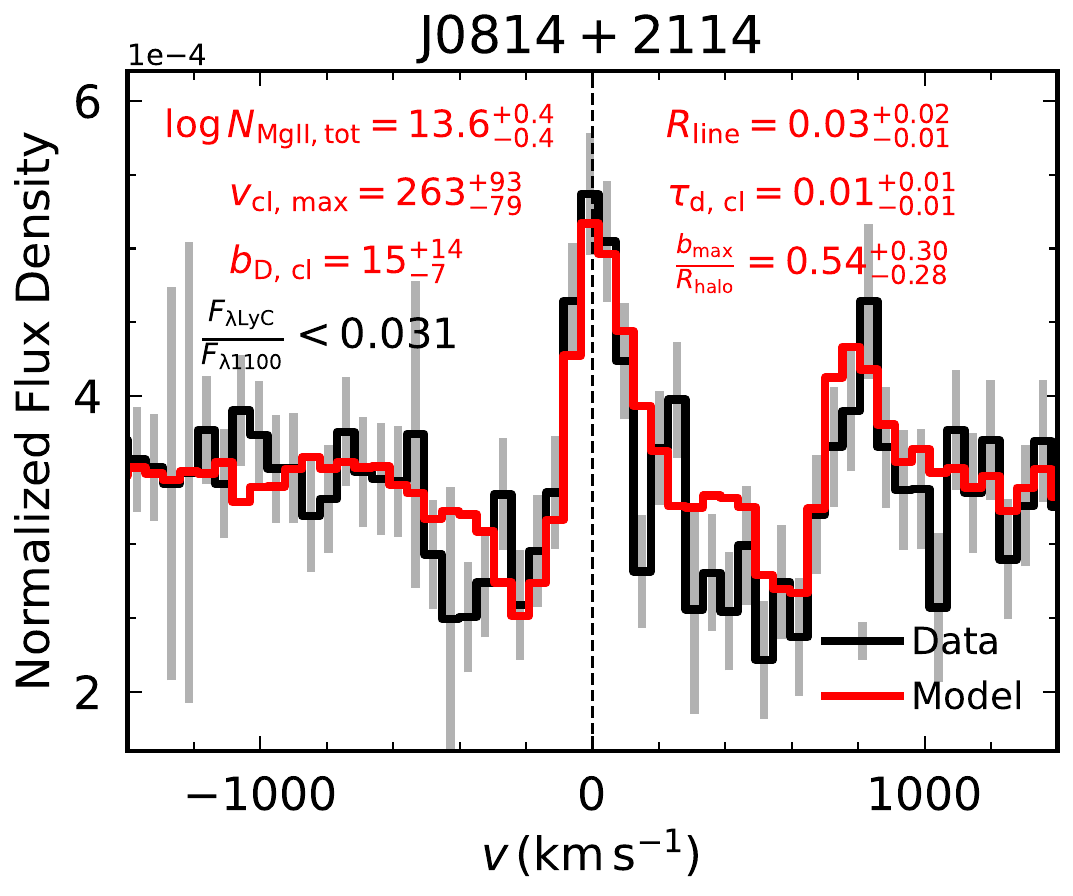}
\includegraphics[width=0.328\textwidth]{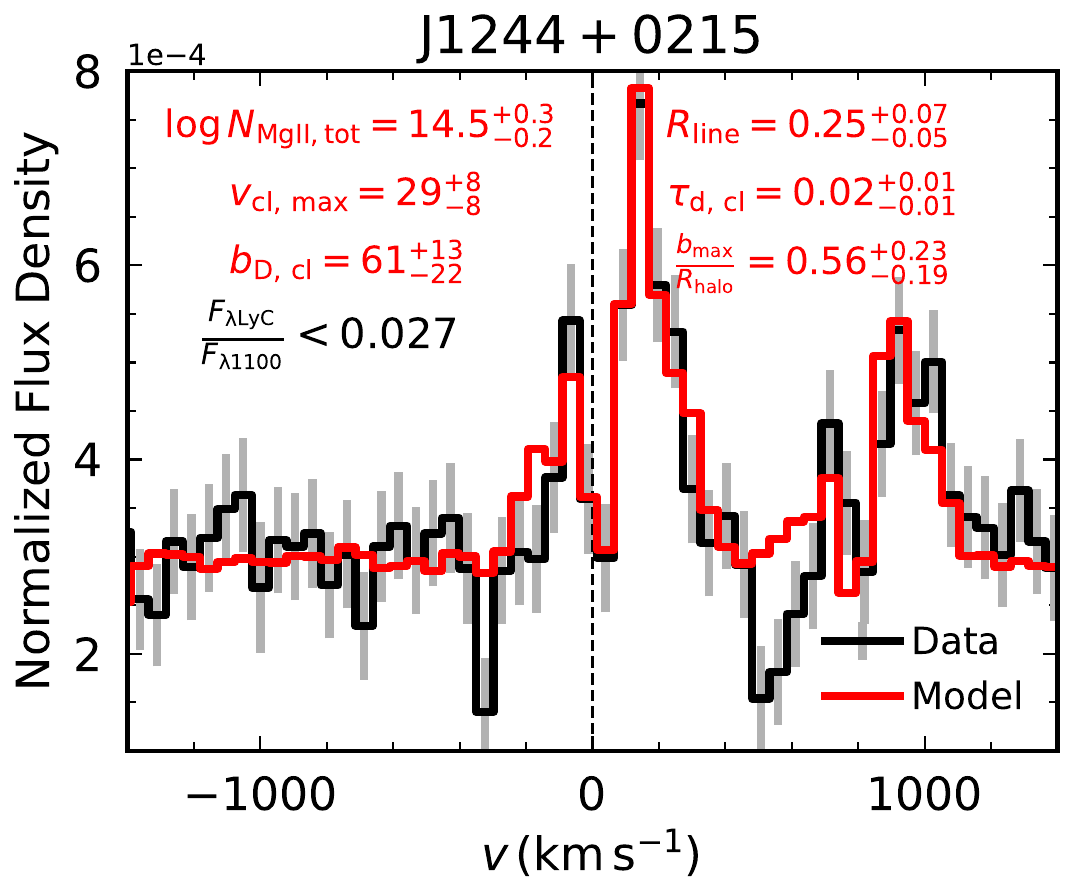}
\includegraphics[width=0.328\textwidth]{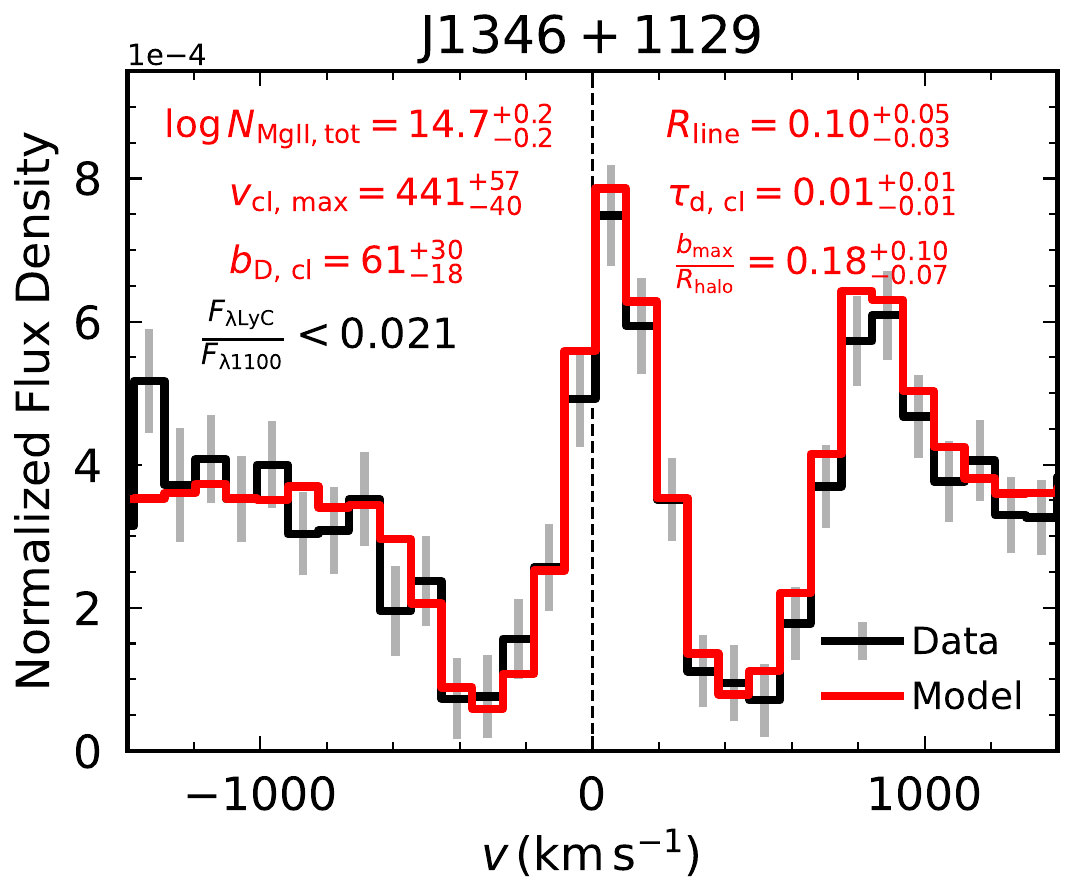}
\includegraphics[width=0.328\textwidth]{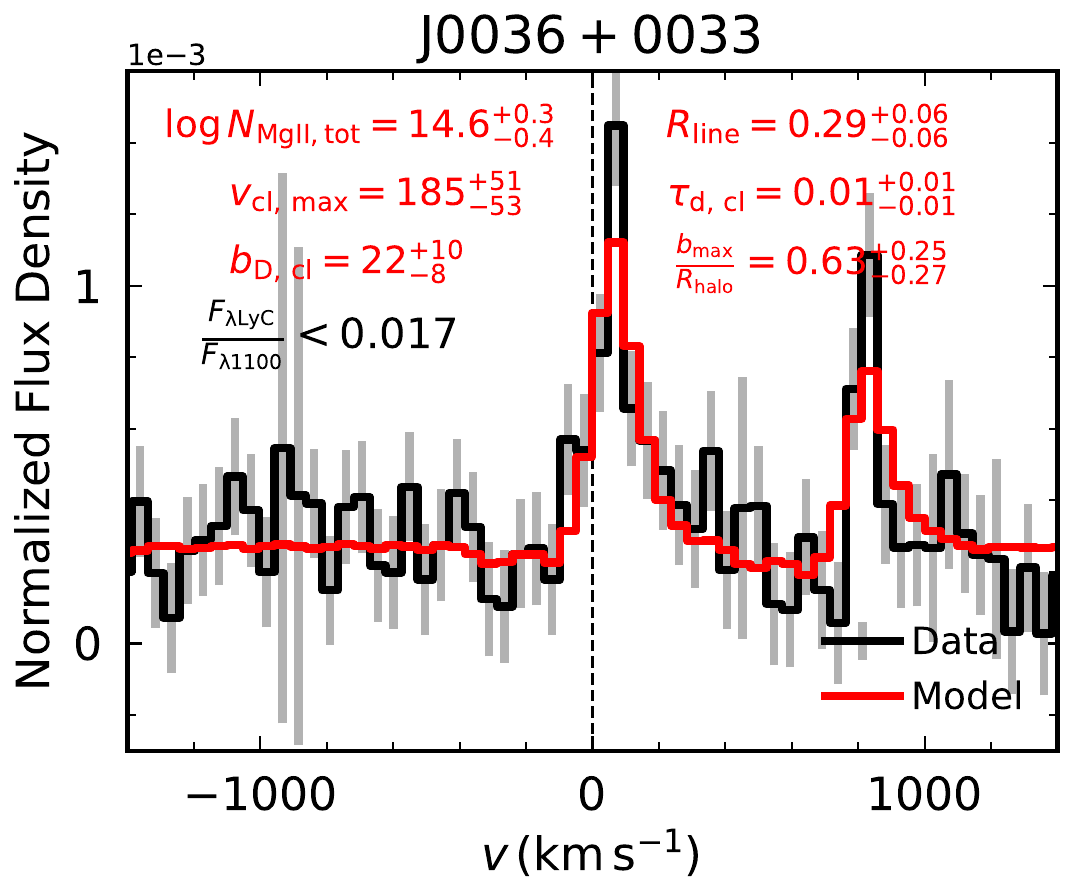}
\includegraphics[width=0.328\textwidth]{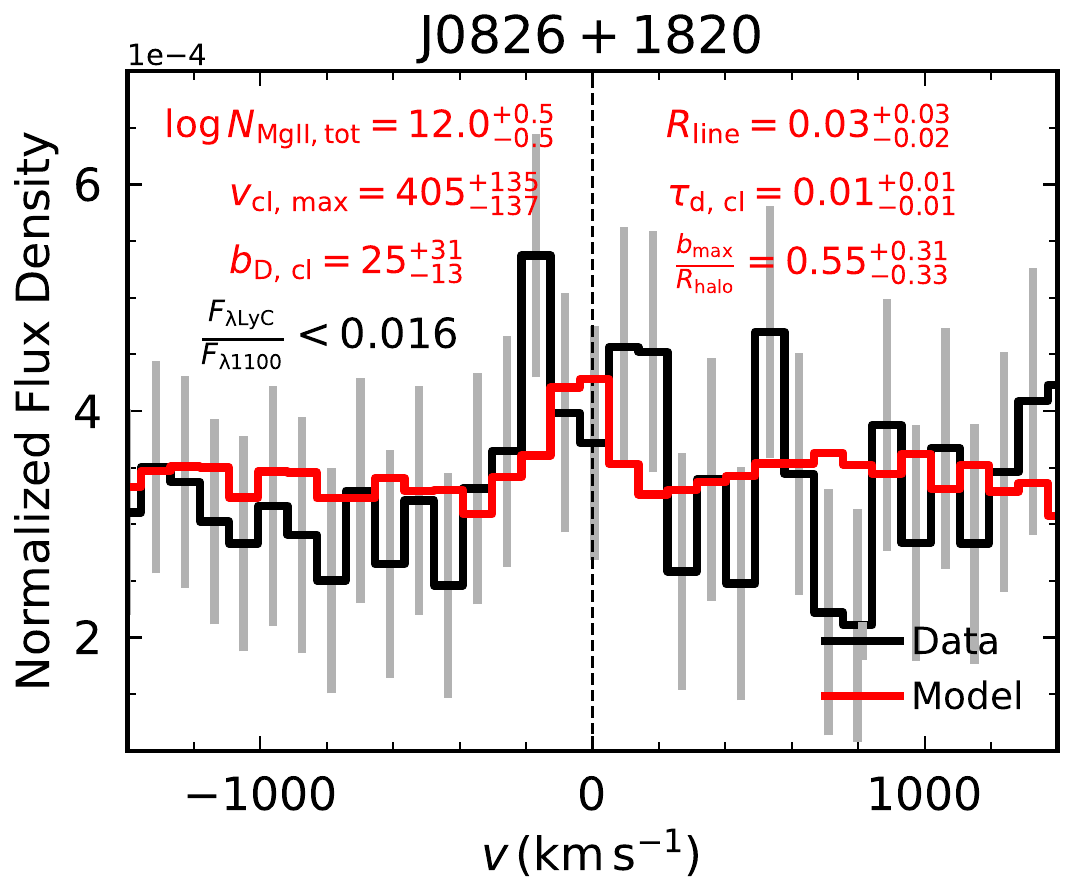}
\includegraphics[width=0.328\textwidth]{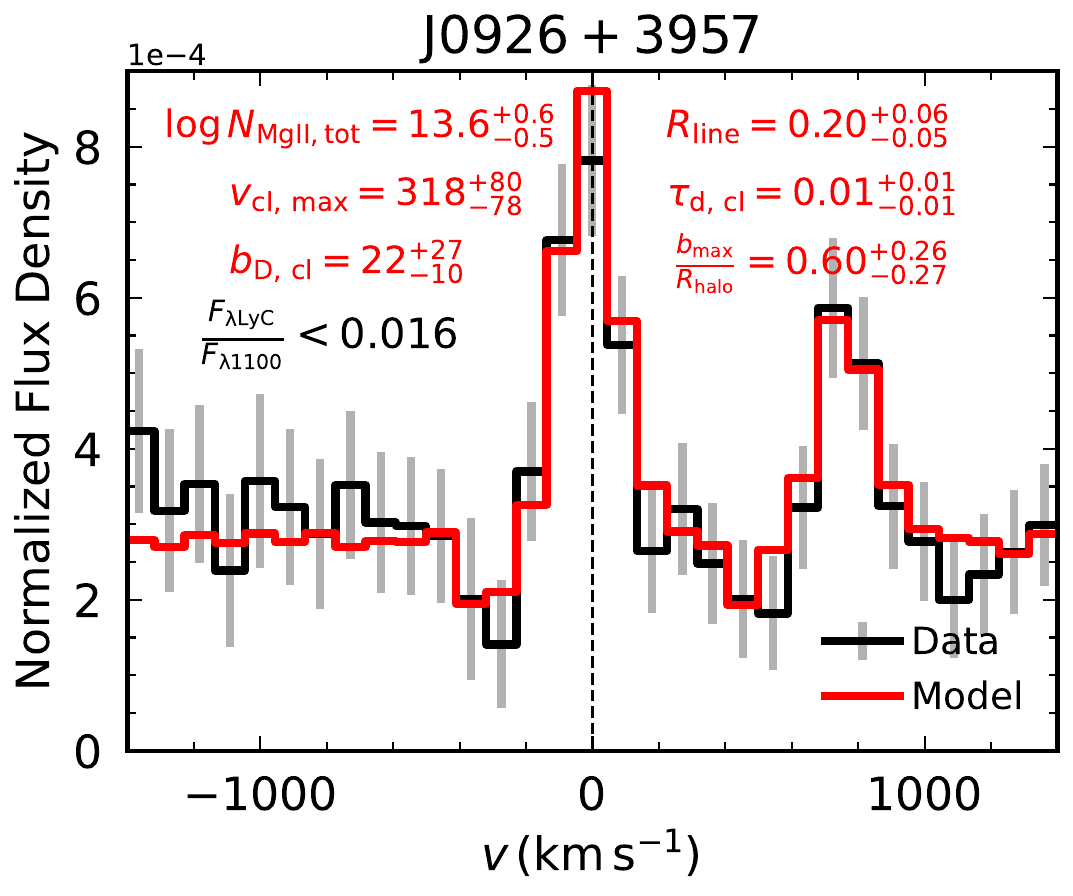}
\includegraphics[width=0.328\textwidth]{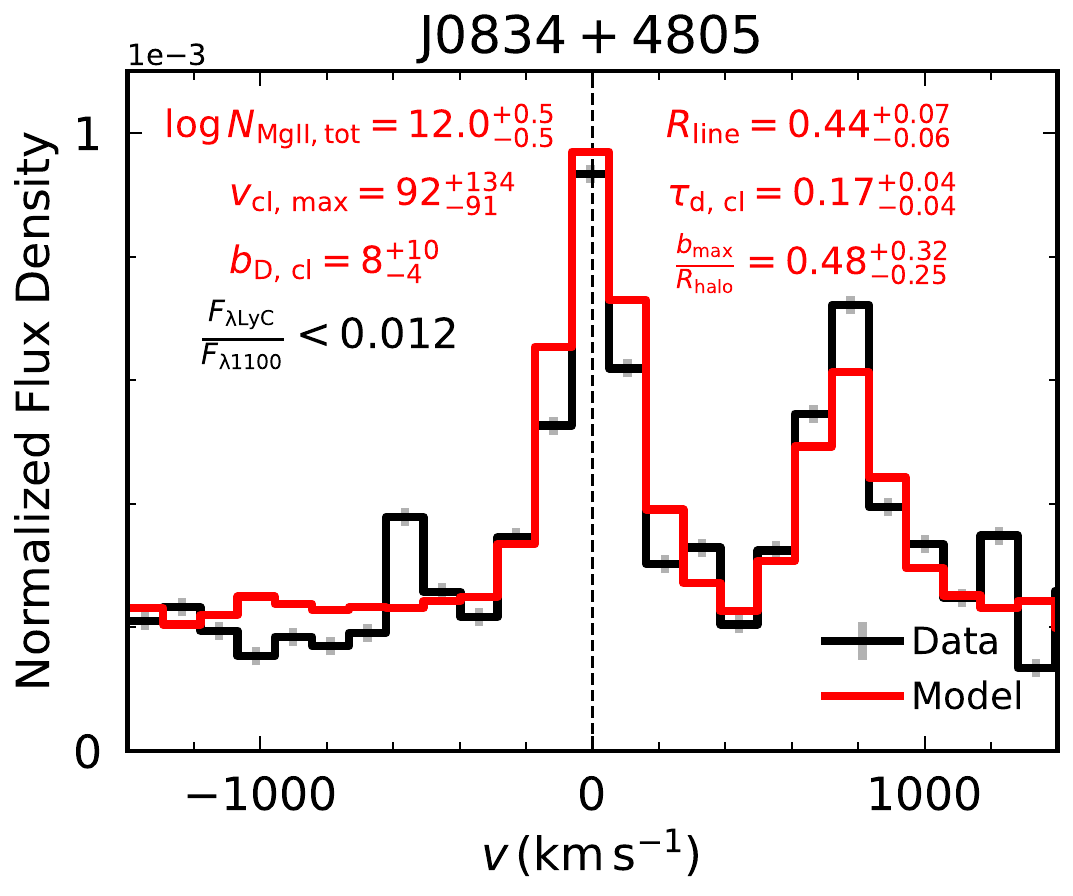}
\includegraphics[width=0.328\textwidth]{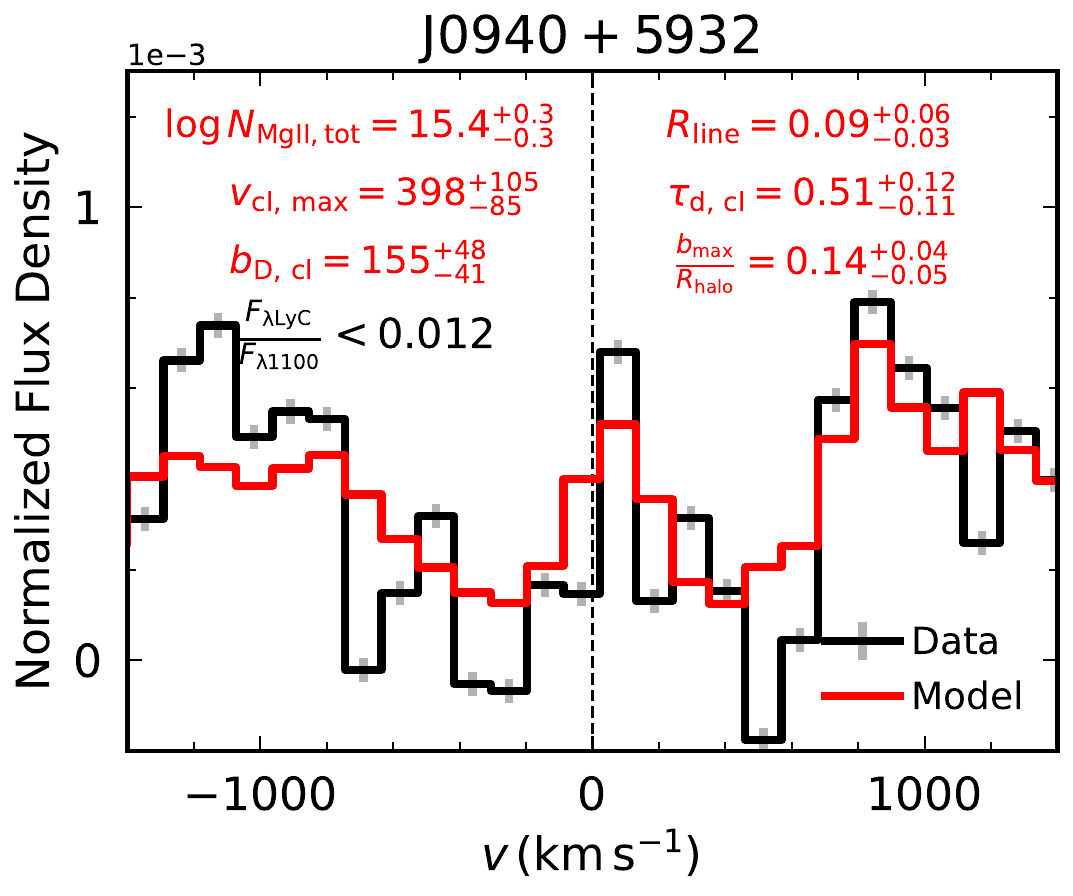}
\includegraphics[width=0.328\textwidth]{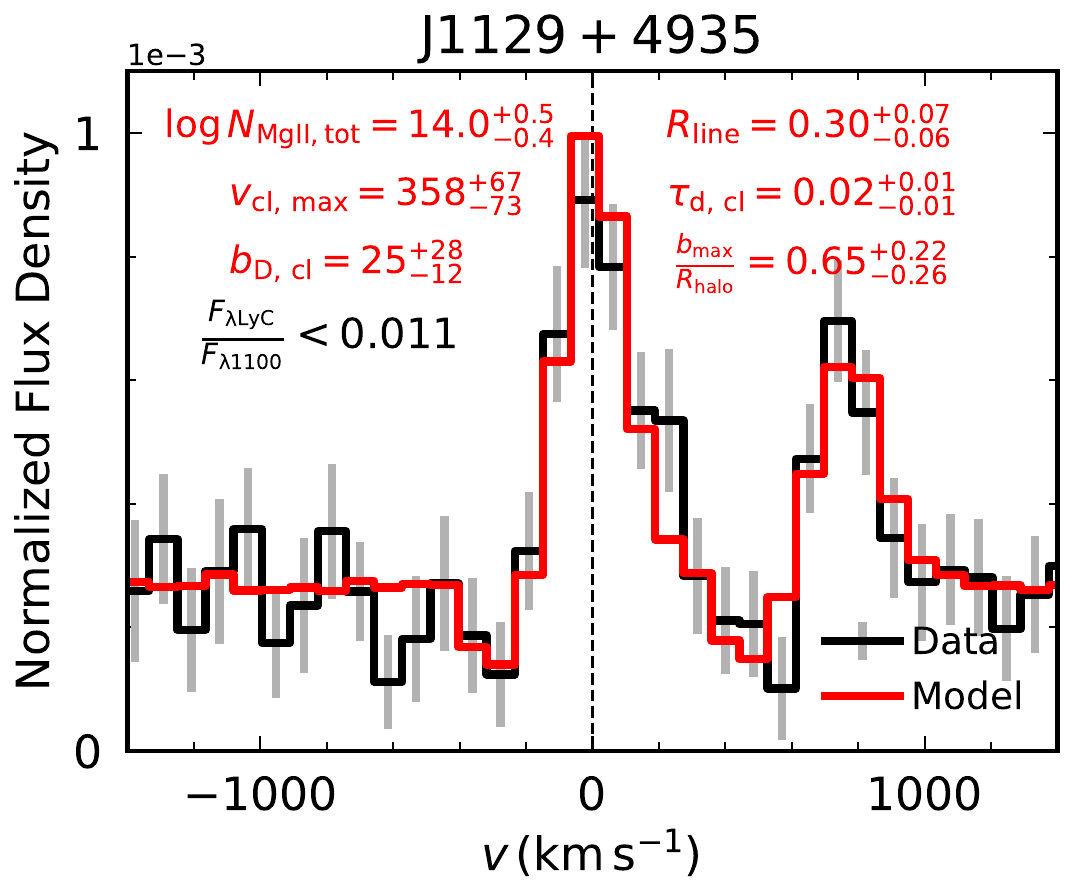}
\includegraphics[width=0.328\textwidth]{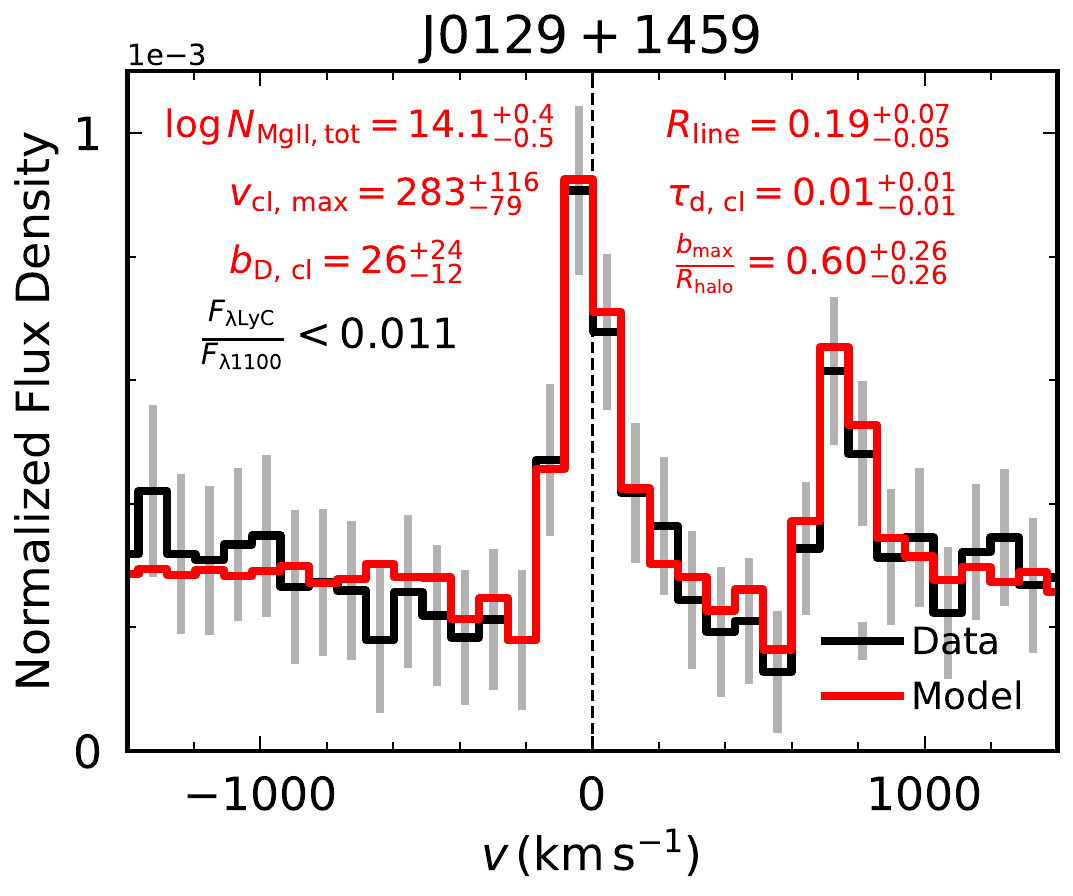}
\includegraphics[width=0.328\textwidth]{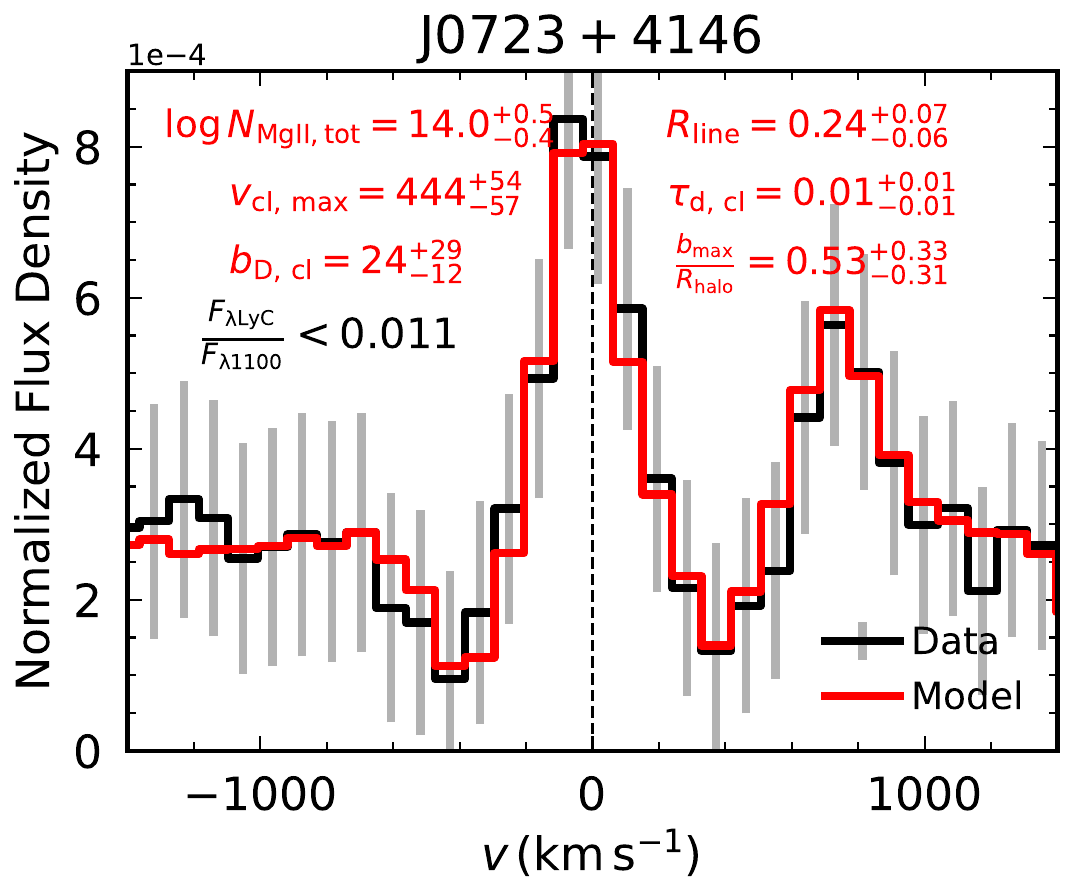}
\includegraphics[width=0.328\textwidth]{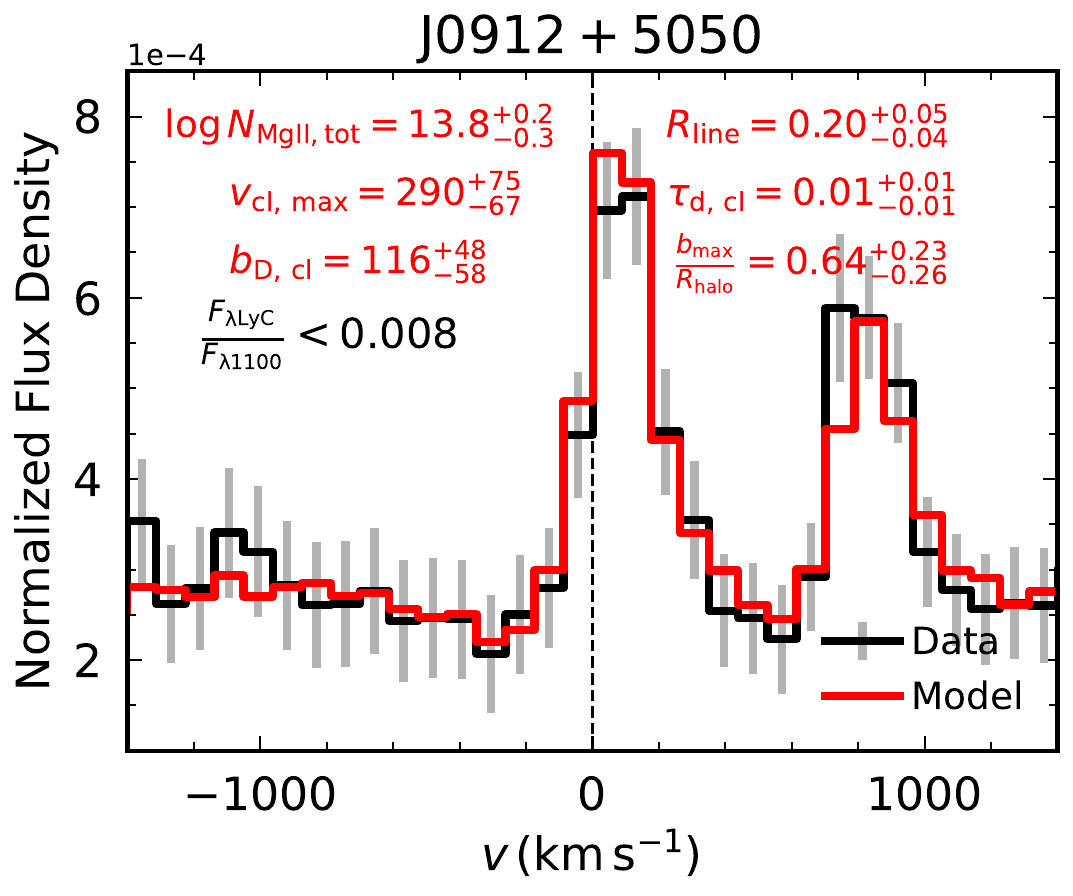}
\includegraphics[width=0.328\textwidth]{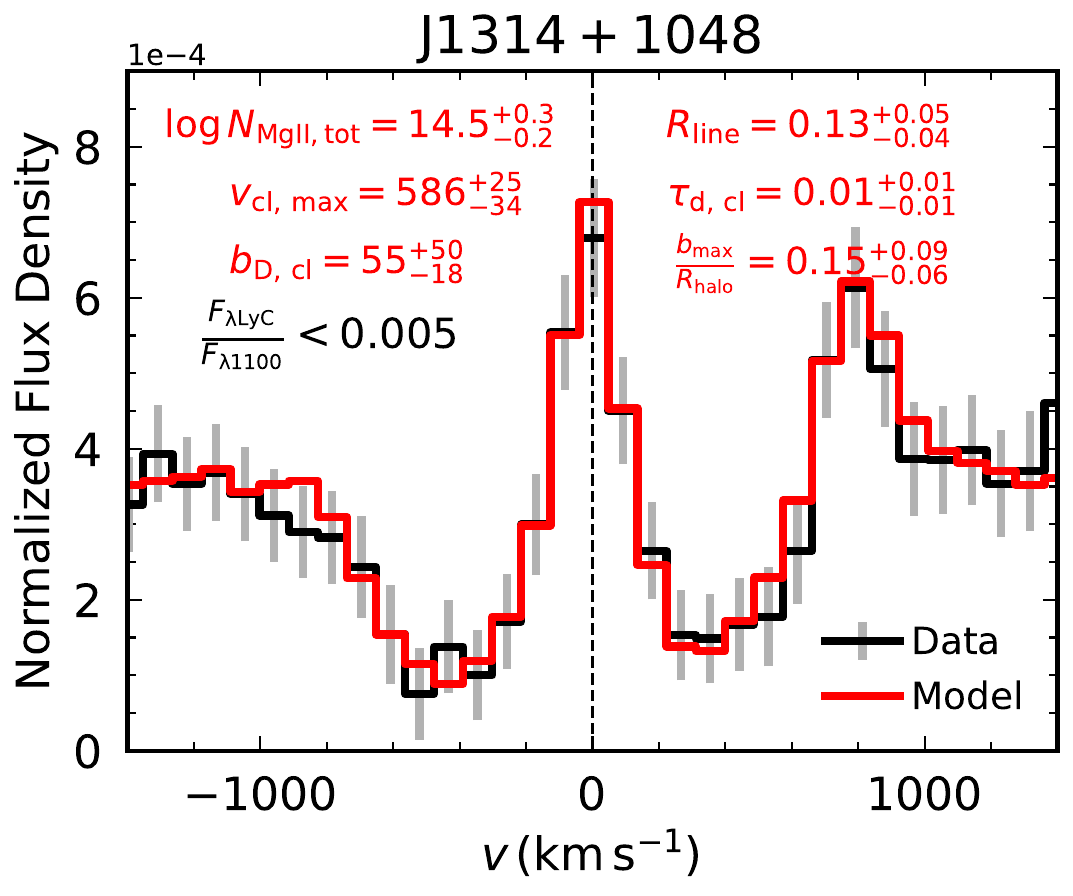}
    \caption{\textbf{Same as Figure \ref{fig:strong_leakers}, but for thirteen non-leakers.} 
    \label{fig:non_leakers}}
\end{figure*}

We have identified an intriguing trend between the best-fit maximum clump radial outflow velocity $v_{\rm MgII,\,max}$ and the total \MgII\ column density $N_{\rm MgII,\,tot}$ inferred from the RT modeling. Here $v_{\rm MgII,\,max}$ is defined as the maximum velocity of the clump radial velocity profile (see Appendix \ref{sec:appendix} for examples), whereas $N_{\rm MgII,\,tot}$ is defined as the average total \MgII\ column density along a sightline, which is the product of the average number of \MgII\ clumps along a sightline and the \MgII\ column density in each clump.

In Figure \ref{fig:vmax_mgII}, we plot all 33 modeled LyC leakers on the $v_{\rm MgII,\,max}$ -- $N_{\rm MgII,\,tot}$ plane, categorized into four distinct colors based on their LyC escape types. We find a necessary (though not sufficient; see discussion below) condition for a LyC leaker to be a strong leaker: a high maximum clump radial outflow velocity ($v_{\rm MgII,\,max} \gtrsim 390\,\rm km\,s^{-1}$) \emph{and} a low total $\rm Mg\,{\textsc {II}}$ column density ($N_{\rm MgII,\,tot} \lesssim 10^{14.3}\,\rm cm^{-2}$).

We now discuss the behavior of each of the four types of LyC leakers and explore the connection between their best-fit parameters and the morphology of their \MgII\ spectra individually (best-fits and parameters presented in Figures \ref{fig:strong_leakers} - \ref{fig:non_leakers}):

\begin{itemize}
    \item Strong leakers: The nine strong leakers are all located in the upper left corner of Figure \ref{fig:vmax_mgII}, defined by $v_{\rm MgII,\,max} \gtrsim 390\,\rm km\,s^{-1}$ and $N_{\rm MgII,\,tot} \lesssim 10^{14.3}\,\rm cm^{-2}$. In terms of spectral morphology, all strong leakers exhibit an absorption trough extending beyond $v \approx -400\,{\rm km\,s}^{-1}$, indicating their high outflow velocities. However, the \MgII\ column density does not appear to be directly associated with any specific spectral feature, making RT modeling necessary for accurate extraction of its value. We note that J1517+3705 was observed with HET/LRS2, which has a particularly low resolution (FWHM $\simeq 125\,\rm km\,s^{-1}$), resulting in a $N_{\rm MgII,\,tot}$ of $\sim 10^{14.3}\,\rm cm^{-2}$ with large uncertainties. Aside from this object, all other strong LyC leakers have $N_{\rm MgII,\,tot} \lesssim 10^{13.7}\rm\,cm^{-2}$. 
    
    \item Moderate leakers: Among the six moderate leakers, two (J1301+5104 and J1038+4527) are located in the same regime as the strong leakers. This suggests that the distinction between moderate and strong leakers may not be clear-cut; instead, they could represent the same galaxy population with similar gas properties, with only slight differences in LyC leakage due to subtle environmental variations and observational uncertainties. The remaining four moderate leakers, however, all exhibit high \MgII\ total column densities with $N_{\rm MgII,\,tot} \gtrsim 10^{14.1}\,\rm cm^{-2}$.

    \item Potential leakers: The five potential leakers exhibit a broad range of total \MgII\ column densities but share a common characteristic: they all have very low maximum clump outflow velocities ($v_{\rm MgII,\,max} \lesssim 320\,\rm km\,s^{-1}$). This is evident from their spectral morphology: the absorption trough at $v < 0$ is either negligible (as seen in J1648+4957 and J1133+6513) or is situated relatively close to the line center ($|v_{\rm trough}| \lesssim 300\,\rm km\,s^{-1}$).

\begin{figure*}
\centering
\includegraphics[width=0.495\textwidth]{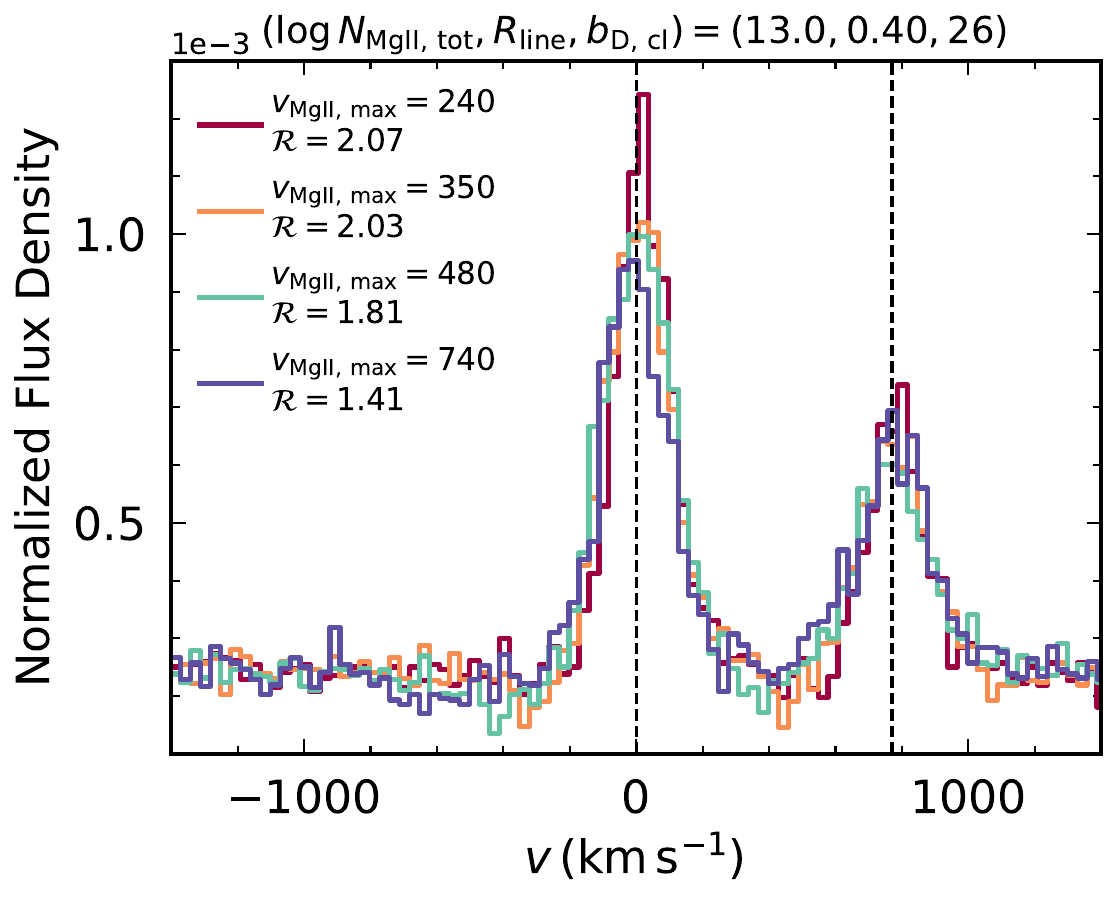}
\includegraphics[width=0.495\textwidth]{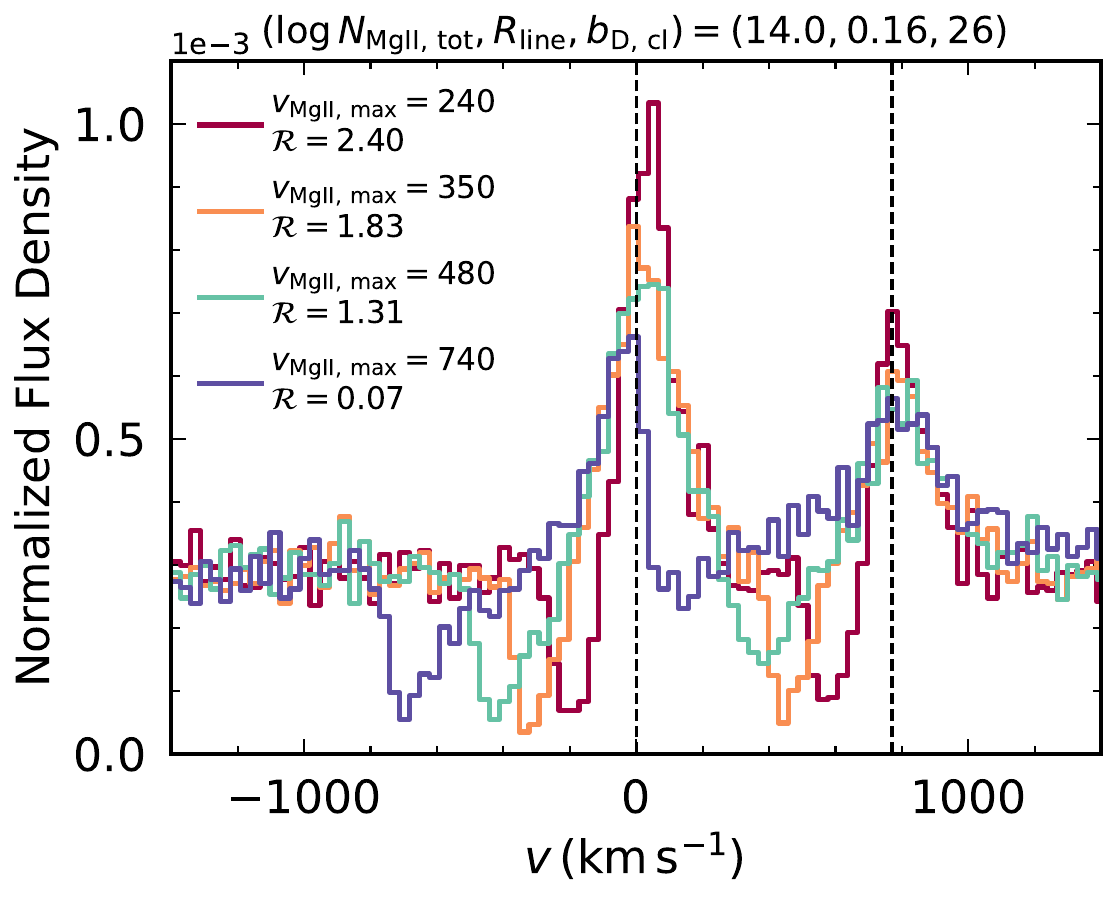}
    \caption{\textbf{Effect of varying the maximum clump outflow velocity on the {\MgII} model spectra.} The left and right panel shows a case with low and high \MgII\ total column density, respectively. The clump optical depth and the aperture correction factor are set to default values, $\tau_{\rm d,\,cl} = 0$ and $b_{\rm max} / R_{\rm halo} = 1.0$. 
    \label{fig:individual_parameter1}}
\end{figure*}

    \item Non-leakers: Among the thirteen non-leakers, one (J0723+4146) is actually situated in the same regime as the strong leakers. It remains unclear whether this non-leaker does share certain characteristics with the strong leakers or this is an artifact caused by its relatively noisy spectrum and/or the low-resolution observation of J1517+3705 (see discussion above). The rest of the non-leakers appear relatively scattered in the $v_{\rm MgII,\,max}$ -- $N_{\rm MgII,\,tot}$ plane but can be categorized into two subgroups. One subgroup has fairly low maximum clump outflow velocities ($v_{\rm MgII,\,max} \lesssim 360\,\rm km\,s^{-1}$), while the other exhibits high total \MgII\ column densities \MgII\ column densities $N_{\rm MgII,\,tot} \gtrsim 10^{14.5}\,\rm cm^{-2}$. Morphologically, the absorption trough is either insignificant or situated relatively close to the line center ($|v_{\rm trough}| \lesssim 300\,\rm km\,s^{-1}$), or it is very deep and extended due to large total \MgII\ column densities (e.g. J1346+1129 and J1314+1048). 
\end{itemize}
    
    Among all 33 modeled objects, we note that the spectrum of J0826+1820 is of poor quality, resulting in a less reliable model fit. We therefore do not fully trust the best-fit parameters for this object and use a hollow black circle in Figure \ref{fig:vmax_mgII} to represent it. In addition, two objects -- J1235+0635 (a moderate leaker) and J1244+0215 (a non-leaker) -- exhibit an intriguing ``quadruple peak'' spectral morphology (see Figures \ref{fig:moderate_leakers} and \ref{fig:non_leakers}) due to strong absorption at the line center of both the \MgII\ H and K transitions. This phenomenon can be attributed to their inferred combination of very high \MgII\ total column density ($N_{\rm MgII,\,tot} \gtrsim 10^{14.4}\,\rm cm^{-2}$) and extremely low clump outflow velocity ($v_{\rm MgII,\,max} \lesssim 30\,\rm km\,s^{-1}$). We will explore this point in more detail in the next section.

\subsection{Effect of Each Individual Parameter on {\MgII} Model Spectra}

\begin{figure*}
\centering
\includegraphics[width=0.485\textwidth]{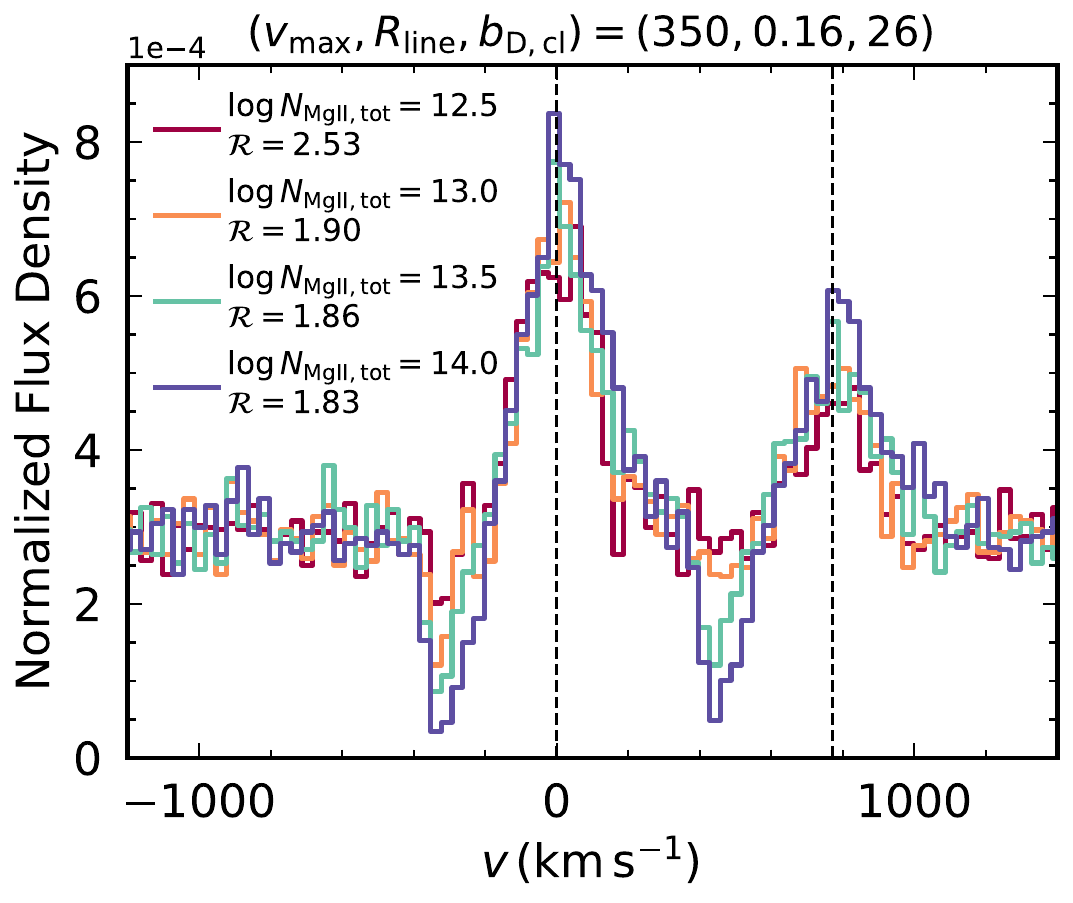}
\includegraphics[width=0.505\textwidth]{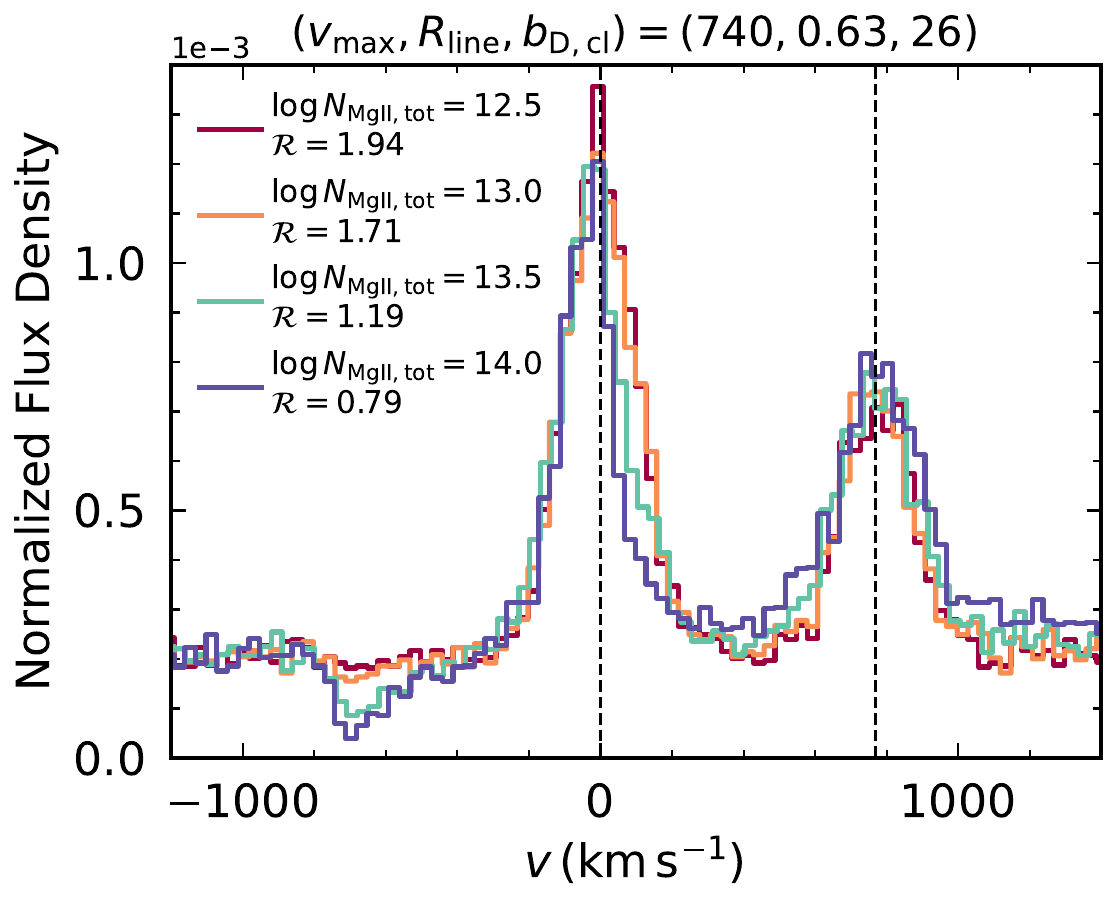}
    \caption{\textbf{Effect of varying the total {\MgII} column density on the {\MgII} model spectra.} The left and right panel shows a case with low and high maximum clump outflow velocity, respectively. The clump optical depth and the aperture correction factor are set to default values, $\tau_{\rm d,\,cl} = 0$ and $b_{\rm max} / R_{\rm halo} = 1.0$.
    \label{fig:individual_parameter2}}
\end{figure*}

\begin{figure*}
    \centering
    \includegraphics[width=0.495\textwidth]{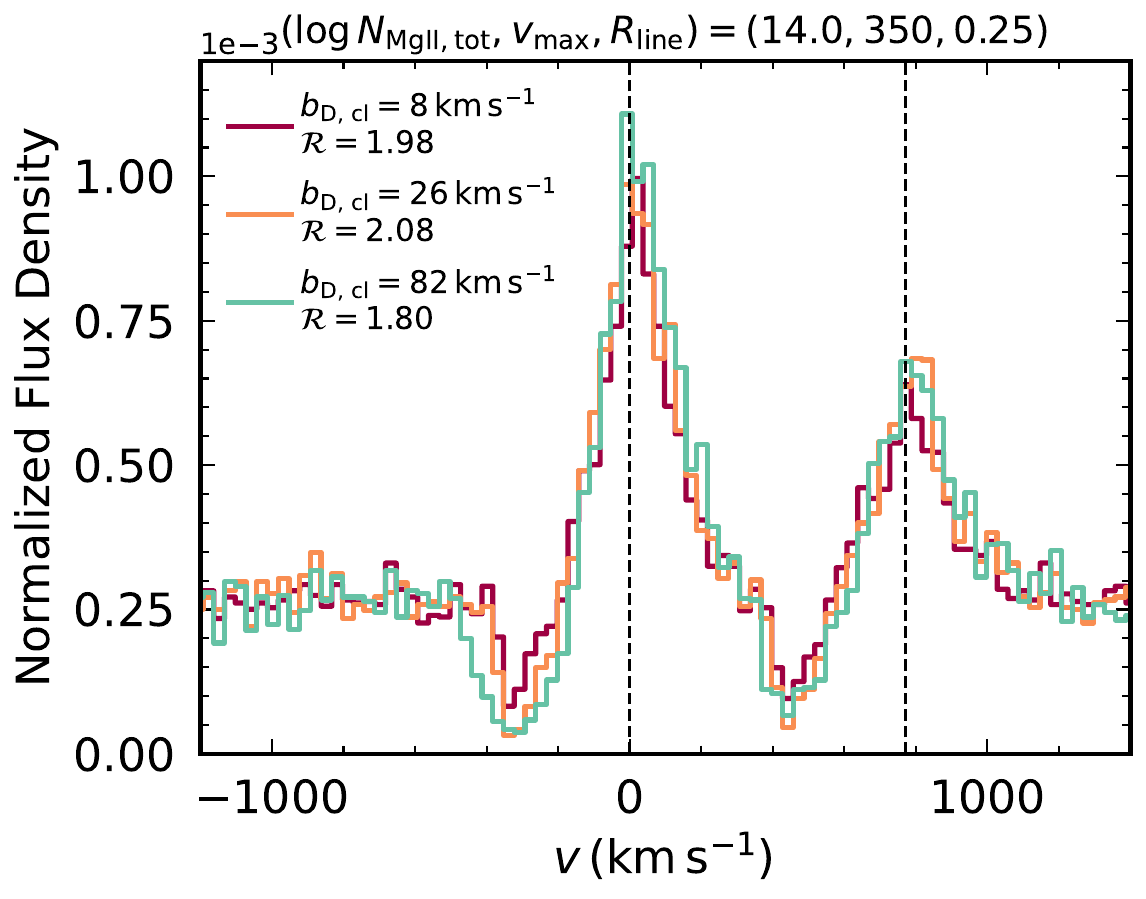}
    \includegraphics[width=0.485\textwidth]{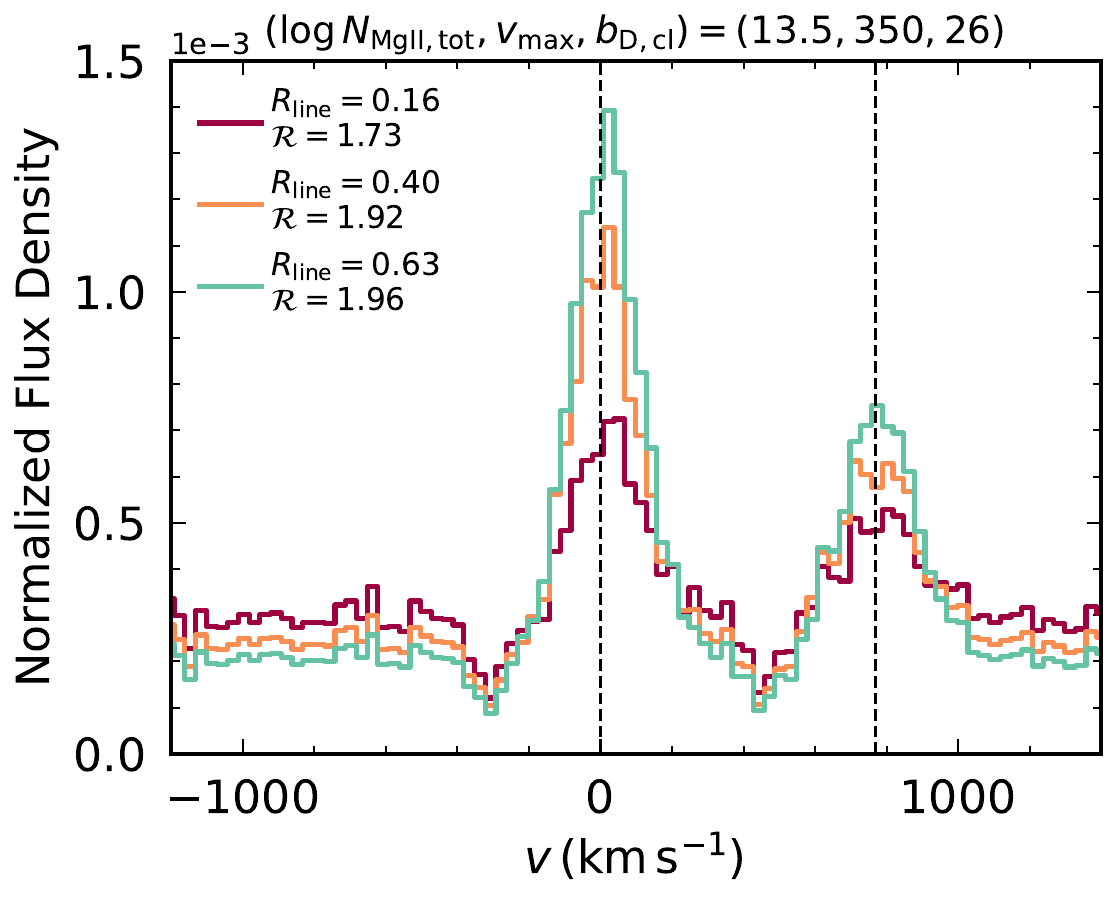}
    \includegraphics[width=0.485\textwidth]{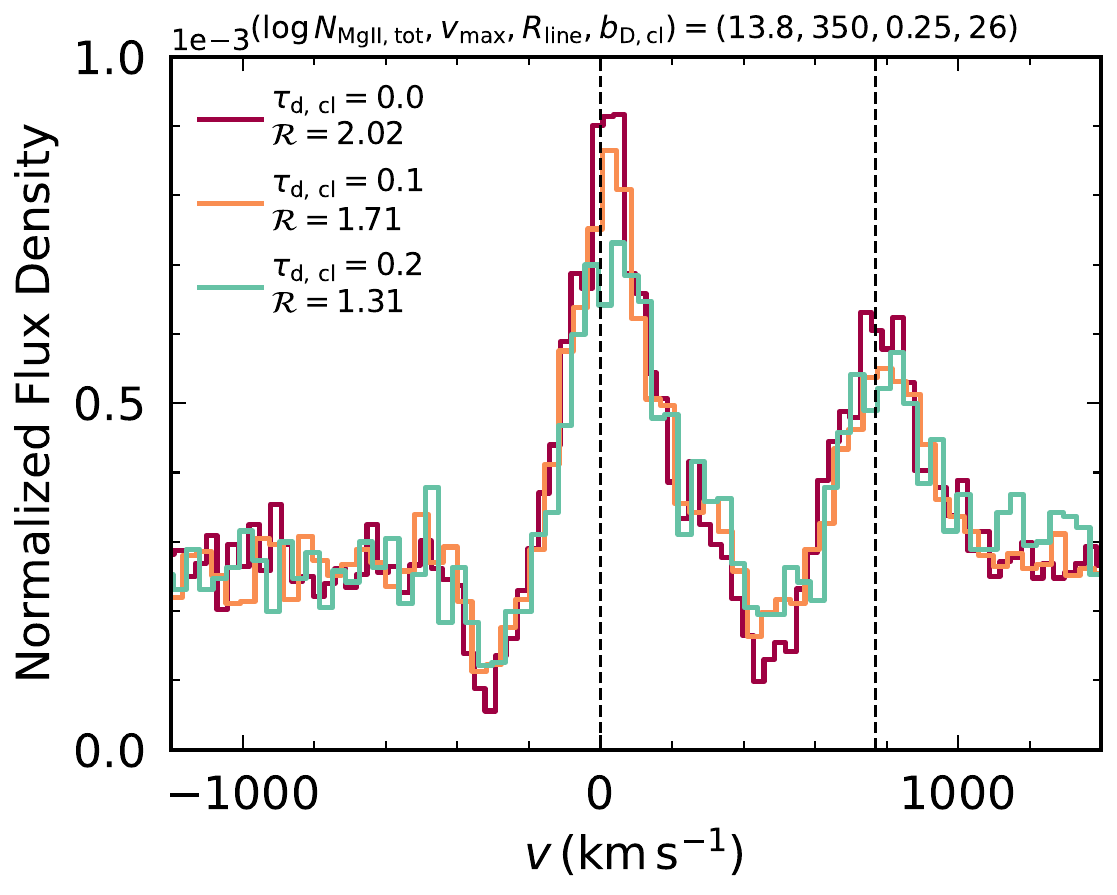}
    \includegraphics[width=0.495\textwidth]{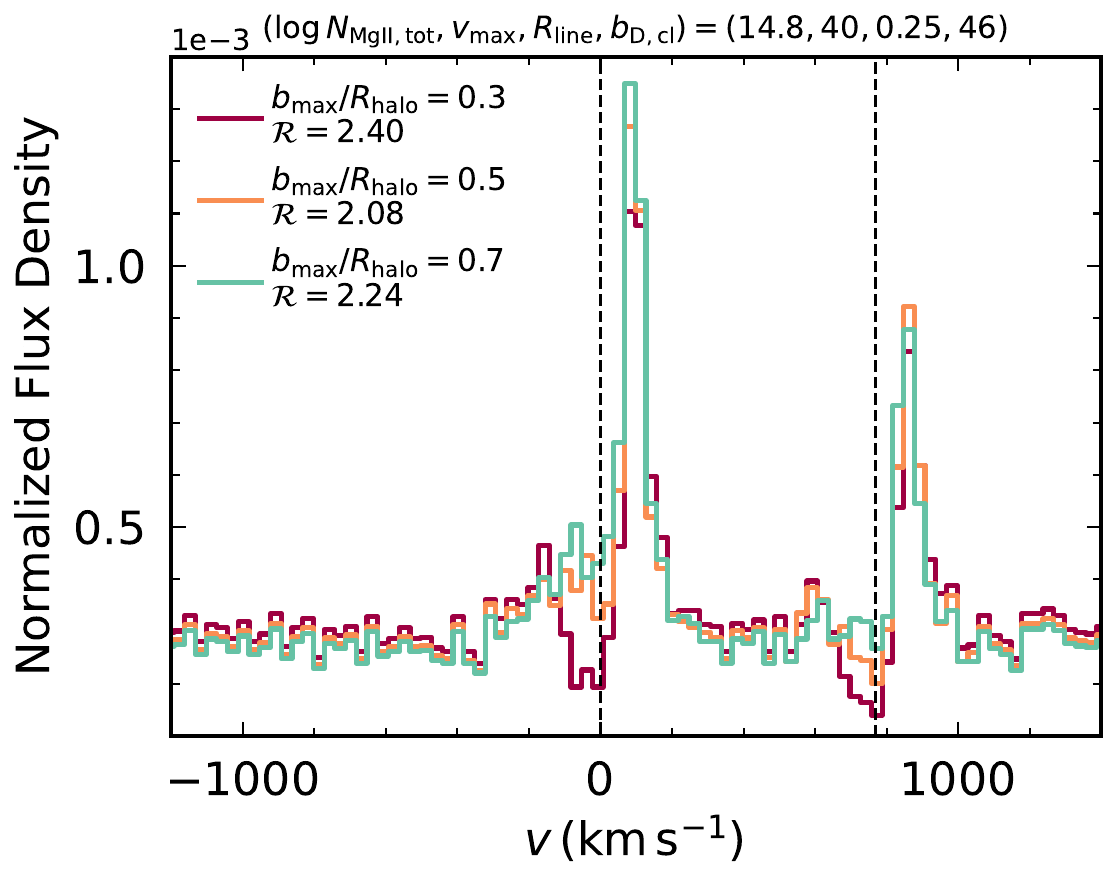}
    \caption{\textbf{Effect of varying additional model parameters on the {\MgII} model spectra, including the clump Doppler parameter $b_{\rm D,\,cl}$, the relative strength of line emission compared to continuum emission $R_{\rm line}$, the clump dust absorption optical depth $\tau_{\rm d,\,cl}$, and the aperture correction factor $b_{\rm max}/R_{\rm halo}$.} 
    \label{fig:individual_parameter3}}
\end{figure*}

To gain a deeper understanding of the results from \MgII\ RT modeling, this section explores the effect of varying individual model parameters on the \MgII\ spectra using a suite of RT simulations. In addition to analyzing the shape of \MgII\ line profiles, we calculate the equivalent width ratio, $\mathcal{R} = \rm EW_{2796}/EW_{2803}$, for each profile to characterize the relative strength of the K and H lines. In the limit where the spectrum is purely emission, $\mathcal{R}$ converges to the doublet flux ratio $F_{\rm 2796}/F_{\rm 2803}$, which has been frequently used in prior studies (e.g., \citealt{Chisholm2020, Chang24, Seon24}).

In the following, we vary each key model parameter individually while keeping the others fixed to examine their effects on the \MgII\ model spectra:

\begin{itemize}
    \item Clump maximum outflow velocity $v_{\rm MgII,\,max}$: We explore the effect of this parameter in two different \MgII\ total column density regimes. The first example has $N_{\rm MgII,\,tot} = 10^{13.0}\,\rm cm^{-2}$, representing strong leakers such as J1158+3125, J1033+6353, J1410+4345, and J1327+4218. The second example has $N_{\rm MgII,\,tot} = 10^{14.0}\,\rm cm^{-2}$, representing non-leakers like J1129+4935, J0129+1459, J0723+4146, and J1314+1048. As shown in Figure \ref{fig:individual_parameter1}, the major difference between the two cases is that the absorption troughs are generally much shallower and narrower in the low column density scenario, where $v_{\rm MgII,\,max}$ is primarily constrained by the relative intensity of the two peaks (see Appendix \ref{sec:special_case}).

    Overall, increasing $v_{\rm MgII,\,max}$ tends to suppress both emission peaks (to a greater extent for the main peak at the K transition) and reduces the equivalent width ratio $\mathcal{R}$. It also shifts the absorption trough away from the line centers of the K and H transitions, with $v_{\rm trough} \sim -v_{\rm MgII,\,max}$, since that is where resonant absorption happens and optical depth is the greatest. In addition, an interesting phenomenon occurs when $v_{\rm MgII,\,max}$ approaches the velocity separation between the K and H transitions ($\sim 770\,\rm km\,s^{-1}$). In this case, the H peak gets enhanced as some K photons are scattered into the H transition, causing a sharp drop in $\mathcal{R}$. Moreover, the absorption trough associated with the H transition becomes blurred and ill-defined (see also \citealt{Chang24}). 
    
    \item Clump \MgII\ total column density: This parameter is defined as the average total \MgII\ along a sightline: 
    $N_{\rm MgII,\,tot} = \frac{4}{3} f_{\rm cl} N_{\rm MgII,\,cl}$, where $f_{\rm cl} = \frac{3}{4} F_{\rm V} R_{\rm halo}/{R_{\rm cl}}$ is the average number of clumps per sightline and $N_{\rm MgII,\,cl}$ is the \MgII\ column density of each clump. Here we also consider two examples with different $v_{\rm MgII,\,max}$ values in Figure \ref{fig:individual_parameter2}: 350 and 740 $\rm km\,s^{-1}$, respectively. In both cases, increasing $N_{\rm MgII,\,tot}$ deepens the absorption troughs without shifting their locations in velocity space, and also decreases the doublet EW ratio $\mathcal{R}$. 
    
    Nevertheless, we have observed a very interesting phenomenon: in the low $v_{\rm MgII,\,max}$ example, increasing $N_{\rm MgII,\,tot}$ enhances both emission peaks, as no photon exchange occurs between the K and H transitions, and the deepening of the absorption trough is compensated by the increase of the associated emission peak. In contrast, in the high $v_{\rm MgII,\,max}$ example, increasing $N_{\rm MgII,\,cl}$ actually suppresses the K peak while enhancing the H peak, due to more K photons being scattered into the H transition.

    \item The Doppler parameter of each clump $b_{\rm D,\,cl}$: This parameter represents the turbulent velocity within each clump. As shown in Figure \ref{fig:individual_parameter3}, increasing $b_{\rm D,\,cl}$ tends to broaden and strengthen both the emission peaks and absorption troughs, as the perturbed velocity field of the \MgII\ gas causes photons across a wider frequency range to participate in scattering. The effect on the doublet EW ratio $\mathcal{R}$ is relatively minor -- interestingly, $\mathcal{R}$ slightly increases as $b_{\rm D,\,cl}$ rises from 8 $\rm km\,s^{-1}$ to 26 $\rm km\,s^{-1}$ due to a more significant decrease in the EW for the H transition. However, as $b_{\rm D,\,cl}$ continues to increase, $\mathcal{R}$ decreases again.
        
    \item The relative strength of line emission versus continuum emission $R_{\rm line}$: This parameter is defined as the ratio of line photons to continuum photons in an RT simulation, reflecting the relative contribution of the two emission mechanisms. For example, $R_{\rm line} = 0$ corresponds to pure continuum emission, while $R_{\rm line} = 1$ indicates equal numbers of line and continuum photons are emitted. As $R_{\rm line}$ increases, the spectrum gradually shifts from being continuum-dominated to line-dominated -- this is evident as the continuum level decreases and the emission peaks become more prominent. The EWs of both the K and H transitions increase, with the K transition experiencing a greater enhancement, leading to an overall increase in the EW ratio $\mathcal{R}$. 

    \item The clump dust absorption optical depth $\tau_{\rm d,\,cl}$: In our modeling, we have assumed the presence of dust within the clumps, which influences the RT of \MgII\ photons through dust scattering and absorption. As $\tau_{\rm d,\,cl}$ increases, the emission peaks diminish due to dust extinction, and the absorption troughs become shallower, as dust extinction weakens the \MgII\ resonant scattering. In general, the effect of dust is more pronounced for the K line than for the H line, leading to a decrease in the doublet EW ratio $\mathcal{R}$.
    
    \item The aperture correction factor $b_{\rm max} / R_{\rm halo}$: This parameter is designed to mimic the aperture loss of \MgII\ photons in real observations, where $b_{\rm max}$ represents the maximum impact parameter of the scattered photons included in the model spectra. The effect of this parameter on the \MgII\ spectra is generally minor, except when the total \MgII\ column density is very high and the clump's maximum outflow velocity is very low. In this regime, the clump’s optical depth is high enough for photons to travel to large impact parameters.
    
    Here we use a ``quadruple peak'' example to illustrate the effect of $b_{\rm max} / R_{\rm halo}$. In this case, photons that are originally near the line center are scattered to large impact parameters, and when $b_{\rm max} / R_{\rm halo}$ is small, these photons are excluded, yielding a deep trough at the line center (even below the continuum level). As $b_{\rm max} / R_{\rm halo}$ increases, the photons with frequencies near the line center are gradually included in the spectra, and the flux at the line center rises above the continuum level again.

\end{itemize}
Combining the analysis from this section and the previous section, we find that the two most important parameters shaping the \MgII\ line profiles are the clump outflow velocity and the total \MgII\ column density. The other four parameters analyzed above ($b_{\rm D,\,cl}, R_{\rm line}, \tau_{\rm d,\,cl},$ and $b_{\rm max} / R_{\rm halo}$) have only subdominant effects on the \MgII\ spectra, and there are no significant correlations between their best-fit values inferred from RT modeling and the amount of LyC leakage. More importantly, it is evident that the effects of different model parameters are highly intertwined, suggesting any empirical indices characterizing specific features of a \MgII\ spectrum may not be sufficient to deduce the underlying gas properties. For instance, a higher doublet EW ratio $\mathcal{R}$ could result from either a lower clump outflow velocity, a lower clump column density, a higher turbulent velocity within the clumps, a higher fraction of line emission, a lower clump dust absorption optical depth, or simply a greater aperture loss. To accurately determine the physical parameters of the scattering gas in the ISM / CGM, systematic RT modeling of the full \MgII\ doublet profile is indispensable, as it is virtually impossible to make any reliable inferences about the underlying gas properties without such simulations.

\begin{table*}
\centering
    \scriptsize \caption{Parameter values of the fiducial model grid used for \lya\ RT modeling.}
    \label{tab:params2}
    \setlength{\tabcolsep}{1pt}
    \begin{tabular}{ccc}
    \hline\hline
    Parameter & Definition & Values\\ 
     (1)  & (2) & (3)\\
    \hline
    ${\rm log}\,n_{\rm HI,\,ICM}$ & ICM residual \HI\ number density & (-8.4, -8.0, ..., -6.8) log cm$^{-3}$\\
    $v_{\rm ICM}$ & ICM outflow velocity & (0, 100, 200) km\,s$^{-1}$\\
    $F_{\rm V}$ & Clump volume filling factor & (0.005, 0.01, 0.02, ..., 0.06)\\
     ${\rm log}\,N_{\rm HI,\,cl}$ & Clump {\rm H\,{\textsc {i}}} column density & (17.5, 18.0, ..., 19.5) log cm$^{-2}$\\
     $b_{\rm D,\,cl}$ & Clump Doppler parameter & (23, 41, 72, 129, 229)\tablenotemark{a} km\,s$^{-1}$ \\
     $\mathcal{V}_{\infty}$ & Clump asymptotic outflow velocity & (300, 500, ..., 900) km\,s$^{-1}$ \\
     $\alpha$ & Clump acceleration power-law index & (1.1, 1.6, ..., 2.6) \\
     $\tau_{\rm d,\,cl}$ & Clump dust absorption optical depth & (0, 0.03, 0.05, 0.1, 0.2, 0.3)\\
     $b_{\rm max}$ & Maximum photon impact parameter& (0.5, 1, 1.5, 2, 4, ..., 10) kpc\\
     $\Delta v$ & Velocity shift relative to systemic $z$ & [-120,\,120] km\,s$^{-1}$  (continuous)\\
    \hline\hline
    \end{tabular}
    \tablenotetext{}{\textbf{Notes.} The parameter values of the fiducial model grid used for fitting the \lya\ profiles. The columns are: (1) parameter name; (2) parameter definition; (3) parameter values on the grid.}
    \tablenotetext{a}{This parameter is varied in increments of 10$^{0.25}$ on the fiducial model grid.}
\end{table*}

\begin{figure*}
\centering
\includegraphics[width=0.325\textwidth]{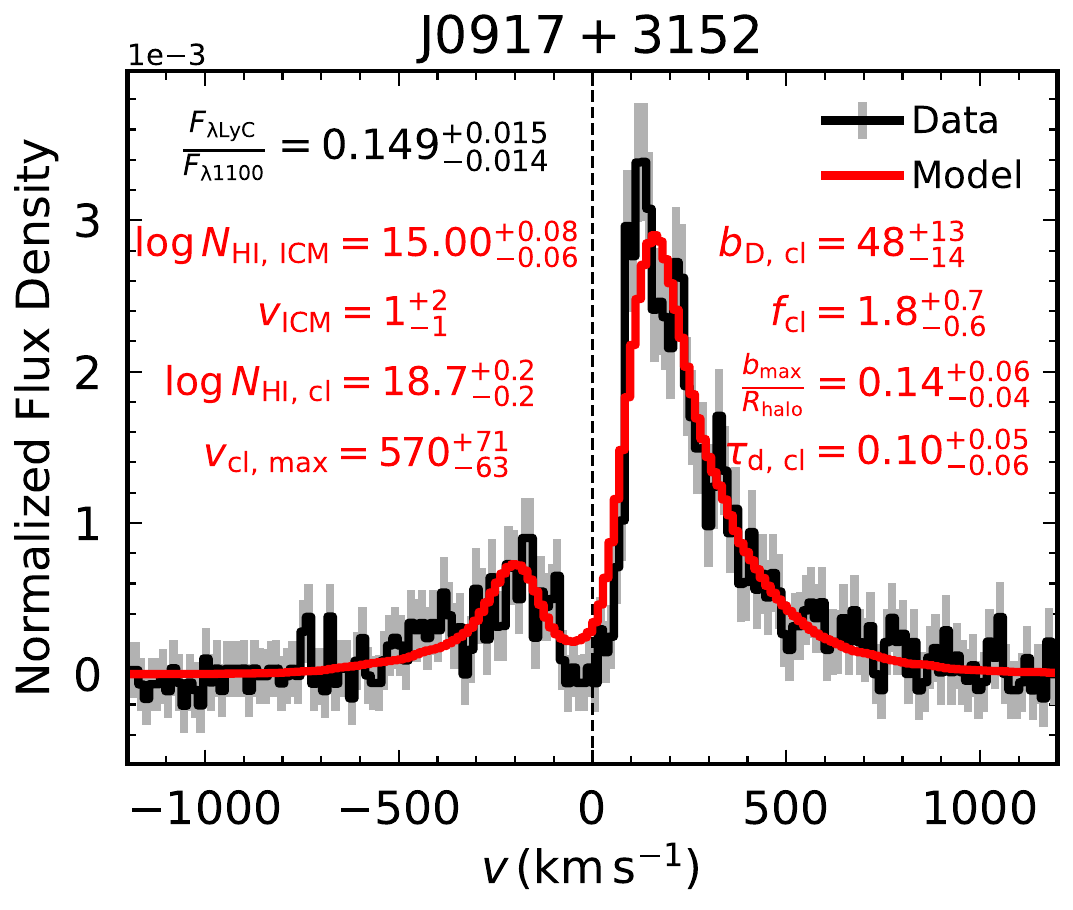}
\includegraphics[width=0.325\textwidth]{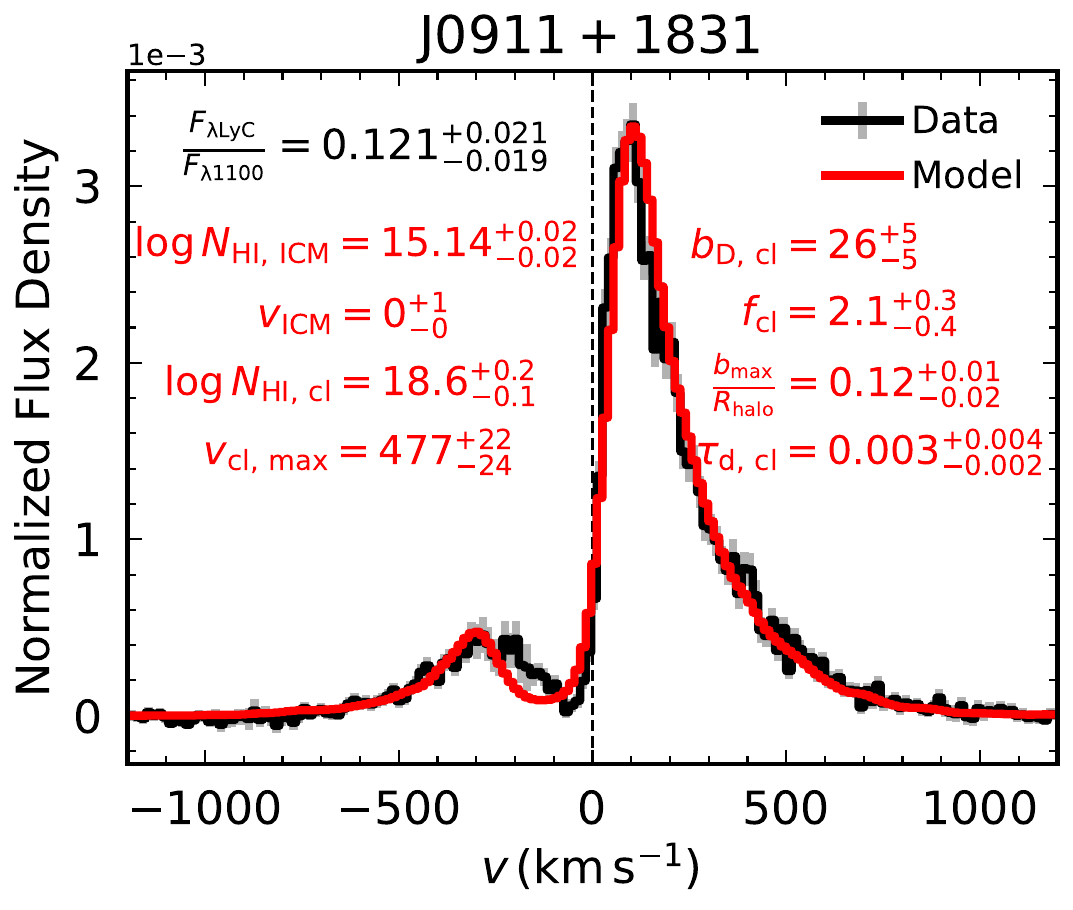}
\includegraphics[width=0.325\textwidth]{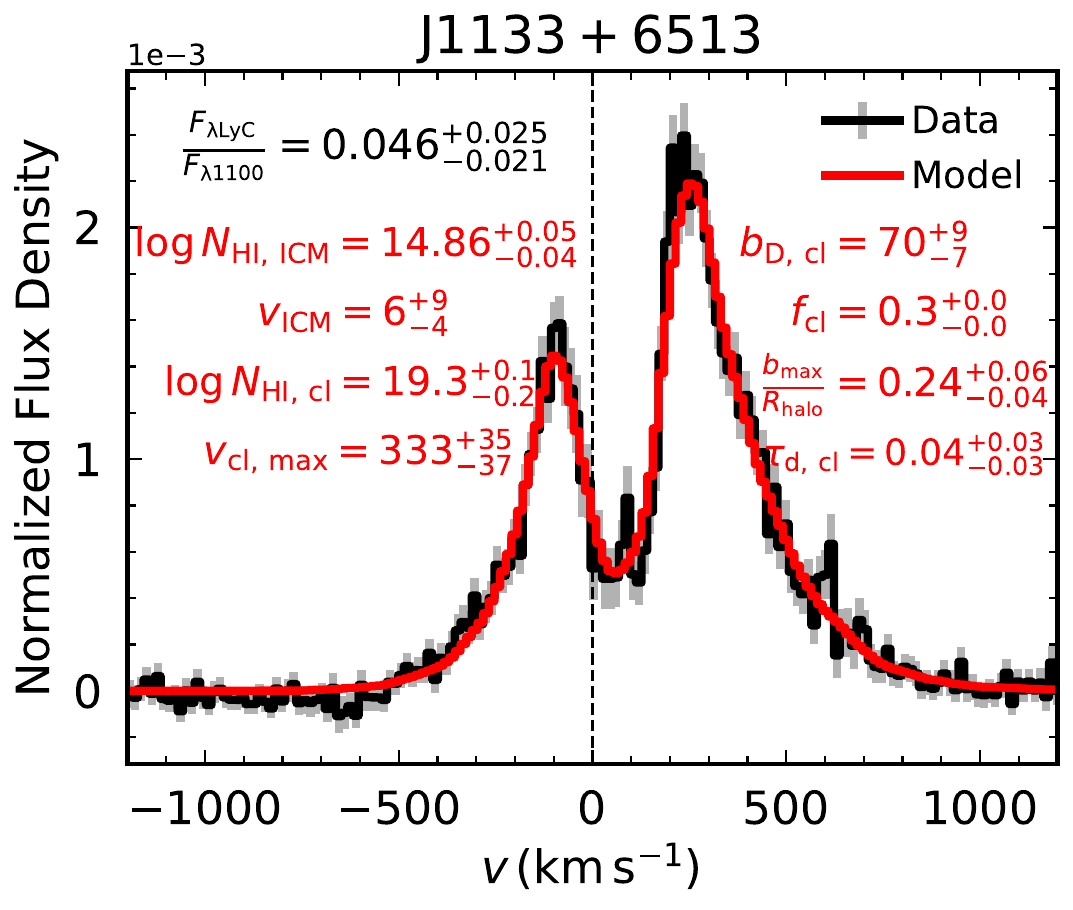}
\includegraphics[width=0.325\textwidth]{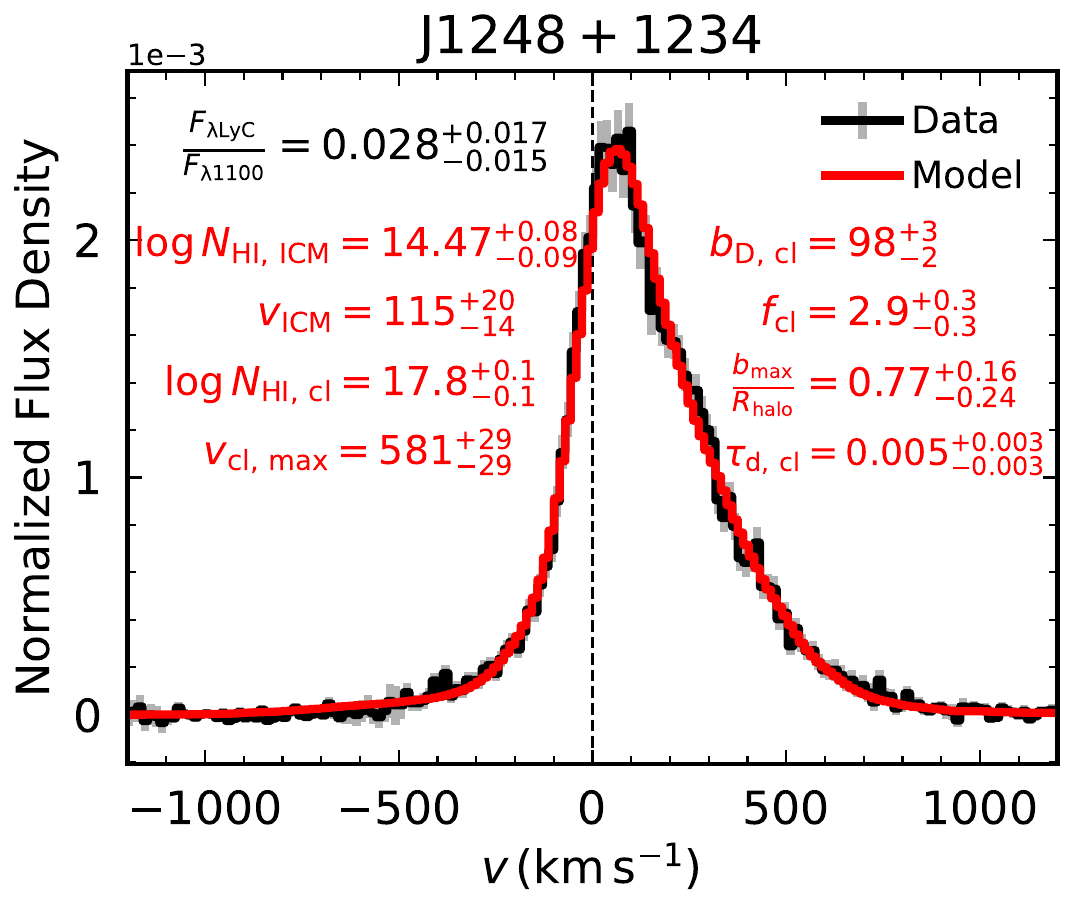}
\includegraphics[width=0.325\textwidth]{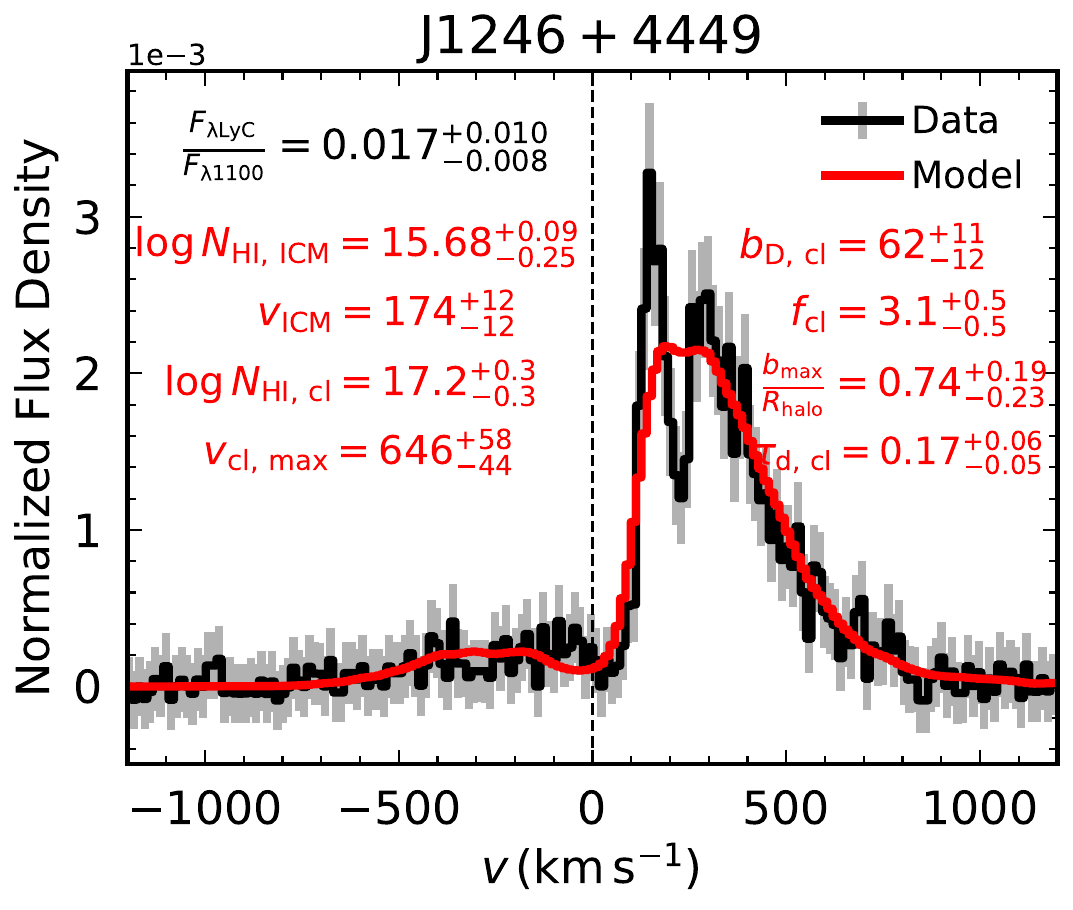}
\includegraphics[width=0.325\textwidth]{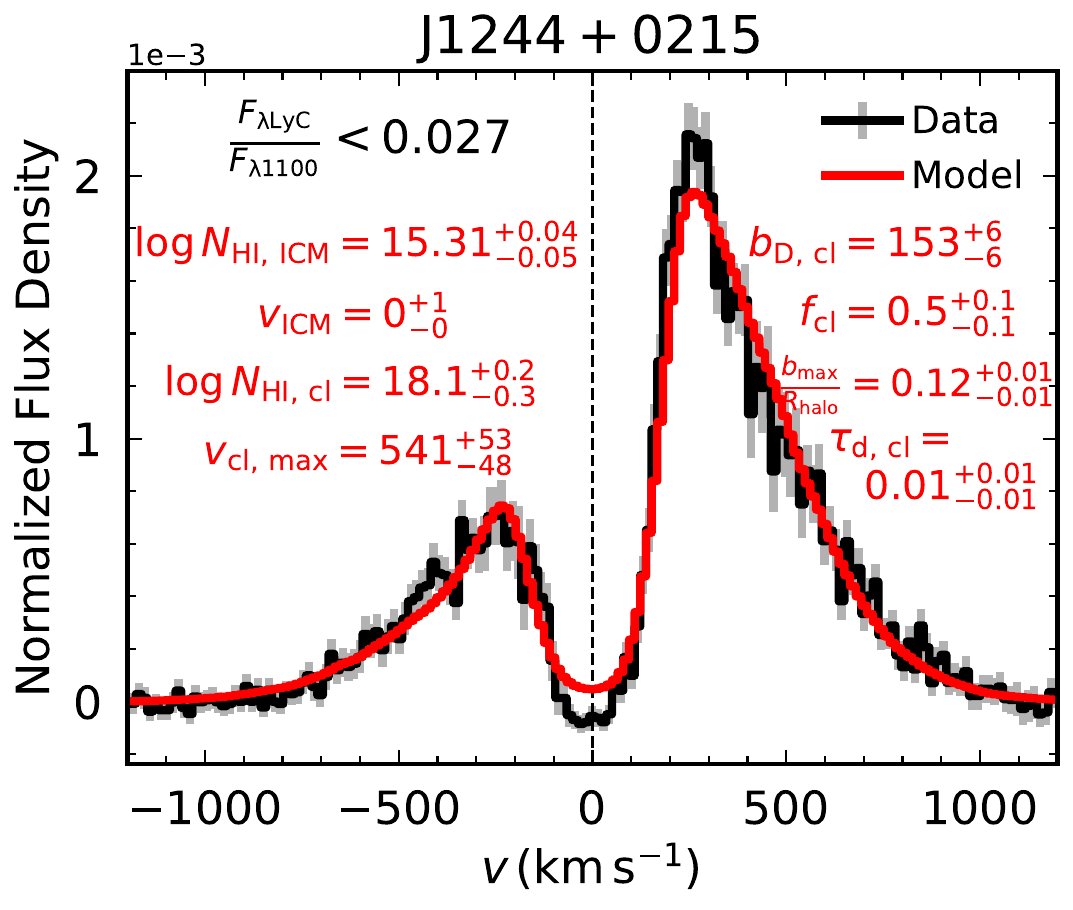}
    \caption{\textbf{\lya\ best-fits for six galaxies in the sample obtained by RT modeling.} Among these six objects, J0917+3152 and J0911+1831 are strong leakers; J1133+6513, J1248+1234, and J1246+4449 are potential leakers; J1244+0215 is a non-leaker. It is clear that significant flux near the \lya\ line center does not necessarily correlate with strong LyC leakage. In fact, both of the strong leakers show virtually no flux near the line center, whereas the two objects with the largest flux near the line center are only potential leakers, with $F_{\rm \lambda LyC} / F_{\rm \lambda 1100} < 0.05$, detected at $\lesssim 2\sigma$ significance.
    \label{fig:lya_best_fits}}
\end{figure*}

\section{Modeling the \lya\ Profiles of the LyC Leakers with A Multiphase, Clumpy Model}\label{sec:lya_modeling}

\subsection{\lya\ RT Modeling}

Although modeling the \MgII\ emission can provide insights into the properties of clumpy gas in the ISM / CGM, these gas properties cannot, in principle, be directly used to infer a galaxy's LyC leakage. This is because \MgII\ emission only constrains the properties of \MgII\ gas, and the relationship between \MgII\ gas and \HI\ gas (which is ultimately responsible for LyC leakage) remains unclear. It is uncertain whether \MgII\ is consistently associated with \HI\ in each gas clump; even if they do co-exist, it is unknown whether they share the same kinematics or spatial extent, or if they have a fixed \MgII-to-\HI\ ratio. Therefore, rather than relying on empirically calibrated \MgII-\HI\ relations, we conduct RT modeling of \lya\ profiles for certain objects in our sample to directly constrain the \HI\ properties in their CGM and further infer their LyC leakage.

We searched for archival HST COS/G160M observations for our sample and found that six galaxies have high-resolution \lya\ profiles published (\citealt{Henry15, Yang17, Henry18}, HST Program ID 12928, 14201, 15865). Among these six galaxies, two are strong leakers (J0917+3152 and J0911+1831), three are potential leakers (J1133+6513, J1248+1234, and J1246+4449), and one is a non-leaker (J1244+0215). The continuum-subtracted \lya\ profiles of these six objects are shown in Figure \ref{fig:lya_best_fits}. At first glance, if we follow the conventional wisdom on inferring LyC leakage from \lya\ profiles (e.g., based on shell model results, \citealt{Verhamme15}; see also \citealt{Naidu22}), one might expect that the galaxy with the most significant flux near the line center (J1248+1234) would be the strongest LyC leaker, with J1133+6513 being the second strongest. However, this is not the case: both are actually potential leakers with $F_{\rm \lambda LyC} / F_{\rm \lambda 1100} < 0.05$ detected at $\lesssim 2\sigma$ significance. Meanwhile, the two strong leakers, J0917+3152 and J0911+1831, exhibit virtually no flux at the line center. This phenomenon poses a challenge to the conventional \lya-LyC connection and prompts us to conduct RT modeling to investigate further.

We use a similar setup for \lya\ modeling as in our \MgII\ modeling, but we also incorporate a hot, inter-clump medium (ICM) that may contain some residual \HI\ capable of scattering \lya\ photons \citep{Li22, Erb23}. This hot gas component, assumed to have a temperature of $\sim10^6$ K, is a volume-filling medium characterized by two parameters: the \HI\ number density, $n_{\rm HI,\,ICM}$, and the radial outflow velocity, $v_{\rm ICM}$ (assumed to be constant). We refer the readers to Figure \ref{fig:schematic} for a schematic representation of the model. The fitting procedure is similar to what we described in Section \ref{sec:mgII_RT}; however, since we do not need $R_{\rm line}$ and $f_{\rm scale}$ for \lya\ modeling, each fitting run still contains a total of 10 free parameters. We present the parameter values for the fiducial \lya\ model grid in Table \ref{tab:params2}. 

We present the best-fit results using the multiphase, clumpy RT model in Figure \ref{fig:lya_best_fits}. While the RT models have successfully reproduced all six \lya\ line profiles, several best-fit parameter values are both intriguing and somewhat puzzling, warranting further discussion. The key findings are as follows:

\begin{itemize}
    \item Clump maximum outflow velocity: We do not observe a clear correlation between the maximum clump outflow velocities inferred from \lya\ and \MgII\ modeling. For J0917+3152, J0911+1831, and J1133+6513, the velocities inferred by both methods are comparable, suggesting the Mg$^+$ and \HI\ gas may track each other kinematically; however, for the other three objects, the outflow velocities inferred from \lya\ are at least twice as large as those inferred from \MgII, and the \MgII-inferred clump maximum outflow velocity lies between the ICM outflow velocity and the \lya-inferred clump maximum outflow velocity (i.e., $v_{\rm ICM} < v_{\rm cl,\,max,\,MgII} < v_{\rm cl,\,max,\,Ly\alpha}$).
    
    \item Clump \HI\ column density: There does not seem to be a clear correlation between the inferred clump \HI\ and \MgII\ column densities either. For the two strong leakers, J0917+3152 and J0911+1831, the ratio of the total column densities of \MgII\ to \HI\ is approximately $10^{-5.5}$, similar to the Mg abundance [Mg/H] in the warm neutral medium of the Milky Way \citep{Jenkins09}. In contrast, J1133+6513 exhibits a much lower Mg-to-\HI\ ratio of $10^{-6.7}$, while J1248+1234, J1246+4449 and J1244+0215 show significantly higher ratios\footnote{Note that a higher Mg-to-\HI\ ratio does not necessarily indicate a higher [Mg/H] abundance, as the column density of H$^+$ in the CGM is unknown from RT modeling (\citealp[see e.g., Section 5.3.2 in][]{Xu2022b}).} of $10^{-4.0}$ to $10^{-2.7}$.

    \item \HI\ in the hot, inter-clump medium: This component primarily contributes to additional \lya\ scattering near the line center. Unsurprisingly, the two objects with the most significant flux near the line center, J1133+6513 and J1248+1234, have the lowest \HI\ column density in the diffuse inter-clump medium ($N_{\rm HI,\,ICM} \lesssim 10^{15}\,\text{cm}^{-2}$). In particular, J1248+1234 has an ICM outflow velocity of $\sim 120\,\text{km}\,\text{s}^{-1}$, which further reduces the \HI\ optical depth at the line center.

    \item Clump volume filling factor (and therefore clump covering factor): For J0917+3152, J0911+1831, J1248+1234 and J1246+4449, the $F_{\rm V}$ values inferred from both \lya\ and \MgII\ modeling are consistent, around 2 -- 4\%. However, for J1133+6513 and J1244+0215, we find large $F_{\rm V}$ values inferred from \MgII\ modeling (3\% and 4\%), but much smaller values inferred from \lya\ modeling (0.4\% and 0.7\%). This discrepancy suggests that the \HI\ gas and Mg$^+$ may not be fully co-spatial; the \HI\ gas may reside in smaller clumps, thereby occupying a smaller volume fraction of the halo.
\end{itemize}

\subsection{Inferring LyC Leakage via \lya\ RT Models}
Having completed the modeling of the observed \lya\ profiles using the multiphase, clumpy RT models, it is now possible to theoretically predict the LyC escape fraction from the best-fit models, as we have determined the spatial distribution and column densities of the \HI\ clumps. To do this, we inject 10,000 photons at wavelengths far from the \lya\ line center (to simulate LyC photons) and record the \HI\ column densities encountered by the photons before they either escape or are absorbed by dust. Since the \HI\ cross-section for LyC photons is $\sim 6.3 \times 10^{-18}\,\rm cm^{-2}$, we classify photons that encounter a total \HI\ column density lower than $\sim 1.6 \times 10^{17}\,\rm cm^{-2}$ and are not absorbed by dust as successfully escaping. The theoretical LyC escape fraction is then calculated by dividing the number of escaped LyC photons by the total number of injected LyC photons. We find that, given the high best-fit clump \HI\ column densities (all greater than $10^{17} \rm cm^{-2}$),  the dominant factor for LyC escape is the clump covering factor, $f_{\rm cl}$. Using the 2-$\sigma$ ranges for the best-fit $F_{\rm V}$ values, along with the corresponding parameter values from the high-likelihood region of the posteriors derived from \lya\ modeling and the relation $f_{\rm cl} = \frac{3}{4} F_{\rm V} R_{\rm halo}/{R_{\rm cl}}$, we infer the value and uncertainties of $f_{\rm cl}$ for all six objects. We then run RT simulations using the derived $f_{\rm cl}$ values to estimate the values and uncertainties of the theoretical LyC escape fractions.

\begin{figure}
\centering
\includegraphics[width=0.5\textwidth]{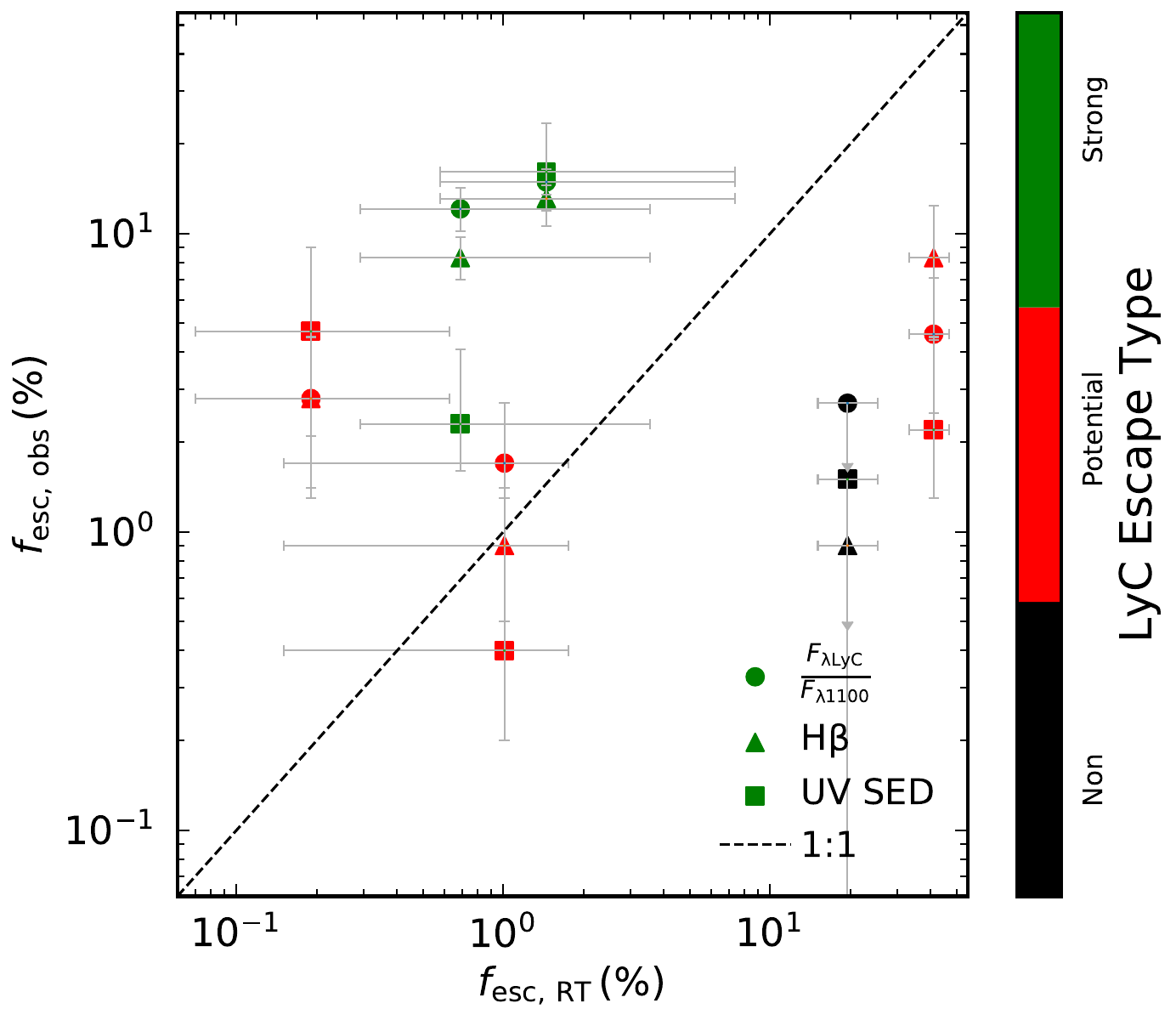}
    \caption{\textbf{Observed LyC escape fractions determined using three different metrics compared to the RT-inferred LyC escape fractions for six galaxies in our sample. } Different point shapes represent the different metrics used to estimate the observed LyC escape fraction (circles for the observed flux near $\lambda_{\rm rest} = 912\,\rm \AA$, triangles for H$\beta$, and squares for the UV SED). Among these six objects, the non-leaker J1244+0215 and the potential leaker J1133+6513 actually exhibit the highest RT-inferred LyC escape fractions, due to their low inferred clump covering factors from \lya\ RT modeling.
    \label{fig:fesc_relation}}
\end{figure}

We plot the RT-inferred LyC escape fractions, $f_{\rm esc,\,RT}$, against the observed escape fraction determined using three different metrics (observed flux near $\lambda_{\rm rest} = 912\,\rm \AA$, H$\rm \beta$, and the UV SED, \citealt{Flury22}) in Figure \ref{fig:fesc_relation}. While the two strong leakers do, on average, exhibit higher $f_{\rm esc,\,RT}$ values compared to two of the potential leakers, it is puzzling to observe that the non-leaker J1244+0215 and the potential leaker J1133+6513 actually exhibit the highest RT-inferred LyC escape fractions. We double-checked these two objects and found that their $f_{\rm cl}$ inference is quite robust, consistently remaining below one. This result indicates that the average number of clumps per sightline is less than one, suggesting a low overall gas covering fraction and the existence of channels that are free of high-$N_{\rm HI}$ clumps. Interestingly, at face value, there seems to be no clear connection between their high RT-inferred LyC escape fractions and their spectral morphology -- J1133+6513 shows significant flux near the line center, whereas J1244+0215 displays almost no flux at the line center.

\begin{figure}
\centering
\includegraphics[width=0.5\textwidth]{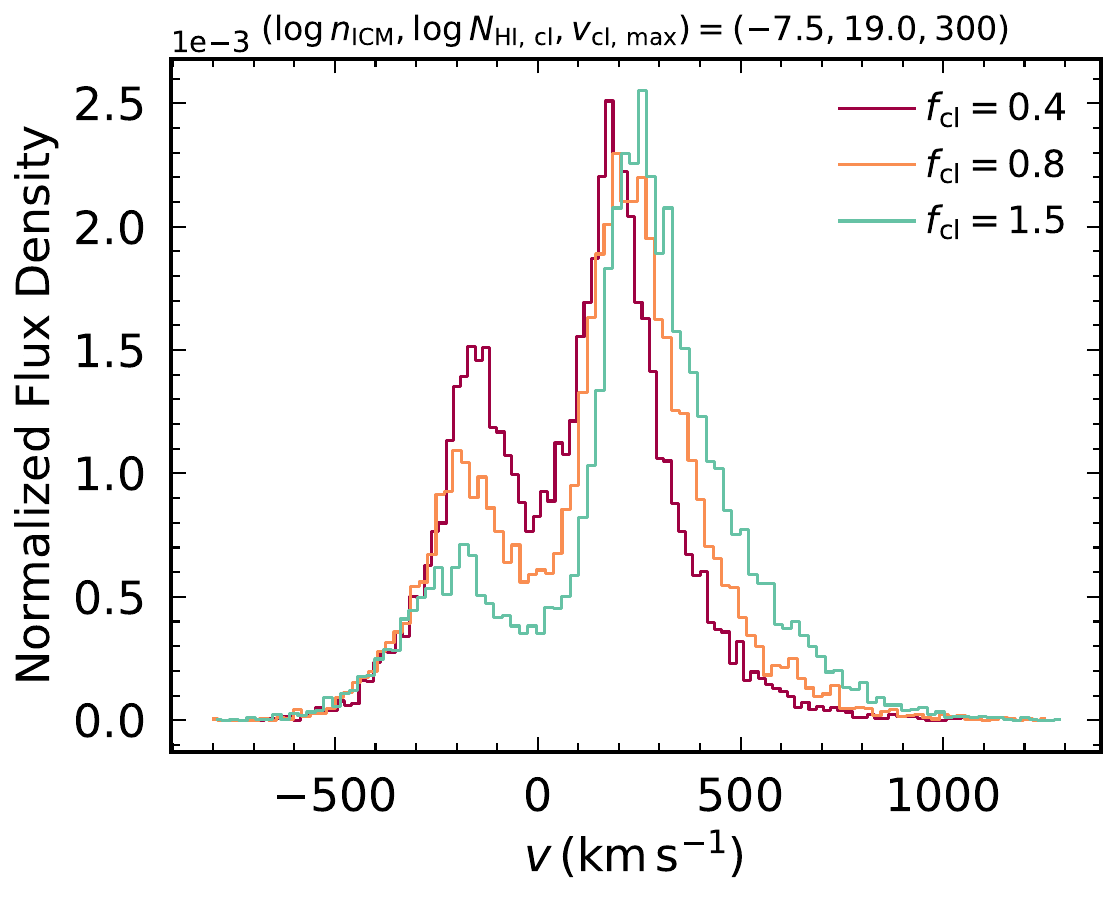}
\includegraphics[width=0.5\textwidth]{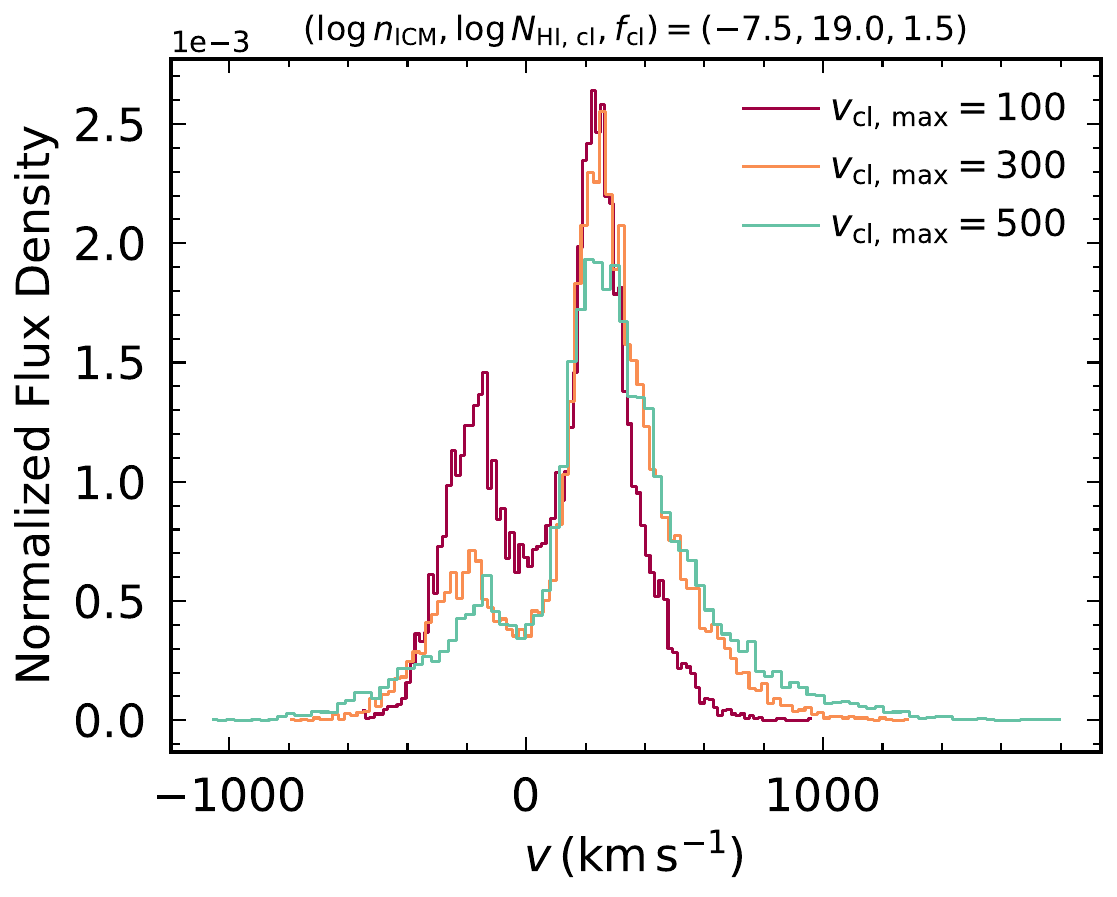}
    \caption{\textbf{Effect of varying $f_{\rm cl}$ and $v_{\rm cl,\,max}$ individually on the model \lya\ spectra.} The upper panel shows that increasing $f_{\rm cl}$ tends to suppress the blue peak relative to the red peak without significantly affecting extent of the the wings, while the lower panel shows that increasing $v_{\rm cl,\,max}$ also suppresses the blue peak but significantly broadens the wings to higher velocities.
    \label{fig:varying_lya}}
\end{figure}

We find that the reason J1133+6513 and J1244+0215 favor low $f_{\rm cl}$  values is due to the relatively strong blue peaks in their \lya\ spectra. From an RT perspective, the relative intensity of the blue peak compared to the red peak is primarily determined by two parameters: the clump covering factor $f_{\rm cl}$ and the clump maximum outflow velocity $v_{\rm cl,\,max}$. Increasing $f_{\rm cl}$ tends to suppress the blue peak relative to the red peak, as it increases the likelihood of scattering, which, in an outflowing medium, favors the escape of red photons over blue photons. Similarly, increasing $v_{\rm cl,\,max}$ also suppresses the blue peak because, in the reference frame of the outflowing clumps, more blue photons are redshifted toward the line center and require additional scattering to escape. In the case of J1133+6513 and J1244+0215, where the red wings extend to around $1000\rm\,km\,s^{-1}$, their $v_{\rm cl,\,max}$ must be reasonably large, requiring $f_{\rm cl}$ to be low in order to produce the strong blue peak. We present several examples to illustrate this point in Figure \ref{fig:varying_lya}, where we vary $f_{\rm cl}$ and $v_{\rm cl,\,max}$ individually while keeping all other model parameters fixed.

At this point, it remains unclear why J1133+6513 and J1244+0215 exhibit particularly strong blue \lya\ peaks, which result in their notably low gas covering factors. Nevertheless, our experiments suggest that modeling spatially integrated \lya\ spectra alone may not be sufficient to accurately infer the LyC escape fractions of LyC leakers. In the future, spatially resolved observations using high-resolution IFU spectrographs may help resolve the degeneracy between model parameters by providing additional spatial information (e.g., \citealt{Erb23}). Additionally, certain assumptions in the RT model may contribute to this puzzle, such as the assumption of angular isotropy, whereas in reality, the escape of \lya\ and LyC photons may be anisotropic (e.g., \citealt{Monter2024}). We plan to explore these possibilities in future work.

\section{Discussion}\label{sec:comparison}

\subsection{Implications of the $v_{\rm MgII,\,max}$ -- $N_{\rm MgII,\,tot}$ Criterion}\label{sec:criteria}

As discussed in Section \ref{sec:plane}, we identified a necessary condition for a LyC leaker to be a strong leaker: a high maximum clump radial outflow velocity ($v_{\rm MgII,\,max} \gtrsim 390\,\rm km\,s^{-1}$) \emph{and} a low total \MgII\ column density ($N_{\rm MgII,\,tot} \lesssim 10^{14.3}\,\rm cm^{-2}$). We now explore the physical reasoning behind this condition.

\begin{figure}
\centering
\hspace{-0.6cm}
\includegraphics[width=0.5\textwidth]{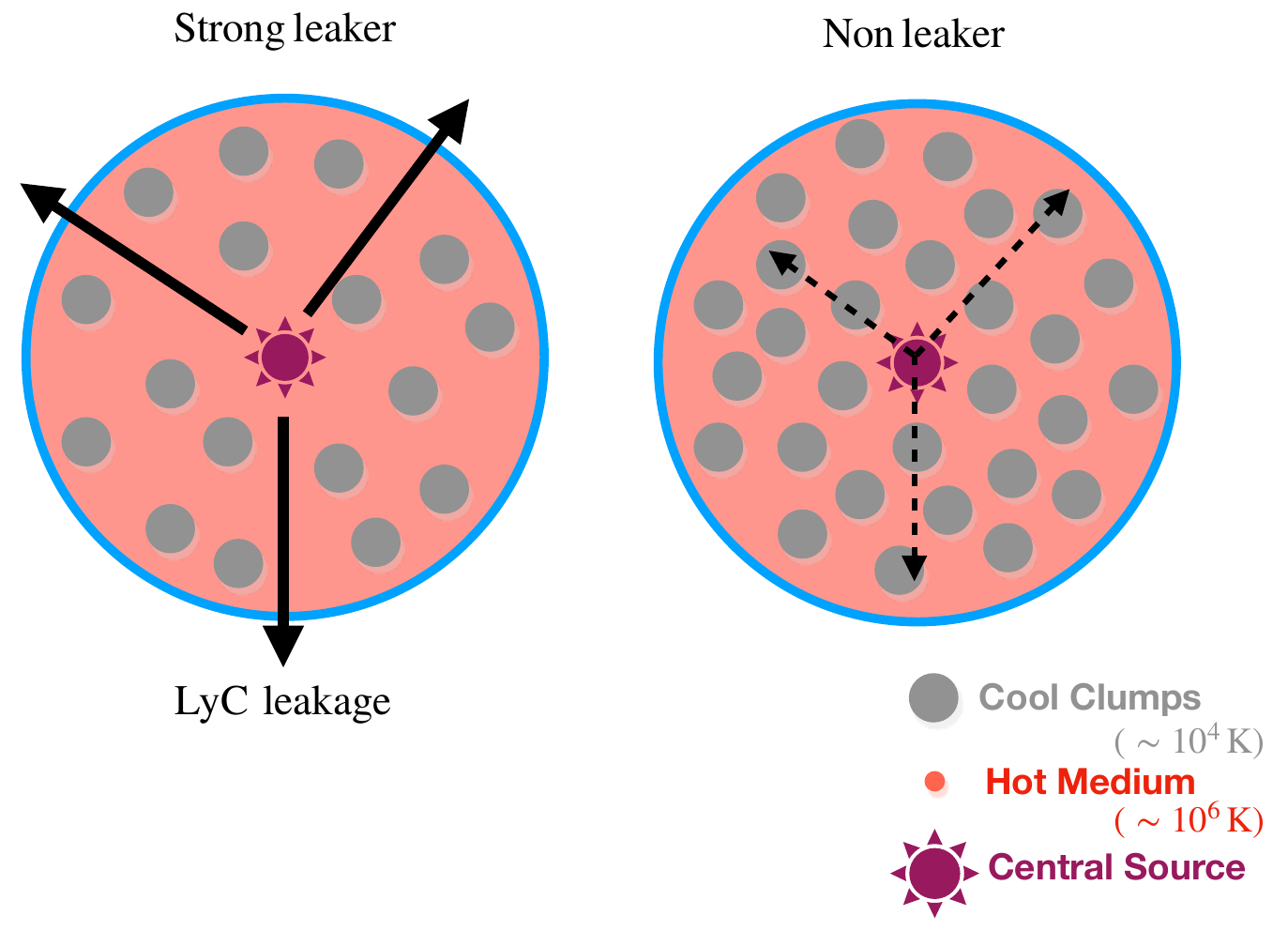}
    \caption{\textbf{Comparison between the CGM structure of a strong LyC leaker and a non-leaker. } In the CGM of the strong leaker, certain sightlines are free of optically thick \HI\ clumps, allowing significant LyC leakage (indicated by solid arrows). In contrast, the CGM of a non-leaker contains no clear sightlines, causing nearly all ionizing photons to be absorbed by optically thick \HI\ clumps (indicated by dashed arrows). High clump outflow velocities ($v_{\rm MgII,\,max}$) and low \MgII\ total column densities ($N_{\rm MgII,\,tot}$) may promote the development of the picket-fence-like structure seen in the strong leaker, thereby facilitating LyC escape.
    \label{fig:leakage_model}}
\end{figure}

Our \lya\ RT modeling of six objects in our sample reveals that the only two galaxies with large ($> 10\%$) theoretical LyC escape fractions exhibit particularly low volume filling factors $F_{\rm V}$ and correspondingly low gas covering factors $f_{\rm cl}$. These objects achieve significant LyC escape not because of lower clump \HI\ column densities (which must exceed $\sim 10^{17} \rm cm^{-2}$ to match the broad peaks of their \lya\ profiles), but because certain sightlines encounter only zero or one clump. In other words, our modeling results support a ``picket fence'' geometry for the CGM \citep[e.g.,][]{Heckman2011, Rivera-Thorsen2017, Gazagnes2018, Gazagnes2020, Saldana-Lopez2022} rather than a ``density-bounded'' scenario, where the clump \HI\ column density is not high enough to prevent the penetration of LyC photons. We illustrate this point with a schematic in Figure \ref{fig:leakage_model}.

The conditions we identified, namely a combination of high $v_{\rm MgII,\,max}$ and low $N_{\rm MgII,\,tot}$, may therefore favor the formation of such a picket-fence-like CGM that facilitates LyC escape. A high \MgII\ outflow velocity likely indicates strong supernova feedback, which can blow out the gas clumps and create low-density \HI\ channels \citep[e.g.,][]{Li2015, Fielding2017, Smith2018, Sarbadhicary2022}. Moreover, statistically speaking, a lower total \MgII\ column density is generally associated with lower \HI\ column densities, further suggesting the presence of these low-density channels. Therefore, galaxies exhibiting both characteristics are understandably more likely to be strong LyC leakers.

While the $v_{\rm MgII,\,max}$ -- $N_{\rm MgII,\,tot}$ criterion we present is relatively straightforward and clear, we recommend exercising caution when using it as an indirect indicator of LyC leakage. A larger sample that has both \MgII\ and \lya\ observations is needed to fully test this criterion. Furthermore, additional high-resolution numerical simulations are essential to deepen our understanding of the physical processes driving this correlation.

\subsection{Comparison to Previous Work}

A recent study by \citet{Carr24} employed the semi-analytical line transfer (SALT) model to analyze \MgII\ line profiles for a similar sample. The SALT model aims to analytically solve the radiative transfer equation using the Sobolev approximation, which simplifies the treatment of radiative transfer. As such, SALT does not simulate the resonant scattering of photons; its setup also differs from the clumpy RT model used in this work -- it assumes a bi-conical, continuous outflowing wind with power-law radial density and velocity profiles, without accounting for turbulent gas motion in the CGM. Given these differences in the modeling approach, it is expected that the SALT model yields different results, such as gas outflow velocities and \MgII\ column densities, compared to our findings.

In their study, \citet{Carr24} modeled 29 galaxies from the sample used in this work, achieving satisfactory fits for 20 galaxies with an outflowing SALT model and for 6 galaxies with a non-outflowing double-Gaussian ISM model. For these six galaxies (J0826+1820, J1033+6353, J1133+6513, J1158+3125, J1235+0635, and J1410+4345), except for J0826+1820 due to its noisy spectrum, our RT model provides a good fit for the remaining five with inferred maximum clump outflow velocities of 533, 272, 432, 30, and 452 $\rm km\,s^{-1}$, respectively. In addition, for two profiles that were not reproduced by the SALT model, our model gives a good fit for J1248+1234 and a decent fit for J1310+2148\footnote{Note that J1310+2148 exhibits an unusual redshifted absorption feature, possibly caused by an additional inflowing absorber along our sightline; nevertheless, our modeling has successfully captured the overall P-Cyngi-like shape of the line profile.}. We find that differences in our data processing procedures may explain the different modeling results -- \citet{Carr24} used a finer binning than the instrumental resolution for their spectra, whereas in this work, we used the resolution-based rebinned version of the \MgII\ spectra as presented in \citet{Xu2023}. These differences in data processing may result in different levels of significance in the blueshifted absorption troughs, particularly for J1033+6353, J1158+3125, and J1410+4345 -- the three galaxies categorized as strong leakers in both studies. This may also explain why no evidence of outflows was found for these galaxies in \citet{Carr24}.

\begin{figure}
\centering
\includegraphics[width=0.5\textwidth]{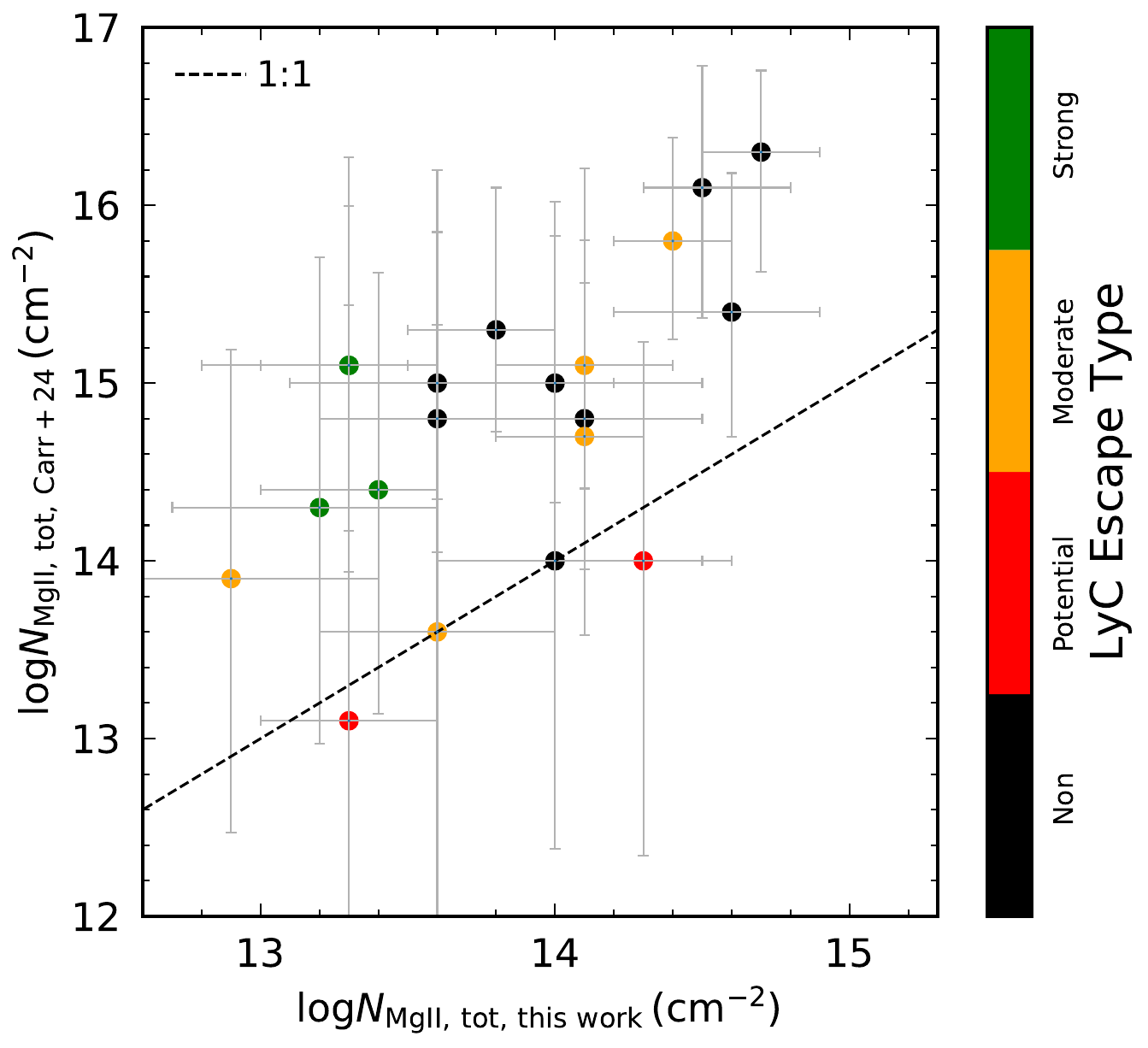}
    \caption{\textbf{Comparison between the total \MgII} \textbf{column densities derived by \cite{Carr24} and this work.} The \MgII\ column densities reported by \cite{Carr24} are systematically higher, with larger uncertainties, compared to those determined in this study. The different colors of the points represent the LyC leakage types as defined in this paper.
    \label{fig:comparison}}
\end{figure}

To do a more detailed comparison, we examined the best-fit total \MgII\ column densities\footnote{We refrain from making a direct comparison of the maximum gas outflow velocities between the two studies, as we have made very different assumptions about the radial velocity and density profiles of the Mg$^+$ gas (e.g., \citealt{Carr24} noted that their terminal velocity $v_{\rm \infty}$  may become unconstrained when the density field drops to an undetectable level before the velocity field reaches its maximum).} derived by \citet{Carr24} for the 20 galaxies successfully modeled with their outflowing SALT framework, as shown in Figure \ref{fig:comparison}. We find that the \MgII\ column densities reported in their work are systematically higher, with larger uncertainties, compared to those derived from our RT model. However, we note that this difference does not necessarily imply that one model outperforms the other; instead, the different $N_{\rm MgII,\,tot}$ values may result from the different model geometries (i.e., spherically symmetric vs. bi-conic), and the different uncertainties may simply reflect the different volumes of the parameter spaces explored.

Given the more comprehensive treatment of resonant scattering physics in our full RT modeling, we suggest that our multiphase, clumpy model may provide valuable constraints on the underlying gas parameters of the CGM. Nonetheless, the SALT model, with its simpler, semi-analytical framework, can also yield useful results, particularly under the assumption of an anisotropic geometric configuration. Future work will be essential to further assess and compare the strengths and limitations of these two models in different contexts. Further \MgII\ observations using high-resolution IFUs, along with comparative studies against cosmological simulations, will also be essential (e.g., \citealt{Zheng2010, Zheng2011, Gronke16_model, Gronke2018, Blaizot2023}, Jennings et al., in prep; Carr et al., in prep).

\section{Conclusions}\label{sec:conclusion}
In this work, we conducted systematic RT modeling of the \MgII\ doublet line profiles for 33 low-$z$ LyC leakers, as well as \lya\ modeling for six of them using a multiphase, clumpy CGM model. The key results are as follows:

\begin{itemize}
    \item Our RT modeling successfully reproduced the \MgII\ line profiles of all 33 galaxies when the data quality was sufficient. From the gas properties derived through \MgII\ RT modeling, we identified a necessary condition for a LyC leaker to be classified as a strong leaker: a high maximum clump radial outflow velocity ($v_{\rm MgII,\,max} \gtrsim 390\,\rm km\,s^{-1}$) {\emph{and}} a low total $\rm Mg\,{\textsc {ii}}$ column density ($N_{\rm MgII,\,tot} \lesssim 10^{14.3}\,\rm cm^{-2}$).

    \item We also explored the effects of individual parameters in the RT model on the \MgII\ spectra. Our findings indicate that the two most influential parameters shaping the \MgII\ line profiles are the clump maximum outflow velocity and the total \MgII\ column density. The other parameters in the model have only subdominant effects on the \MgII\ spectra, and there are no significant correlations between their best-fit values inferred from RT modeling and the amount of LyC leakage. Overall, the effect of these parameters is highly complex, which necessitates full RT modeling to reliably extract the underlying gas properties of the CGM.

    \item Using archival HST COS/G160M data, we performed RT modeling on six objects and successfully reproduced their \lya\ profiles as well. Interestingly, we find that their \lya\ spectral properties, such as fluxes near the line center, do not fully align with conventional criteria typically used to infer LyC leakage from \lya\ profiles. However, we have yet to identify any clear correlation between the parameters derived from our \MgII\ and \lya\ modeling, such as the gas outflow velocities and column densities.

    \item We inferred the theoretical LyC escape fractions based on the \HI\ properties derived from \lya\ RT modeling. Our findings suggest that the amount of RT-inferred LyC leakage is primarily governed by the average number of optically thick \HI\ clumps per sightline, $f_{\rm cl}$. Interestingly, J1133+6513, a potential leaker, and J1244+0215, a non-leaker, exhibit the lowest $f_{\rm cl}$ values that lead to their highest inferred LyC escape fractions. We find that this is driven by their strong blue peaks of their \lya\ profiles, though the reason for these pronounced blue peaks remains unclear. We highlight the need for high-resolution, spatially resolved IFU observations in the future to further break model degeneracies and address this puzzle.

    \item Our \lya\ RT modeling supports a ``picket fence'' geometry for the CGM rather than a ``density-bounded'' scenario. In other words, the galaxies have high theoretical LyC escape fractions due to the presence of sightlines free of high-$N_{\rm HI}$ clumps, not because the clump \HI\ column densities are insufficient to block the penetration of LyC photons. A combination of high $v_{\rm MgII,\,max}$ and low $N_{\rm MgII,\,tot}$, may therefore favor the formation of such a picket-fence-like CGM that facilitates LyC escape. While promising, the $v_{\rm MgII,\,max}$ -- $N_{\rm MgII,\,tot}$ criterion we discovered from our \MgII\ RT modeling should be applied cautiously, since it requires larger datasets and high-resolution simulations to confirm its robustness.
    
\end{itemize}

\begin{acknowledgments}
We acknowledge the contributions of the LzLCS team members who made this project possible. This work was carried out at the Advanced Research Computing at Hopkins (ARCH) core facility  (rockfish.jhu.edu), which is supported by the National Science Foundation (NSF) grant number OAC 1920103. MG thanks the Max Planck Society for support through the Max Planck Research Group. ZL has been supported in part by grant AST-2009278 from the U.S. National Science Foundation. Numerical calculations were run on the Caltech compute cluster ``Wheeler,'' allocations from XSEDE TG-AST130039 and PRAC NSF.1713353 supported by the NSF, and NASA HEC SMD-16-7592. 
\end{acknowledgments}

%

\vspace{5mm}
\facilities{HST (COS), MMT (Blue channel), VLT (X-Shooter), HET (LRS2).}





\appendix

\section{Example clump radial outflow velocity profiles}\label{sec:appendix}

\begin{figure}
\centering
\includegraphics[width=0.48\textwidth]{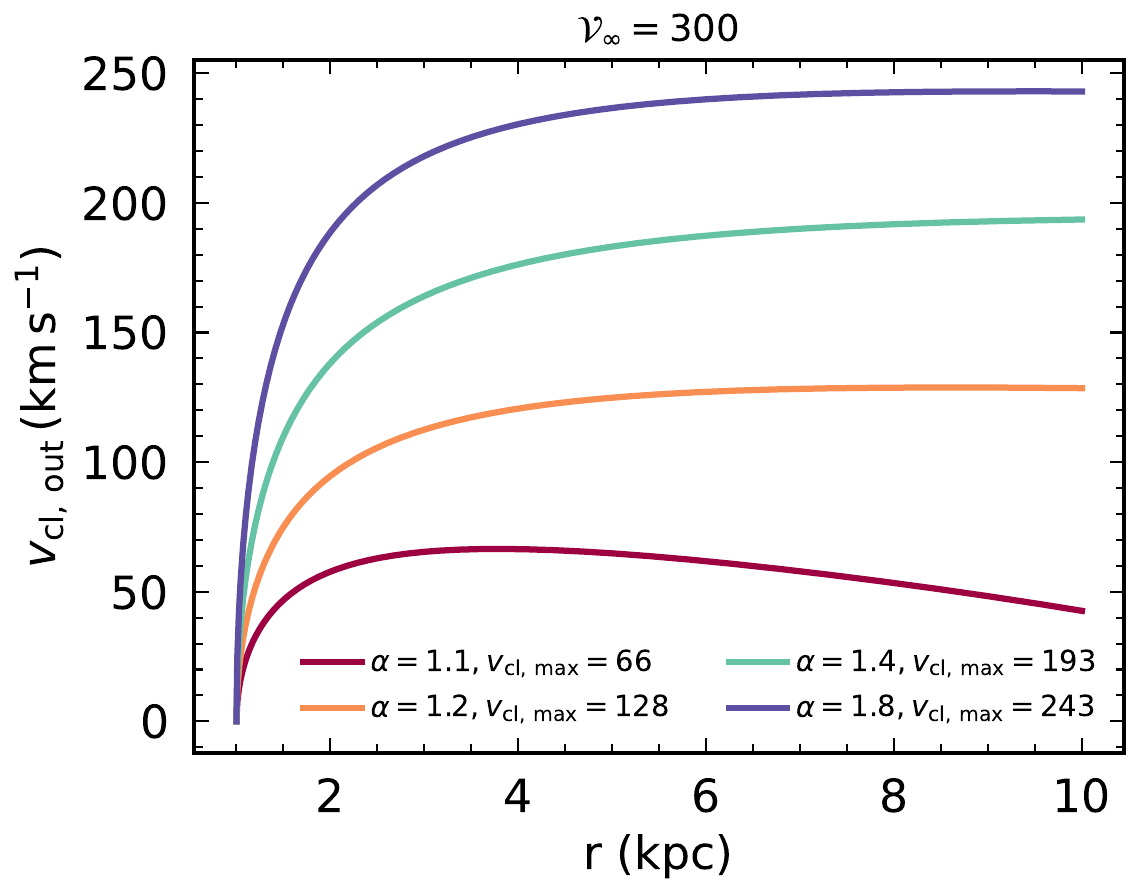}
\includegraphics[width=0.48\textwidth]{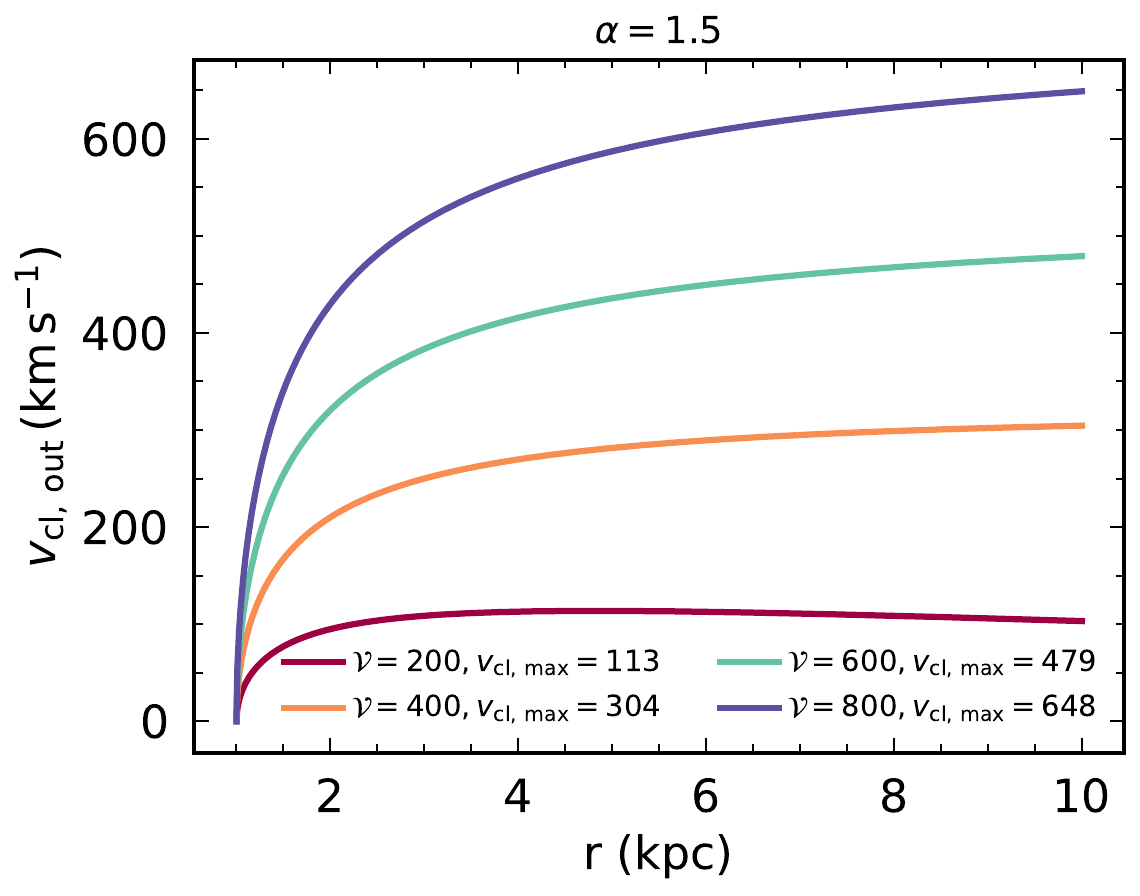}
    \caption{\textbf{Example clump radial outflow velocity profiles with varying $\mathcal{V}_{\rm \infty}$ and $\alpha$.} The maximum clump outflow velocity $v_{\rm cl,\,max}$ has been indicated on each label.}
    \label{fig:vout_examples}
\end{figure}

Here we present several example profiles of clump radial outflow velocity (see Eq. \ref{eq:v}), varying the asymptotic clump outflow velocity $\mathcal{V_{\infty}}$ and the clump acceleration power-law index $\alpha$ individually. We assume a dark matter halo mass of $M_{\rm vir} = 10^{11} M_{\odot}$ and a redshift of $z = 0.2$. As shown in Figure \ref{fig:vout_examples}, adjusting $\mathcal{V_{\infty}}$ and $\alpha$ alters both the shape and amplitude of the $v_{\rm cl,\,out}$ curves, with $v_{\rm cl,\,max}$ (defined as the maximum value of $v_{\rm cl,\,out}$) generally remaining below $\mathcal{V_{\infty}}$ due to the presence of gravity.

\section{Special Case: The Low Total \MgII\ Column Density Regime}\label{sec:special_case}

We note that when the total \MgII\ column density is particularly low ($N_{\rm MgII,\,tot} \lesssim 10^{12.5}\,\rm cm^{-2}$), the blueshifted absorption in the \MgII\ line profiles may become insignificant. Nonetheless, in this regime, it remains possible to constrain the maximum clump outflow velocity, $v_{\rm MgII,\,max}$, by using the relative intensity of the K and H peaks.

\begin{figure}[htbp!]
\centering
\includegraphics[width=0.48\textwidth]{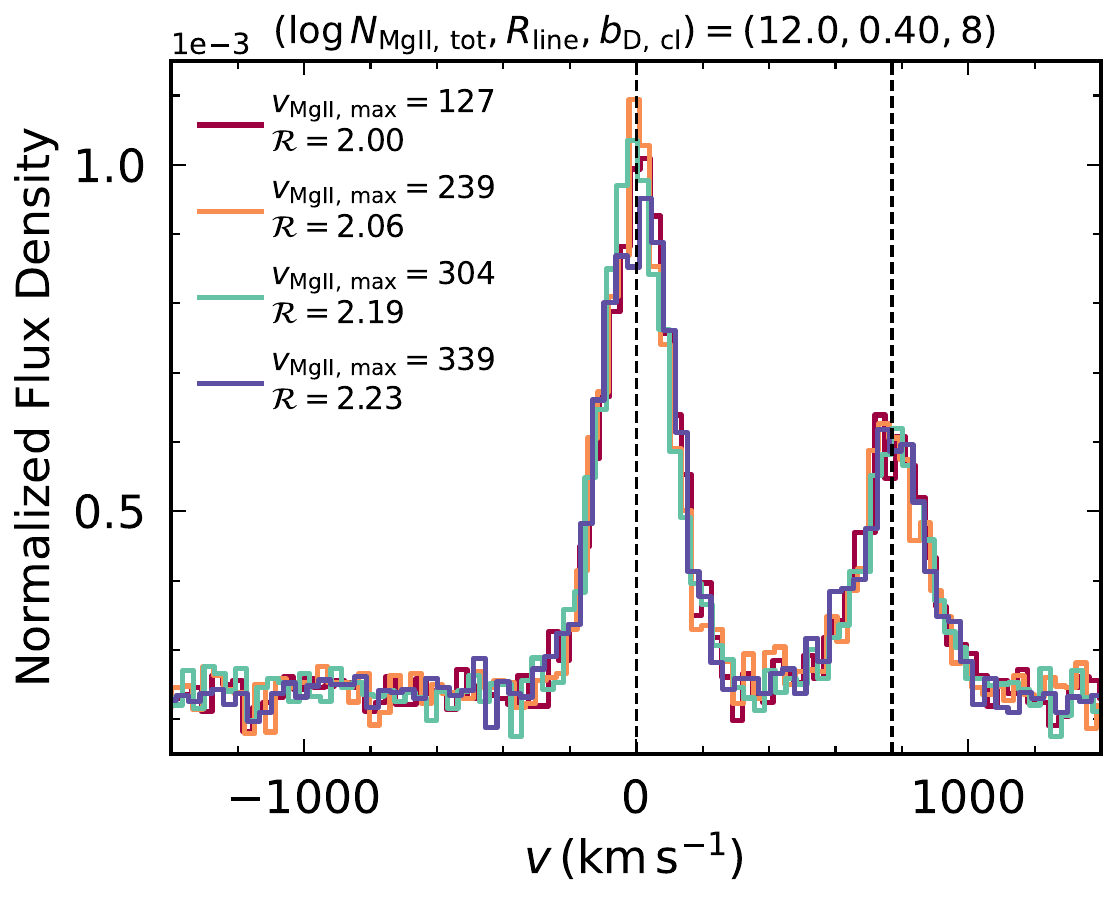}
    \caption{\textbf{$\rm Mg\,{\textsc {II}}$ model spectra with $N_{\rm MgII,\,tot} = 10^{12.0}\,\rm cm^{-2}$ and varying $v_{\rm MgII,\,max}$.} The other model parameters are chosen to be close to the best-fit parameters of J1133+6353.}
    \label{fig:N12}
\end{figure}

We illustrate this point using galaxy J1133+6353 as an example, which has a best-fit total \MgII\ column density of $10^{12.0}\,\rm cm^{-2}$ and maximum clump outflow velocity of 272 $\rm km\,s^{-1}$. Even at such a low \MgII\ column density, some \MgII\ resonant scattering has already occurred (with the line center optical depth around 0.05; see Figure 3 in \citealt{Chang24}). In Figure \ref{fig:N12}, we present several \MgII\ model spectra with parameters close to the best-fit values for J1133+6353, varying only $v_{\rm MgII,\,max}$. Although the absorption trough is insignificant, as $v_{\rm MgII,\,max}$ increases, the EWs of both the H and K peaks decrease, with the EW of the H peak decreasing more significantly. Consequently, the doublet EW ratio $\mathcal{R}$ slightly increases, allowing $v_{\rm MgII,\,max}$ to still be constrained by the data.

\FloatBarrier
\bibliography{references}{}

\begin{thebibliography}{}
\expandafter\ifx\csname natexlab\endcsname\relax\def\natexlab#1{#1}\fi
\providecommand{\url}[1]{\href{#1}{#1}}
\providecommand{\dodoi}[1]{doi:~\href{http://doi.org/#1}{\nolinkurl{#1}}}
\providecommand{\doeprint}[1]{\href{http://ascl.net/#1}{\nolinkurl{http://ascl.net/#1}}}
\providecommand{\doarXiv}[1]{\href{https://arxiv.org/abs/#1}{\nolinkurl{https://arxiv.org/abs/#1}}}

\bibitem[{{Almada Monter} \& {Gronke}(2024)}]{Monter2024}
{Almada Monter}, S., \& {Gronke}, M. 2024, \mnras, 534, L7,
  \dodoi{10.1093/mnrasl/slae074}

\bibitem[{{Blaizot} {et~al.}(2023){Blaizot}, {Garel}, {Verhamme}, {Katz},
  {Kimm}, {Michel-Dansac}, {Mitchell}, {Rosdahl}, \& {Trebitsch}}]{Blaizot2023}
{Blaizot}, J., {Garel}, T., {Verhamme}, A., {et~al.} 2023, \mnras, 523, 3749,
  \dodoi{10.1093/mnras/stad1523}

\bibitem[{{Burchett} {et~al.}(2021){Burchett}, {Rubin}, {Prochaska}, {Coil},
  {Vaught}, \& {Hennawi}}]{Burchett2021}
{Burchett}, J.~N., {Rubin}, K. H.~R., {Prochaska}, J.~X., {et~al.} 2021, \apj,
  909, 151, \dodoi{10.3847/1538-4357/abd4e0}

\bibitem[{{Carr} {et~al.}(2024){Carr}, {Cen}, {Scarlata}, {Xu}, {Henry},
  {Marques-Chaves}, {Schaerer}, {Amor{\'\i}n}, {Oey}, {Komarova}, {Flury},
  {Jaskot}, {Saldana-Lopez}, {Ji}, {Huberty}, {Heckman}, {Ostlin}, {Bait},
  {Hayes}, {Thuan}, {Berg}, {Giavalisco}, {Borthakur}, {Chisholm}, {Ferguson},
  {Michel-Dansac}, {Verhamme}, \& {Worseck}}]{Carr24}
{Carr}, C.~A., {Cen}, R., {Scarlata}, C., {et~al.} 2024, arXiv e-prints,
  arXiv:2409.05180, \dodoi{10.48550/arXiv.2409.05180}

\bibitem[{{Chang} \& {Gronke}(2024)}]{Chang24}
{Chang}, S.-J., \& {Gronke}, M. 2024, \mnras, 532, 3526,
  \dodoi{10.1093/mnras/stae1664}

\bibitem[{{Chisholm} {et~al.}(2020){Chisholm}, {Prochaska}, {Schaerer},
  {Gazagnes}, \& {Henry}}]{Chisholm2020}
{Chisholm}, J., {Prochaska}, J.~X., {Schaerer}, D., {Gazagnes}, S., \& {Henry},
  A. 2020, \mnras, 498, 2554, \dodoi{10.1093/mnras/staa2470}

\bibitem[{{Dijkstra} \& {Kramer}(2012)}]{Dijkstra2012}
{Dijkstra}, M., \& {Kramer}, R. 2012, \mnras, 424, 1672,
  \dodoi{10.1111/j.1365-2966.2012.21131.x}

\bibitem[{{Draine}(2003{\natexlab{a}})}]{Draine03}
{Draine}, B.~T. 2003{\natexlab{a}}, \apj, 598, 1017, \dodoi{10.1086/379118}

\bibitem[{{Draine}(2003{\natexlab{b}})}]{Draine03b}
---. 2003{\natexlab{b}}, \araa, 41, 241,
  \dodoi{10.1146/annurev.astro.41.011802.094840}

\bibitem[{{Dutta} {et~al.}(2023){Dutta}, {Fossati}, {Fumagalli}, {Revalski},
  {Lofthouse}, {Nelson}, {Papini}, {Rafelski}, {Cantalupo}, {Arrigoni Battaia},
  {Dayal}, {Longobardi}, {P{\'e}roux}, {Prichard}, \& {Prochaska}}]{Dutta2023}
{Dutta}, R., {Fossati}, M., {Fumagalli}, M., {et~al.} 2023, \mnras, 522, 535,
  \dodoi{10.1093/mnras/stad1002}

\bibitem[{{Erb} {et~al.}(2023){Erb}, {Li}, {Steidel}, {Chen}, {Gronke},
  {Strom}, {Trainor}, \& {Rudie}}]{Erb23}
{Erb}, D.~K., {Li}, Z., {Steidel}, C.~C., {et~al.} 2023, \apj, 953, 118,
  \dodoi{10.3847/1538-4357/acd849}

\bibitem[{{Erb} {et~al.}(2012){Erb}, {Quider}, {Henry}, \& {Martin}}]{Erb2012}
{Erb}, D.~K., {Quider}, A.~M., {Henry}, A.~L., \& {Martin}, C.~L. 2012, \apj,
  759, 26, \dodoi{10.1088/0004-637X/759/1/26}

\bibitem[{{Feltre} {et~al.}(2018){Feltre}, {Bacon}, {Tresse}, {Finley},
  {Carton}, {Blaizot}, {Bouch{\'e}}, {Garel}, {Inami}, {Boogaard},
  {Brinchmann}, {Charlot}, {Chevallard}, {Contini}, {Michel-Dansac}, {Mahler},
  {Marino}, {Maseda}, {Richard}, {Schmidt}, \& {Verhamme}}]{Feltre2018}
{Feltre}, A., {Bacon}, R., {Tresse}, L., {et~al.} 2018, \aap, 617, A62,
  \dodoi{10.1051/0004-6361/201833281}

\bibitem[{{Fielding} {et~al.}(2017){Fielding}, {Quataert}, {McCourt}, \&
  {Thompson}}]{Fielding2017}
{Fielding}, D., {Quataert}, E., {McCourt}, M., \& {Thompson}, T.~A. 2017,
  \mnras, 466, 3810, \dodoi{10.1093/mnras/stw3326}

\bibitem[{{Finley} {et~al.}(2017){Finley}, {Bouch{\'e}}, {Contini}, {Paalvast},
  {Boogaard}, {Maseda}, {Bacon}, {Blaizot}, {Brinchmann}, {Epinat}, {Feltre},
  {Marino}, {Muzahid}, {Richard}, {Schaye}, {Verhamme}, {Weilbacher}, \&
  {Wisotzki}}]{Finley2017}
{Finley}, H., {Bouch{\'e}}, N., {Contini}, T., {et~al.} 2017, \aap, 608, A7,
  \dodoi{10.1051/0004-6361/201731499}

\bibitem[{{Flury} {et~al.}(2022){Flury}, {Jaskot}, {Ferguson}, {Worseck},
  {Makan}, {Chisholm}, {Saldana-Lopez}, {Schaerer}, {McCandliss}, {Wang},
  {Ford}, {Heckman}, {Ji}, {Giavalisco}, {Amorin}, {Atek}, {Blaizot},
  {Borthakur}, {Carr}, {Castellano}, {Cristiani}, {De Barros}, {Dickinson},
  {Finkelstein}, {Fleming}, {Fontanot}, {Garel}, {Grazian}, {Hayes}, {Henry},
  {Mauerhofer}, {Micheva}, {Oey}, {Ostlin}, {Papovich}, {Pentericci},
  {Ravindranath}, {Rosdahl}, {Rutkowski}, {Santini}, {Scarlata}, {Teplitz},
  {Thuan}, {Trebitsch}, {Vanzella}, {Verhamme}, \& {Xu}}]{Flury22}
{Flury}, S.~R., {Jaskot}, A.~E., {Ferguson}, H.~C., {et~al.} 2022, \apjs, 260,
  1, \dodoi{10.3847/1538-4365/ac5331}

\bibitem[{{Gazagnes} {et~al.}(2020){Gazagnes}, {Chisholm}, {Schaerer},
  {Verhamme}, \& {Izotov}}]{Gazagnes2020}
{Gazagnes}, S., {Chisholm}, J., {Schaerer}, D., {Verhamme}, A., \& {Izotov}, Y.
  2020, \aap, 639, A85, \dodoi{10.1051/0004-6361/202038096}

\bibitem[{{Gazagnes} {et~al.}(2018){Gazagnes}, {Chisholm}, {Schaerer},
  {Verhamme}, {Rigby}, \& {Bayliss}}]{Gazagnes2018}
{Gazagnes}, S., {Chisholm}, J., {Schaerer}, D., {et~al.} 2018, \aap, 616, A29,
  \dodoi{10.1051/0004-6361/201832759}

\bibitem[{{Giavalisco} {et~al.}(2011){Giavalisco}, {Vanzella}, {Salimbeni},
  {Tripp}, {Dickinson}, {Cassata}, {Renzini}, {Guo}, {Ferguson}, {Nonino},
  {Cimatti}, {Kurk}, {Mignoli}, \& {Tang}}]{Giavalisco2011}
{Giavalisco}, M., {Vanzella}, E., {Salimbeni}, S., {et~al.} 2011, \apj, 743,
  95, \dodoi{10.1088/0004-637X/743/1/95}

\bibitem[{{Gronke} \& {Dijkstra}(2014)}]{Gronke14}
{Gronke}, M., \& {Dijkstra}, M. 2014, \mnras, 444, 1095,
  \dodoi{10.1093/mnras/stu1513}

\bibitem[{{Gronke} \& {Dijkstra}(2016)}]{Gronke16_model}
---. 2016, \apj, 826, 14, \dodoi{10.3847/0004-637X/826/1/14}

\bibitem[{{Gronke} {et~al.}(2018){Gronke}, {Girichidis}, {Naab}, \&
  {Walch}}]{Gronke2018}
{Gronke}, M., {Girichidis}, P., {Naab}, T., \& {Walch}, S. 2018, \apjl, 862,
  L7, \dodoi{10.3847/2041-8213/aad286}

\bibitem[{{Heckman} {et~al.}(2011){Heckman}, {Borthakur}, {Overzier},
  {Kauffmann}, {Basu-Zych}, {Leitherer}, {Sembach}, {Martin}, {Rich},
  {Schiminovich}, \& {Seibert}}]{Heckman2011}
{Heckman}, T.~M., {Borthakur}, S., {Overzier}, R., {et~al.} 2011, \apj, 730, 5,
  \dodoi{10.1088/0004-637X/730/1/5}

\bibitem[{{Henry} {et~al.}(2018){Henry}, {Berg}, {Scarlata}, {Verhamme}, \&
  {Erb}}]{Henry18}
{Henry}, A., {Berg}, D.~A., {Scarlata}, C., {Verhamme}, A., \& {Erb}, D. 2018,
  \apj, 855, 96, \dodoi{10.3847/1538-4357/aab099}

\bibitem[{{Henry} {et~al.}(2015){Henry}, {Scarlata}, {Martin}, \&
  {Erb}}]{Henry15}
{Henry}, A., {Scarlata}, C., {Martin}, C.~L., \& {Erb}, D. 2015, \apj, 809, 19,
  \dodoi{10.1088/0004-637X/809/1/19}

\bibitem[{{Huang} {et~al.}(2021){Huang}, {Chen}, {Shectman}, {Johnson},
  {Zahedy}, {Helsby}, {Gauthier}, \& {Thompson}}]{Huang2021}
{Huang}, Y.-H., {Chen}, H.-W., {Shectman}, S.~A., {et~al.} 2021, \mnras, 502,
  4743, \dodoi{10.1093/mnras/stab360}

\bibitem[{{Izotov} {et~al.}(2022){Izotov}, {Chisholm}, {Worseck}, {Guseva},
  {Schaerer}, \& {Prochaska}}]{Izotov2022}
{Izotov}, Y.~I., {Chisholm}, J., {Worseck}, G., {et~al.} 2022, \mnras, 515,
  2864, \dodoi{10.1093/mnras/stac1899}

\bibitem[{{Jenkins}(2009)}]{Jenkins09}
{Jenkins}, E.~B. 2009, \apj, 700, 1299, \dodoi{10.1088/0004-637X/700/2/1299}

\bibitem[{{Kornei} {et~al.}(2013){Kornei}, {Shapley}, {Martin}, {Coil}, {Lotz},
  {Weiner}, \& {Newman}}]{Kornei2013}
{Kornei}, K.~A., {Shapley}, A.~E., {Martin}, C.~L., {et~al.} 2013, \apj, 774,
  50, \dodoi{10.1088/0004-637X/774/1/50}

\bibitem[{{Li} {et~al.}(2015){Li}, {Ostriker}, {Cen}, {Bryan}, \&
  {Naab}}]{Li2015}
{Li}, M., {Ostriker}, J.~P., {Cen}, R., {Bryan}, G.~L., \& {Naab}, T. 2015,
  \apj, 814, 4, \dodoi{10.1088/0004-637X/814/1/4}

\bibitem[{{Li} \& {Gronke}(2022)}]{Li22}
{Li}, Z., \& {Gronke}, M. 2022, \mnras, 513, 5034,
  \dodoi{10.1093/mnras/stac1207}

\bibitem[{{Li} {et~al.}(2024){Li}, {Gronke}, \& {Steidel}}]{Li24}
{Li}, Z., {Gronke}, M., \& {Steidel}, C.~C. 2024, \mnras, 529, 444,
  \dodoi{10.1093/mnras/stae469}

\bibitem[{{Martin} {et~al.}(2012){Martin}, {Shapley}, {Coil}, {Kornei},
  {Mostardi}, \& {Newman}}]{Martin2012}
{Martin}, C.~L., {Shapley}, A.~E., {Coil}, A.~L., {et~al.} 2012, \apj, 760,
  127, \dodoi{10.1088/0004-637X/760/2/127}

\bibitem[{{Naidu} {et~al.}(2022){Naidu}, {Matthee}, {Oesch}, {Conroy},
  {Sobral}, {Pezzulli}, {Hayes}, {Erb}, {Amor{\'\i}n}, {Gronke}, {Schaerer},
  {Tacchella}, {Kerutt}, {Paulino-Afonso}, {Calhau}, {Llerena}, \&
  {R{\"o}ttgering}}]{Naidu22}
{Naidu}, R.~P., {Matthee}, J., {Oesch}, P.~A., {et~al.} 2022, \mnras, 510,
  4582, \dodoi{10.1093/mnras/stab3601}

\bibitem[{{Nikolis} \& {Gronke}(2024)}]{Nikolis2024}
{Nikolis}, C., \& {Gronke}, M. 2024, \mnras, 530, 4597,
  \dodoi{10.1093/mnras/stae1169}

\bibitem[{{Pessa} {et~al.}(2024){Pessa}, {Wisotzki}, {Urrutia}, {Pharo},
  {Augustin}, {Bouch{\'e}}, {Feltre}, {Guo}, {Kozlova}, {Krajnovic},
  {Kusakabe}, {Leclercq}, {Salas}, {Schaye}, \& {Verhamme}}]{Pessa2024}
{Pessa}, I., {Wisotzki}, L., {Urrutia}, T., {et~al.} 2024, arXiv e-prints,
  arXiv:2408.16067, \dodoi{10.48550/arXiv.2408.16067}

\bibitem[{{Prochaska} {et~al.}(2011){Prochaska}, {Kasen}, \&
  {Rubin}}]{Prochaska2011}
{Prochaska}, J.~X., {Kasen}, D., \& {Rubin}, K. 2011, \apj, 734, 24,
  \dodoi{10.1088/0004-637X/734/1/24}

\bibitem[{{Rigby} {et~al.}(2014){Rigby}, {Bayliss}, {Gladders}, {Sharon},
  {Wuyts}, \& {Dahle}}]{Rigby2014}
{Rigby}, J.~R., {Bayliss}, M.~B., {Gladders}, M.~D., {et~al.} 2014, \apj, 790,
  44, \dodoi{10.1088/0004-637X/790/1/44}

\bibitem[{{Rivera-Thorsen} {et~al.}(2017){Rivera-Thorsen}, {Dahle}, {Gronke},
  {Bayliss}, {Rigby}, {Simcoe}, {Bordoloi}, {Turner}, \&
  {Furesz}}]{Rivera-Thorsen2017}
{Rivera-Thorsen}, T.~E., {Dahle}, H., {Gronke}, M., {et~al.} 2017, \aap, 608,
  L4, \dodoi{10.1051/0004-6361/201732173}

\bibitem[{{Rubin} {et~al.}(2011){Rubin}, {Prochaska}, {Koo}, {Phillips}, \&
  {Weiner}}]{Rubin2011}
{Rubin}, K.~H.~R., {Prochaska}, J.~X., {Koo}, D.~C., {Phillips}, A.~C., \&
  {Weiner}, B.~J. 2011, \apj, 728, 55, \dodoi{10.1088/0004-637X/728/1/55}

\bibitem[{{Rubin} {et~al.}(2010){Rubin}, {Weiner}, {Koo}, {Martin}, \&
  {Prochaska}}]{Rubin2010}
{Rubin}, K.~H.~R., {Weiner}, B.~J., {Koo}, D.~C., {Martin}, C.~L., \&
  {Prochaska}, J.~X. 2010, \apj, 719, 1503,
  \dodoi{10.1088/0004-637X/719/2/1503}

\bibitem[{{Rupke} {et~al.}(2019){Rupke}, {Coil}, {Geach}, {Tremonti},
  {Diamond-Stanic}, {George}, {Hickox}, {Kepley}, {Leung}, {Moustakas},
  {Rudnick}, \& {Sell}}]{Rupke2019}
{Rupke}, D. S.~N., {Coil}, A., {Geach}, J.~E., {et~al.} 2019, \nat, 574, 643,
  \dodoi{10.1038/s41586-019-1686-1}

\bibitem[{{Saldana-Lopez} {et~al.}(2022){Saldana-Lopez}, {Schaerer},
  {Chisholm}, {Flury}, {Jaskot}, {Worseck}, {Makan}, {Gazagnes}, {Mauerhofer},
  {Verhamme}, {Amor{\'\i}n}, {Ferguson}, {Giavalisco}, {Grazian}, {Hayes},
  {Heckman}, {Henry}, {Ji}, {Marques-Chaves}, {McCandliss}, {Oey},
  {{\"O}stlin}, {Pentericci}, {Thuan}, {Trebitsch}, {Vanzella}, \&
  {Xu}}]{Saldana-Lopez2022}
{Saldana-Lopez}, A., {Schaerer}, D., {Chisholm}, J., {et~al.} 2022, \aap, 663,
  A59, \dodoi{10.1051/0004-6361/202141864}

\bibitem[{{Sarbadhicary} {et~al.}(2022){Sarbadhicary}, {Martizzi},
  {Ramirez-Ruiz}, {Koch}, {Auchettl}, {Badenes}, \&
  {Chomiuk}}]{Sarbadhicary2022}
{Sarbadhicary}, S.~K., {Martizzi}, D., {Ramirez-Ruiz}, E., {et~al.} 2022, \apj,
  928, 54, \dodoi{10.3847/1538-4357/ac3094}

\bibitem[{{Scarlata} \& {Panagia}(2015)}]{Scarlata2015}
{Scarlata}, C., \& {Panagia}, N. 2015, \apj, 801, 43,
  \dodoi{10.1088/0004-637X/801/1/43}

\bibitem[{{Schroetter} {et~al.}(2015){Schroetter}, {Bouch{\'e}}, {P{\'e}roux},
  {Murphy}, {Contini}, \& {Finley}}]{Schroetter2015}
{Schroetter}, I., {Bouch{\'e}}, N., {P{\'e}roux}, C., {et~al.} 2015, \apj, 804,
  83, \dodoi{10.1088/0004-637X/804/2/83}

\bibitem[{{Seive} {et~al.}(2022){Seive}, {Chisholm}, {Leclercq}, \&
  {Zeimann}}]{Seive2022}
{Seive}, T., {Chisholm}, J., {Leclercq}, F., \& {Zeimann}, G. 2022, \mnras,
  515, 5556, \dodoi{10.1093/mnras/stac2180}

\bibitem[{{Seon}(2024)}]{Seon24}
{Seon}, K.-i. 2024, \apj, 971, 184, \dodoi{10.3847/1538-4357/ad58bd}

\bibitem[{{Shaban} {et~al.}(2022){Shaban}, {Bordoloi}, {Chisholm}, {Sharma},
  {Sharon}, {Rigby}, {Gladders}, {Bayliss}, {Barrientos}, {Lopez}, {Tejos},
  {Ledoux}, \& {Florian}}]{Shaban2022}
{Shaban}, A., {Bordoloi}, R., {Chisholm}, J., {et~al.} 2022, \apj, 936, 77,
  \dodoi{10.3847/1538-4357/ac7c65}

\bibitem[{{Skilling}(2004)}]{Skilling04}
{Skilling}, J. 2004, in American Institute of Physics Conference Series, Vol.
  735, American Institute of Physics Conference Series, ed. R.~{Fischer},
  R.~{Preuss}, \& U.~V. {Toussaint}, 395--405, \dodoi{10.1063/1.1835238}

\bibitem[{{Skilling}(2006)}]{Skilling06}
{Skilling}, J. 2006, Bayesian Anal., 1, 833, \dodoi{10.1214/06-BA127}

\bibitem[{{Smith} {et~al.}(2018){Smith}, {Sijacki}, \& {Shen}}]{Smith2018}
{Smith}, M.~C., {Sijacki}, D., \& {Shen}, S. 2018, \mnras, 478, 302,
  \dodoi{10.1093/mnras/sty994}

\bibitem[{{Speagle}(2020)}]{Speagle20}
{Speagle}, J.~S. 2020, \mnras, 493, 3132, \dodoi{10.1093/mnras/staa278}

\bibitem[{{Verhamme} {et~al.}(2015){Verhamme}, {Orlitov{\'a}}, {Schaerer}, \&
  {Hayes}}]{Verhamme15}
{Verhamme}, A., {Orlitov{\'a}}, I., {Schaerer}, D., \& {Hayes}, M. 2015, \aap,
  578, A7, \dodoi{10.1051/0004-6361/201423978}

\bibitem[{{Weiner} {et~al.}(2009){Weiner}, {Coil}, {Prochaska}, {Newman},
  {Cooper}, {Bundy}, {Conselice}, {Croton}, {Davis}, {Koo}, {Noeske}, \&
  {Yan}}]{Weiner2009}
{Weiner}, B.~J., {Coil}, A.~L., {Prochaska}, J.~X., {et~al.} 2009, \apj, 692,
  187, \dodoi{10.1088/0004-637X/692/1/187}

\bibitem[{{Xu} {et~al.}(2022{\natexlab{a}}){Xu}, {Henry}, {Heckman},
  {Chisholm}, {Worseck}, {Gronke}, {Jaskot}, {McCandliss}, {Flury},
  {Giavalisco}, {Ji}, {Amor{\'\i}n}, {Berg}, {Borthakur}, {Bouche}, {Carr},
  {Erb}, {Ferguson}, {Garel}, {Hayes}, {Makan}, {Marques-Chaves}, {Rutkowski},
  {{\"O}stlin}, {Rafelski}, {Saldana-Lopez}, {Scarlata}, {Schaerer},
  {Trebitsch}, {Tremonti}, {Verhamme}, \& {Wang}}]{Xu2022}
{Xu}, X., {Henry}, A., {Heckman}, T., {et~al.} 2022{\natexlab{a}}, \apj, 933,
  202, \dodoi{10.3847/1538-4357/ac7225}

\bibitem[{{Xu} {et~al.}(2022{\natexlab{b}}){Xu}, {Heckman}, {Henry}, {Berg},
  {Chisholm}, {James}, {Martin}, {Stark}, {Aloisi}, {Amor{\'\i}n},
  {Arellano-C{\'o}rdova}, {Bordoloi}, {Charlot}, {Chen}, {Hayes}, {Mingozzi},
  {Sugahara}, {Kewley}, {Ouchi}, {Scarlata}, \& {Steidel}}]{Xu2022b}
{Xu}, X., {Heckman}, T., {Henry}, A., {et~al.} 2022{\natexlab{b}}, \apj, 933,
  222, \dodoi{10.3847/1538-4357/ac6d56}

\bibitem[{{Xu} {et~al.}(2023){Xu}, {Henry}, {Heckman}, {Chisholm},
  {Marques-Chaves}, {Leclercq}, {Berg}, {Jaskot}, {Schaerer}, {Worseck},
  {Amor{\'\i}n}, {Atek}, {Hayes}, {Ji}, {{\"O}stlin}, {Saldana-Lopez}, \&
  {Thuan}}]{Xu2023}
{Xu}, X., {Henry}, A., {Heckman}, T., {et~al.} 2023, \apj, 943, 94,
  \dodoi{10.3847/1538-4357/aca89a}

\bibitem[{{Yang} {et~al.}(2017){Yang}, {Malhotra}, {Gronke}, {Rhoads},
  {Leitherer}, {Wofford}, {Jiang}, {Dijkstra}, {Tilvi}, \& {Wang}}]{Yang17}
{Yang}, H., {Malhotra}, S., {Gronke}, M., {et~al.} 2017, \apj, 844, 171,
  \dodoi{10.3847/1538-4357/aa7d4d}

\bibitem[{{Zabl} {et~al.}(2021){Zabl}, {Bouch{\'e}}, {Wisotzki}, {Schaye},
  {Leclercq}, {Garel}, {Wendt}, {Schroetter}, {Muzahid}, {Cantalupo},
  {Contini}, {Bacon}, {Brinchmann}, \& {Richard}}]{Zabl2021}
{Zabl}, J., {Bouch{\'e}}, N.~F., {Wisotzki}, L., {et~al.} 2021, \mnras, 507,
  4294, \dodoi{10.1093/mnras/stab2165}

\bibitem[{{Zheng} {et~al.}(2010){Zheng}, {Cen}, {Trac}, \&
  {Miralda-Escud{\'e}}}]{Zheng2010}
{Zheng}, Z., {Cen}, R., {Trac}, H., \& {Miralda-Escud{\'e}}, J. 2010, \apj,
  716, 574, \dodoi{10.1088/0004-637X/716/1/574}

\bibitem[{{Zheng} {et~al.}(2011){Zheng}, {Cen}, {Trac}, \&
  {Miralda-Escud{\'e}}}]{Zheng2011}
---. 2011, \apj, 726, 38, \dodoi{10.1088/0004-637X/726/1/38}

\end{thebibliography}
\bibliographystyle{aasjournal}



\end{document}